\documentclass[12pt]{article}

\usepackage{amssymb,amsmath,amsfonts,eurosym,geometry,ulem,graphicx,caption,color,setspace,sectsty,comment,footmisc,caption,pdflscape,subfigure,array,placeins,dcolumn}

\usepackage{appendix}

\usepackage[english]{babel}  
\usepackage[colorlinks=true,linkcolor=blue,citecolor=blue]{hyperref}

\addto\extrasenglish{

}

\usepackage[utf8]{inputenc}
\usepackage{enumitem}
\usepackage{tabularx}
\usepackage{caption}
\usepackage{booktabs} 
\usepackage[super]{nth}
\usepackage{titlesec}
\usepackage[flushleft]{threeparttable}
\usepackage{soul}
\usepackage{txfonts}
  
\usepackage[ 
backend=biber,%
maxcitenames=2,%
mincitenames=1,%
style=authoryear,%
citestyle=authoryear,%
natbib=true,%
url=false,%
uniquename=false,%
uniquelist=false,%
doi=true,%
eprint=false%
]{biblatex}
\appto{\citesetup}{\color{blue}}

\newcolumntype{L}[1]{>{\raggedright\let\newline\\arraybackslash\hspace{0pt}}m{#1}}
\newcolumntype{C}[1]{>{\centering\let\newline\\arraybackslash\hspace{0pt}}m{#1}}
\newcolumntype{R}[1]{>{\raggedleft\let\newline\\arraybackslash\hspace{0pt}}m{#1}}

\newcommand*\dd{\text{d\kern 0.03em}}
\newcommand*\dz{\text{d\kern 0.05em}z_t}
\newcommand*\dq{\text{d\kern 0.05em}z_t^{\mathbb{Q}}}

\definecolor{ao(english)}{rgb}{0.0, 0.5, 0.0}
\newcommand{\hi}[1]{} %needs review
\newcolumntype{d}{D{.}{.}{-3}}
\newcommand{\specialcell}[2][c]{\begin{tabular}[#1]{@{}c@{}}#2\end{tabular}} % used for multi-line cells in the tables exported from R

\titleformat{\section}{\normalfont\large\centering}{\bfseries {\bfseries\thesection.}}{0.5em}{\MakeUppercase}
\titlespacing*{\section}{0pt}{2em}{0.25em}

\titleformat{\subsection}{\normalfont\bfseries\centering}{\thesubsection.}{0.5em}{}
\titlespacing*{\subsection}{0pt}{1em}{0.3em}{}

\titleformat{\subsubsection}{\normalfont\itshape\normalsize\centering}{\thesubsubsection.}{0.5em}{}
\titlespacing*{\subsubsection}{0pt}{1em}{0.3em}{}

\def\checkversion{1}
\newcommand\versioned[2]
{%
    \ifnum\checkversion=#1
        #2%
    \else%
        \textbf{{\color{red}[#2]}}%
    \fi%
}

\defcitealias{defradata}{DEFRA}

\newcommand\citepwithalias[1]{(\citetalias{#1}, \citeyear{#1})}

\newcommand{\pvarDescriptivesCbPopulationEnglandShare}{35\ } % In percent. 1951 census. Share of population in CBs relative to total population in England. (descriptives_cb.r Tir Jun 18 09:45:38 2024)
\newcommand{\pvarConstructGroupsNumberOfCBsWithBorderIrregularity}{15\ } % Number of CBs that we drop as Nanna classified them as having border irregularities and hence may not have collected the boundaries for the SCAs inside. (construct_groups.r Tue Jan 23 16:18:15 2024)
\newcommand{\pvarConstructGroupsNumberOfCBsWithoutBorderIrregularity}{70\ } % Number of CBs that do not have border irregularities. (construct_groups.r Tue Jan 23 16:18:15 2024)
\newcommand{\pvarConstructGroupsNumberOfCBsAdopting}{55\ } % Number of CBs that had a SCA before 1974. (construct_groups.r Tue Jan 23 16:18:15 2024)
\newcommand{\pvarConstructGroupsNamesOfCBsAdopting}{Barnsley, Birkenhead, Birmingham, Blackburn, Bolton, Bootle, Bradford, Bristol, Burnley, Burton Upon Trent, Bury, Canterbury, Coventry, Darlington, Dewsbury, Doncaster, Exeter, Gateshead, Gloucester, Halifax, Huddersfield, Kingston Upon Hull, Leeds, Leicester, Lincoln, Liverpool, Luton, Manchester, Newcastle Upon Tyne, Northampton, Norwich, Nottingham, Oldham, Oxford, Portsmouth, Preston, Reading, Rochdale, Rotherham, Salford, Sheffield, Solihull, South Shields, Southampton, Southport, St Helens, Stockport, Stoke on Trent, Sunderland, Tynemouth, Wakefield, Wallasey, Warrington, Wigan, York\ } % Names of CBs that had a SCA before 1974. (construct_groups.r Tue Jan 23 16:18:15 2024)
\newcommand{\pvarConstructGroupsNumberOfCBsNonAdopting}{15\ } % Number of CBs that did not have a SCA before 1974. (construct_groups.r Tue Jan 23 16:18:15 2024)
\newcommand{\pvarConstructGroupsNamesOfCBsNonAdopting}{Barrow-in-Furness, Bath, Blackpool, Bournemouth, Brighton, Carlisle, Chester, Eastbourne, Great Yarmouth, Grimsby, Hastings, Ipswich, Plymouth, Southend-on-Sea, Worcester\ } % Names of CBs that did not have a SCA before 1974. (construct_groups.r Tue Jan 23 16:18:15 2024)
\newcommand{\pvarConstructGroupsNamesOfCBsWithBorderIrregularity}{Croydon, Derby, Dudley, East Ham, Hartlepool, Middlesbrough, Smethwick, Teesside, Torbay, Walsall, Warley, West Bromwich, West Ham, West Hartlepool, Wolverhampton\ } % Names of CBs that we drop as Nanna classified them as having border irregularities and hence may not have collected the boundaries for the SCAs inside. (construct_groups.r Tue Jan 23 16:18:15 2024)
\newcommand{\pvarConstructStationsObsPollutionBSAllEnglandWales}{160,833\ } % Total BS obs. in pollution panel for England/Wales 1954--1973. (construct_stations.r Thu May 16 17:39:25 2024)
\newcommand{\pvarConstructStationsObsPollutionSOAllEnglandWales}{156,395\ } % Total SO obs. in pollution panel for England/Wales 1954--1973. (construct_stations.r Thu May 16 17:39:25 2024)
\newcommand{\pvarConstructStationsObsPollutionBSCBs}{52,234\ } % Total BS obs. in pollution panel for CBs in England/Wales 1954--1973. (construct_stations.r Thu May 16 17:39:25 2024)
\newcommand{\pvarConstructStationsObsPollutionSOCBs}{50,803\ } % Total SO2 obs. in pollution panel for CBs in England/Wales 1954--1973. (construct_stations.r Thu May 16 17:39:25 2024)
\newcommand{\pvarConstructStationsObsPollutionTCPostSixtyTwoBS}{35,145\ } % Total obs. in pollution panel for black smoke in treated/control CBs 1962--1973. (construct_stations.r Thu May 16 17:39:25 2024)
\newcommand{\pvarConstructStationsObsPollutionStationsTCPostSixtyTwoBS}{401\ } % Total stations in pollution panel for black smoke in treated/control CBs 1962--1973. (construct_stations.r Thu May 16 17:39:25 2024)
\newcommand{\pvarConstructStationsObsPollutionTCPostSixtyTwoSulphur}{34,951\ } % Total obs. in pollution panel for so2 in treated/control CBs 1962--1973. (construct_stations.r Thu May 16 17:39:25 2024)
\newcommand{\pvarConstructStationsObsPollutionStationsTCPostSixtyTwoSulphur}{399\ } % Total stations in pollution panel for so2 in treated/control CBs 1962--1973. (construct_stations.r Thu May 16 17:39:25 2024)
\newcommand{\pvarConstructIndividualsObsIndividualsAll}{444,707\ } % All UKB individuals (construct_individuals.r Tue May 14 13:20:39 2024)
\newcommand{\pvarConstructIndividualsObsIndividualsPostRosla}{113,704\ } % All UKB individuals born after 1957 (RoSLA) and by end of 1969 (construct_individuals.r Tue May 14 13:20:45 2024)
\newcommand{\pvarConstructIndividualsObsIndividualsPostRoslaTC}{41,329\ } % All UKB individuals born after 1957 (RoSLA) and by end of 1969 in treated/control CBs (construct_individuals.r Tue May 14 13:20:45 2024)
\newcommand{\pvarConstructIndividualsObsIndividualsPostRoslaTCNoBunch}{16,573\ } % All UKB individuals born after 1957 (RoSLA) and by end of 1969 in treated/control CBs that do NOT have a bunched birth location. (construct_individuals.r Tue May 14 13:20:45 2024)
\newcommand{\pvarConstructIndividualsObsIndividualsPostRoslaTCLeast}{5,749\ } % All UKB individuals born after 1957 (RoSLA)  and by end of 1969 in treated/control CBs that do NOT have a bunched birth location. for outcome with least non-missing observations. (construct_individuals.r Tue May 14 13:20:45 2024)
\newcommand{\pvarConstructIndividualsObsIndividualsPostRoslaTCMost}{16,535\ } % All UKB individuals born after 1957 (RoSLA) and by end of 1969 in treated/control CBs that do NOT have a bunched birth location. for outcome with most non-missing observations. (construct_individuals.r Tue May 14 13:20:45 2024)
\newcommand{\pvarAnalysisStationsEventObsAverageWait}{16\ } % Rounded average number of months between submission and operation for the SCAs in which we have at least one station in our pollution analysis sample. (analysis_stations_event.r Tir Jun 11 21:27:43 2024)
\newcommand{\pvarAnalysisRandomnessNumberCbsNotReachTenPctCoverage}{3\ } % Number of districts that do not reach the 10 percent of area covered milestone and is thus dropped in the randomness appendix (analysis_randomness.r Tir Jun 11 23:01:33 2024)
\newcommand{\pvarAnalysisRandomnessNamesCbsNotReachTenPctCoverage}{Norwich, Portsmouth, Southport\ } % Names of districts that do not reach the 10 percent of area covered milestone and is thus dropped in the randomness appendix (analysis_randomness.r Tir Jun 11 23:01:33 2024)
\newcommand{\SCA}{\text{SCA}}
\newcommand{\ijt}{i \! jt}

\begin{document}

\title{The long-term human capital and health impacts of a pollution reduction programme}
\author{Nanna Fukushima\thanks{VTI, The Swedish National Road and Transport Research Institute. E-mail: \href{mailto:Nanna.Fukushima@vti.se}{Nanna.Fukushima@vti.se}} \and Stephanie von Hinke\thanks{School of Economics, University of Bristol; Institute for Fiscal Studies. E-mail: \href{mailto:S.vonHinke@bristol.ac.uk}{S.vonHinke@bristol.ac.uk}} \and Emil N. S\o{}rensen\thanks{School of Economics, University of Bristol. E-mail: \href{mailto:E.Sorensen@bristol.ac.uk}{E.Sorensen@bristol.ac.uk}} }

\date{\today\thanks{We gratefully acknowledge financial support from the European Research Council (Starting Grant Reference 851725). We thank Samuel Baker for digitizing historical data on the nationwide introduction dates of smoke control areas, and Pravin Piramanayagam and Arani Heiyanthuduwa for excellent research assistance. We also thank seminar participants at the 2024 European Social Science Genetics Network (ESSGN) Conference and the 2024 UAM-UJI `Workshop on the Economics of Health and Human Capital' for their helpful comments and suggestions. This work is based on data provided through www.visionofbritain.org.uk and uses historical material which is copyright of the Great Britain Historical GIS Project and the University of Portsmouth. This work also uses data from Department for Environment, Food, and Rural Affairs (DEFRA) and the Devolved Administrations which are Crown copyright and licensed under the Open Government Licence \url{http://www.nationalarchives.gov.uk/doc/open-government-licence/version/2/}.}}
\maketitle

\begin{abstract}
\singlespacing
\noindent 
This paper investigates the effects of the staggered roll-out of a pollution reduction programme introduced in the UK in the 1950s. The policy allowed local authorities to introduce so-called ``Smoke Control Areas'' (SCAs) which banned smoke emissions. We start by digitizing historical pollution data to show that the policy led to an immediate reduction in black smoke concentrations. 
We then merge data on the exact location, boundary and month of introduction of SCAs to individual-level outcomes in older age using individuals' year-month and location of birth. We show that exposure to the programme increased individuals' birth weights as well as height in adulthood. We find no impact on their years of education or fluid intelligence.

\vspace{3mm}

\noindent \textbf{Keywords:} Smoke control areas; Developmental origins; Staggered treatment effects
\vspace{1mm}\newline
\textbf{JEL Classifications:} I18, I15, C21, C22
\end{abstract}

\clearpage
\doublespacing

\section{Introduction}
The early-life environment is crucial in shaping individuals' outcomes, with potentially life-long, irreversible impacts in older age. 
The existing ``Developmental Origins'' literature has mostly focused on the long-term impacts of adverse nutritional, health and economic conditions during the prenatal and early childhood period \citep[for reviews, see e.g.,][]{AlmondCurrie2011a, 
almond2018childhood}. There is relatively little empirical evidence however, on the prolonged and cumulative consequences of early-life pollution exposure, despite the well-known adverse contemporaneous effects on infants' health \citep[see e.g.,][for a review]{graff2013environment}. One of the main reasons for this is a lack of high-quality historical pollution data. Evidence on the long-term effects however, is important, since ignoring these substantially underestimates the total welfare effects caused by exposure to environmental toxins. 

This paper addresses this directly, estimating the immediate and long-term impacts of a UK pollution reduction programme introduced in the mid 1950s. We deal with the general lack of historical pollution data by digitising local monthly measurements of black smoke and sulphur dioxide for a 20 year period, covering 1954 to 1973. We then exploit the staggered introduction of so-called ``Smoke Control Areas'' (SCAs) -- i.e., zones introduced by local authorities that banned smoke emissions from residential as well as non-residential dwellings.  

We study both (i) the effect of the introduction of SCAs on local pollution levels, and (ii) their (long-term) impact on individuals' human capital and health. The former relies on our digitization of historical monthly pollution measurements. The latter uses UK Biobank data: a large population-based cohort of approximately 500,000 individuals living in the United Kingdom for whom we observe their year-month of birth, as well as their eastings and northings of birth. The data also include rich information on individuals' later life health and economic outcomes, linked to administrative records. We focus on those born in England and merge in the exact location, boundary, and month of introduction of all SCAs introduced in English County Boroughs (CBs), obtained from \citet{fukushima2021uk}. CBs are local authorities with administrative autonomy due to their population size or historical significance (i.e., relatively urban areas).  
Our identification strategy exploits the spatial and time variation in the staggered roll-out of Smoke Control Areas using an event study approach and we show that our results are robust to using group-time average effects that account for staggered treatment \citep{sunabraham2021, callawaySantanna2021, borusyakJaravelSpiess2024}. In other words, we compare measurement stations (individuals) located (born) in or downwind of smoke control areas before and after its creation, relative to those of a control group of never-treated stations (individuals), whilst controlling for weather variation, local area-specific trends in the outcomes of interest (and individual-level controls). 

With that, we provide one of the first estimates of the dynamic and time-varying impact of the introduction of Smoke Control Areas in the 1950s and 1960s on local pollution levels. We also explore the determinants of the \textit{timing} of SCA introduction, and investigate whether areas selected to become smokeless were systematically different from those not selected. We show that -- as expected -- more densely populated CBs introduced SCAs earlier. We find no evidence however, that the timing of SCA introduction is related to the socio-economic composition of local areas, nor to pre-programme pollution levels, suggesting that the \textit{timing} of SCA introduction is largely unrelated to local area characteristics.  We do find that areas that were selected to become smokeless during this period were different from areas not selected within the same CB, in that they were slightly less polluted and of higher SES prior to SCA introduction. We describe these differences in detail and directly account for them in our analyses. 

We then examine the \textit{immediate} as well as \textit{long-term} impacts of the introduction of Smoke Control Areas on individuals. Whilst there is a large literature on the contemporaneous effects of pollution on children's birth outcomes, we are aware of only a handful of papers that empirically examine the causal impacts of early-life pollution exposure on outcomes in older age.\footnote{There is also an increasing literature studying the effects of pollution on outcomes in childhood or early adulthood, showing negative impacts on e.g., human capital formation, labour force participation and wages \citep{sanders2012doesn, isen2017every, bharadwaj2017gray, persico2020can, persico2021effects, heissel2022does}.} All of these exploit the 1952 London smog as an exogenous pollution event and show that those exposed to the smog are less likely to have a degree, work fewer hours, have lower fluid intelligence and are more likely to have developed respiratory disease \citep[][]{bharadwaj2016early, Ball2018, von2023long, martin2024effect}. 

Compared to this literature that exploits a single extreme event, we investigate the impact of a large-scale pollution reduction programme, the aim of which was to improve air quality and with that, public health. This is similar to \citet{isen2017every}, who exploit the introduction of the 1970 Clean Air Act Amendments that forced US counties with pollution levels that exceeded maximum concentrations to reduce their emissions but left others unaffected.
As we discuss in more detail in \autoref{sec:background}, the 1956 UK Clean Air Act that allowed for the introduction of SCAs was one of the most comprehensive and influential pieces of legislation of its time and marked a significant step in national air pollution control. It was among the first to address industrial as well as domestic sources of air pollution on a national scale. The introduction of SCAs required all coal burning to be replaced with smokeless fuels, which was considered an innovative approach to reducing pollution from residential dwellings; a major source of pollution at the time. The Clean Air Act was also one of the first pieces of legislation to clearly link air pollution to public health and to take steps to mitigate its impacts. It subsequently served as a model for other countries to tackle air pollution and protect public health (e.g., the 1963 US Clean Air Act). We provide the first empirical analysis of the long-term consequences of the Act.

Our findings show that the roll-out of Smoke Control Areas substantially and persistently reduced black smoke emissions for at least five years post-introduction. We find no consistent impacts on sulphur dioxide. This is not surprising since, first, SCAs targeted black smoke only and second, bituminous coal (the predominant form of heating at the time) and smokeless fuel (its replacement) are known to release similar amounts of sulphur dioxide \citep{mitchell2016impact}.\footnote{The vast majority ($>$98$\%$) of particles emitted from coal combustion are smaller than 2.5 micrometers in diameter (also referred to as PM$_{2.5}$); sufficiently small to penetrate the lung system and reach the blood circulation and is therefore considered particularly harmful. Black Smoke consists of fine particulate matter and is emitted mainly from fuel combustion. Since pollution from road traffic was still minimal at the time, the vast majority of PM$_{2.5}$ would have come from coal. } When studying the impacts on individuals' health and human capital outcomes, we find that the introduction of SCAs led to an increase in weight at birth as well as height in adulthood. In contrast, we find no consistent evidence of impacts on years of education and intelligence, though our results are suggestive of some heterogeneity by SES, with these outcomes more likely to improve in high SES areas. These findings are robust to a host of sensitivity checks, but they are in contrast with the studies above that find long-term deteriorations of human capital in response to pollution exposure. 

We highlight two possible reasons for this. First, rather than exploring a huge but transitory pollution spike (i.e., the London smog), we examine the long-term consequences of smaller but persistent changes in pollution exposure. It is possible that such different patterns of exposure differentially impact individuals' short as well as longer-term outcomes. Second, the different estimates may reflect differences in exposure to specific pollutants. Indeed, the 1952 London smog dispersed a range of toxins, including tar, carbon monoxide, carbon dioxide and sulphuric acid \citep{wilkins1954air}. 
In contrast, the introduction of SCAs targeted black smoke \textit{only}, since a reduction in gaseous pollutants was not considered feasible at the time. The evidence suggests that the main change in moving from bituminous coal to smokeless fuel is a drop in particulate matter, with no changes in e.g., nitrogen oxides (including nitrogen dioxide), sulphur oxides (including sulphur dioxide) and carbon monoxide \citep{mitchell2016impact}. Indeed, one of the strengths of this study is that we directly estimate this ``first stage'', showing clear reductions in black smoke concentrations but not sulphur dioxide. 
Our findings therefore suggest that pollutants other than black smoke may be responsible for the adverse effects on human capital, whereas black smoke affects fetal and child growth.

We explore three sources of heterogeneity in the estimated impact of SCAs. In addition to heterogeneity by gender and SES, we examine whether the impact of SCAs varies by individuals' genetic ``endowments'', as measured by one's polygenic score (also known as a polygenic index) that is specific to the outcome of interest. We consider genetic heterogeneity for three reasons. First, we are interested in whether the significant public health investment in smoke control affected genetic inequalities. Gene-by-environment analysis allows us to explore this empirically. Although we cannot change our genetic make-up, we \textit{can} change the environment. Hence, understanding the extent to which (local) government policies affect population subgroups differently informs us about potential impacts on inequalities in relevant (health and economic) outcomes. 
Second, exploring gene-by-environment interplay helps us improve our understanding of the health and human capital production function. This literature emphasizes the role of complementarities between endowments and investments, suggesting that individuals with higher endowments benefit more from subsequent investments \citep{becker1986human, cunha2007technology}. \citet{muslimova2020nature} suggest the use of genetic information to capture such endowments, exploring its interaction with (a proxy for) exogenous parental investments.\footnote{Such complementarity analysis would ideally use parent-child trio or sibling data \citep[as in][]{muslimova2020nature} to allow one to also isolate \textit{exogenous} endowments. However, although there are around 40,000 siblings in the UK Biobank, our analysis sample only contains about 160 sibling pairs, making within-sibling analysis infeasible.} We follow this literature, but instead examine the interplay with \textit{public health} investments. As such, positive gene-by-environment interactions are consistent with such complementarities. Third, finding evidence of gene-by-environment interplay provides evidence against arguments of genetic or environmental determinism. This in turn is crucial in the debate about whether one's success in life is due to efforts or circumstance \citep[see e.g.,][]{roemer1993pragmatic, roemer1996theories}. 

Our genetic heterogeneity analyses provide suggestive evidence that the introduction of SCAs increased inequalities in population health but not in economic outcomes. Indeed, we find that the introduction of SCAs led to larger increases in birth weight and height among those with a high polygenic score for these outcomes. 
These results are consistent with the existence of complementarities in the health production function \citep[see also][]{berg2023early}, but not the human capital production function \citep[as in][]{muslimova2020nature}.

The rest of the paper is structured as follows. \autoref{sec:background} provides the background to roll-out of Smoke Control Areas in England and \autoref{sec:data} describes the data sources used in our analyses. We set out the empirical strategy in \autoref{sec:empirical_strategy}, and discuss the results in \autoref{sec:results}. We explore the sensitivity of our findings in \autoref{sec:overview_robustness} and conclude in \autoref{sec:conclusion}.

\section{Background}\label{sec:background}
One of the oldest accounts on the impact of smoke on health was addressed to King Charles II in 1661 in a treatise called ``Fumifugium; or The Inconvenience of the Aer and Smoake of London Dissipated'' \citep{Evelyn1661}. This suggested that smoke pollution shortened the lives of Londoners. Despite this, 
the industrial revolution, the massive migration of workers to urban areas, and the overall increase in population meant that the smoke problem would continue to escalate for many centuries. Only after the Great London Smog in December 1952, which brought premature death to thousands of citizens, lawmakers and the public became fully aware of the potential damage of smoke, which led to the swift passing of the Clean Air Act in 1956 \citep{cleanAirAct1956}.

Before the Clean Air Act, the UK's primary source of air pollution was emission from burning bituminous coal, with coal fires being the predominant form of heating in most dwellings far into the 1960s. The 1956 Clean Air Act consisted of two main parts. In the first part, the Act prohibits the emission of dark smoke from all buildings. The Act defines smoke as fly ash, grit, and gritty particles, and the shade of the smoke was determined by comparing the colour of the smoke against the Ringelmann Chart. The fine for breaching the law was a maximum of ten pounds for a private dwelling and 100 pounds for all other buildings. While the regulation contributed to improving the air quality in the nation, many exemptions to the law and generous lead times of up to seven years for industries hampered its full potential. In contrast, the second part of the law gave local authorities across the UK the mandate to introduce so-called Smoke Control Areas (SCAs), which banned any smoke emissions inside the area. At the start, the fine for violating a smoke control order carried a penalty of up to ten pounds per offense which increased to 20 pounds in the revised 1968 Clean Air Act. 
It is important to note that, while the Act regulates the emission of smoke, it does not target gaseous pollutants that are present in coal such as sulfur dioxide (SO$_{2}$). 
In fact, the reason for \textit{not} targeting SO$_2$ emissions was that its elimination was thought unattainable at the time since SO$_{2}$ is equally present in bituminous and smokeless fuel.\footnote{SO$_2$ is formed by the oxidation of sulfur in fuel combustion. It can cause direct harm to health by damaging the lung capacity and indirectly as a secondary particulate matter when reacting to other airborne particulate matters. }

Local authorities had to apply to the Ministry of Housing and Local Government to implement a Smoke Control Area. This application process had a number of `milestone dates' \citep{smokeless1956}. First, the local council would survey the area, propose a boundary and operation date, and announce it to the public in a local newspaper as well as in the \textit{London Gazette}.\footnote{Local authorities were free to decide SCA's boundaries and operation dates but were required to provide a minimum of six months' notice before operation.} The `submission date' refers to the date on which the proposal was submitted for review to the Ministry of Housing and Local Government, usually leading to the Ministry confirming, perhaps with adjustments, the agreed date of operation and boundary for the proposed SCA. The SCA would then come into effect at the `operation date', after which the monitoring and enforcement measures would start. 

To comply with a smoke control order, the dwelling owner could substitute bituminous coal for (manufactured) smokeless fuel such as anthracite or gas (if available). For older dwellings, this typically required owners to adapt their appliances to new fuel types; 70\% of the costs associated with such conversions were reimbursed by the local council, as long as the adjustment work was done before the `operation date' \citetext{\citealp{smokeless1958}; \citealp{smokeless1960}}. 
The local council would in turn receive a 40\% (four-seventh of 70\%) contribution from the exchequer to cover these additional costs; the other 30\% was borne by the council. The total costs for these adaptations were not negligible, highlighting the willingness of local authorities and central government to control pollution levels despite significant expenses associated with them.\footnote{For example, Sheffield local authority spent approximately 2.3 million pounds (26 million pounds in 2023 prices) on reimbursement of adjustment work to complete the district's smoke control programme during 1957--1972 \citep{mohSheffield1972}.}

\section{Data and descriptive statistics}\label{sec:data}
We combine four sources of data for our main analysis: (1) exact locations, boundaries and `milestone dates' for all smoke control areas introduced in County Boroughs between 1957 and 1973, (2) weather data for the same period, (3) monthly measurements of black smoke and sulphur dioxide for the years 1954-1973, and (4) individual-level data from the UK Biobank that includes residential location at birth as well as individuals' human capital and health outcomes at older ages for cohorts born before and after the introduction of smoke control areas. We discuss each of these in more detail below. To differentiate between the data used to explore the impact on \textit{levels of pollution} and the effect on \textit{individual outcomes}, we distinguish between the ``pollution panel'' and the ``individual-level data'' respectively.

\subsection{Smoke Control Areas}\label{sec:Data_SCA}
We use the data from \citet{fukushima2021uk} that record the exact location, boundaries and year-month of submission and operation for all 1,027 smoke control areas that were introduced in so-called County Boroughs (CBs) across England by 1973. These data come from multiple sources: local historical archives, 
notices in historical editions of the \emph{London Gazette}, relevant \textit{Medical Officer of Health} reports, as well as historical issues of ``\textit{Smokeless Air}''; a quarterly publication by the National Smoke Abatement Society \citep[see, e.g.,][]{smokeless1958}. We restrict our attention to SCAs in CBs only, as these are predominantly residential areas. Indeed, although they only make up $\sim$3\% of the total land area, they cover over a third of the English population (\pvarDescriptivesCbPopulationEnglandShare\!\% according to the 1951 Census). 

\begin{figure}[!h]\caption{\label{uni_sca_statuses}Smoke control areas in England by year and status.}\centering\includegraphics[width=0.7\textwidth]{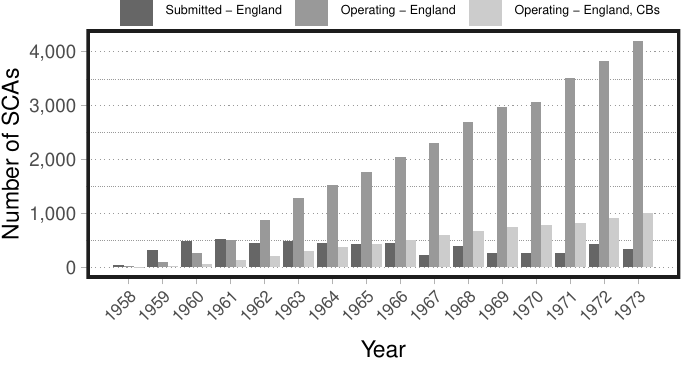}\caption*{\emph{Plots the number of submitted and operating smoke control areas in England by year. Also shows the subset of smoke control areas operating in CBs in England, as these are the ones we use in our analysis. The totals for England have been digitised from reports by the National Smoke Abatement Society.}}\end{figure}

We further collect data on the \textit{universe} of SCAs (rather than those in CBs only) introduced from September 1958 to December 1973. These are reported by National Smoke Abatement Society, but they are only available at a more aggregate level (year-quarter rather than year-month). More importantly, they do not provide SCA \textit{boundaries}, meaning we cannot pinpoint their exact location and shape within a CB. Our main empirical analysis therefore focuses on the exact boundaries and year-\textit{month} of submission and operation of SCAs in County Boroughs only.\footnote{We use the information on the \textit{universe} of SCAs in our robustness analysis to identify geographical areas outside CBs that never introduced any SCA within our period of interest; see \autoref{sec:definition_of_control_group}.} 
The more aggregated data however, are useful for descriptive purposes, showing the extent to which local governments engaged with the new legislation and introduced areas of smoke control. Indeed, \autoref{uni_sca_statuses} uses these data \citep{Baker2024a} to present the number of submitted SCAs in England by year.\footnote{Although the first smoke control areas were introduced in 1957, shortly after the passing of the 1956 Clean Air Act, their exact `milestone dates' were not recorded. We therefore only report the number of SCAs from 1958 onwards.} This shows a steady increase in the number of SCAs during this time, reaching over 4,000 by 1973. \autoref{uni_sca_spatial} in \autoref{sec:additional} plots the spatial distribution of SCAs across England in 1973. Although any local authority could submit an application for a new SCA, the figure shows that they are concentrated in and around urban areas such as County Boroughs. While this may suggest that more densely populated areas were more likely to implement SCAs, and to do so earlier than other districts, we show in \autoref{sec:appendix_timing} that conditional on population density, the timing of SCA implementation across CBs is not strongly or systematically driven by predetermined district-level characteristics (i.e., socioeconomic composition), nor by pre-treatment pollution levels.

When estimating the impact of the introduction of SCAs, we drop \pvarConstructGroupsNumberOfCBsWithBorderIrregularity\,out of the total 85 CBs because of significant boundary changes during our study period (e.g., total/partial splits and merges with neighbouring local authorities that make it impossible to consistently observe them over time). Of the remaining \pvarConstructGroupsNumberOfCBsWithoutBorderIrregularity\,CBs, \pvarConstructGroupsNumberOfCBsAdopting\,adopted at least one SCA before 1974, and \pvarConstructGroupsNumberOfCBsNonAdopting\,did not. We refer to these as ``adopting'' and ``non-adopting'' CBs, respectively.\footnote{The following \pvarConstructGroupsNumberOfCBsWithBorderIrregularity\,CBs are dropped due to significant boundary changes: \pvarConstructGroupsNamesOfCBsWithBorderIrregularity. \autoref{tab:cb_classification} lists the CBs remaining in our analyses by adoption status. The ``non-adopting'' CBs include those that either never adopted a SCA or only adopted one in greenfield land areas within their jurisdictions (i.e., undeveloped land used for agriculture or landscape design).}

\subsection{Weather}
Given the importance of the weather for pollution dispersion as well as individuals' health \citep[see e.g.][]{hanlon2021temperature}, we merge in daily data on wind speed, wind direction, precipitation, and temperature from the ERA5 reanalysis data \citep{era5}. These have an approximate grid resolution of 25km and we assign weather measurements to SCAs, stations, and individuals by linking them to the nearest grid point. For SCAs, we link to its centroid, while we use the actual measurement and birth locations for stations and individuals, respectively.

We use these weather data in two ways. First, we directly control for weather conditions in our analyses, averaging individuals' exposure to the weather conditions during the prenatal period and the first two years of life. Second, we use the historical weather data to identify stations that are \textit{downwind} from an SCA, and therefore potentially indirectly affected by its introduction. We discuss this in more detail below.

\subsection{Pollution}\label{sec:data_pollution}
To construct our ``pollution panel'', assessing the impact of the introduction of SCAs on local air pollution, we digitise six years of monthly pollution measurements and station locations from the \citet{DSIR}.\footnote{The pollution data from 1954-1960 comes from \citet{fukushima2021uk}; the 1961 pollution data are from \citet{Baker2024b}. The data were collected by the world's first coordinated national air pollution monitoring network: the UK Investigation of Atmospheric Pollution, run by the Warren Spring Laboratory. Historically, it monitored and compiled data on two main pollutants: black smoke and sulfur dioxide. Black smoke was measured using smoke samplers, drawing 50 cubic meters of air through a white filter paper over 24 hours. The density of the deposit was then assessed using a reflectometer. Sulphur dioxide was measured by drawing the same sample of air through a chemical solution that reacts with sulphur dioxide to form sulphuric acid. The measured acidity of the solution was then used to approximate the concentration of sulphur dioxide in the air sample. \citet{smokenetworkmanual} gives a detailed description of these processes in the context of the monitoring network.} We combine these with measurements and station locations for 1961--1973 \citepwithalias{defradata}, allowing us to construct a panel of monthly black smoke and sulphur dioxide levels taken at measurement stations across England between 1954 and 1973. Note that this is an unbalanced panel, with fewer measurement stations at the start of the observation period and new ones being introduced over time. Indeed, we do not observe any pollution measurements in non-adopting CBs until 1962 when the first measurement stations in these CBs were introduced; we come back to this below. 

We assign pollution stations to SCAs as well as CBs by projecting the stations' locations onto our digitised SCA boundaries and the 1971 CB boundaries \citep{GBHDGIS2011,southall2011rebuilding}, respectively. We define stations located inside a SCA as `treated', and stations in non-adopting CBs as `never-treated' (controls). When a CB is only partly covered by one (or more) SCA, we add stations that are \textit{inside} the CB but \textit{outside} the SCA boundaries to the group of `never-treated' stations. We show the robustness of our results to dropping these altogether in \autoref{sec:overview_robustness} below.

Starting from \pvarConstructStationsObsPollutionBSAllEnglandWales station-year-month observations for black smoke and \pvarConstructStationsObsPollutionSOAllEnglandWales for sulphur dioxide, we restrict our pollution panel as follows. First, since our SCA \textit{boundary data} are restricted to stations located in CBs, we drop those located outside CBs, reducing our sample to \pvarConstructStationsObsPollutionBSCBs and \pvarConstructStationsObsPollutionSOCBs observations for black smoke and sulphur dioxide respectively. Second, we include only stations in adopting or non-adopting CBs, dropping those with border irregularities (see \autoref{sec:Data_SCA}). Furthermore, since pollution is not measured in any of the non-adopting CBs until after 1961, the pollution panel covers 1962 onwards, leaving us with \pvarConstructStationsObsPollutionTCPostSixtyTwoBS station-year-month observations (\pvarConstructStationsObsPollutionStationsTCPostSixtyTwoBS stations) for black smoke and \pvarConstructStationsObsPollutionTCPostSixtyTwoSulphur station-year-month observations (\pvarConstructStationsObsPollutionStationsTCPostSixtyTwoSulphur stations) for sulphur dioxide.

\begin{figure}[!h]\caption{\label{fig:pollution_timeseries_treated_control}Historical measurements of pollution (black smoke and $\text{SO}_2$) by station treatment status.}\centering\includegraphics[width=\textwidth]{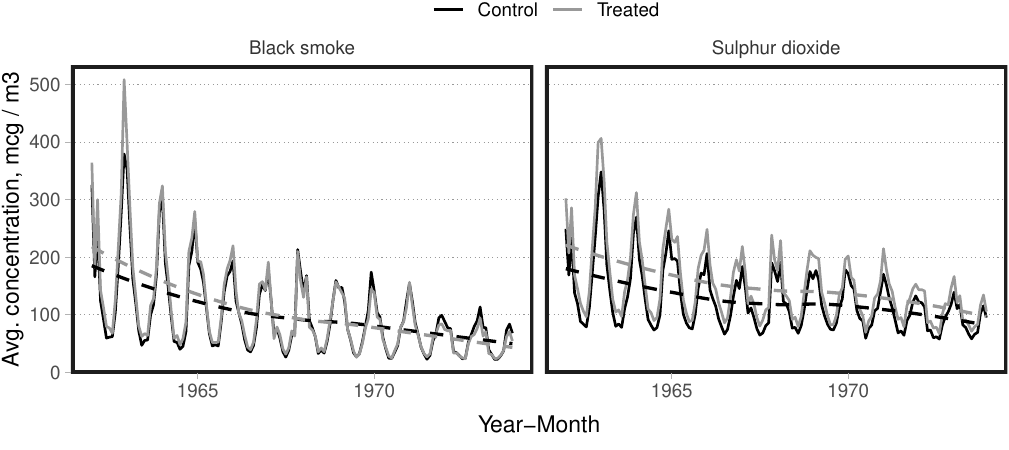}\caption*{\emph{Historical measurements of pollution (black smoke and $\text{SO}_2$) from 1962 to 1973 averaged over stations by treatment status and year-month. Average for treated stations in grey and average for control stations in black. Dashed lines show long term trends by group using locally weighted polynomial regression (LOESS).}}\end{figure}

\autoref{fig:pollution_timeseries_treated_control} plots the trends in air pollution during the roll-out of the smoke control programme, plotting the monthly black smoke (left panel) and sulphur dioxide (right panel) levels as measured at treated and control stations; that is, stations located respectively inside or outside the \textit{exact} boundaries of all smoke control areas in English County Boroughs.
The graphs highlight three key points. First, it shows clear seasonality in pollution, with peaks in both black smoke and sulphur dioxide during the winter months, and lower levels during the summer. Second, there is a reduction in air pollution levels measured at both treated and control stations over time, as shown by the smoothed averages (dashed lines). And third, the decrease in levels of black smoke was larger at treated stations compared to controls, resulting in the average black smoke levels converging over time. For sulphur dioxide, the gap between treated and control stations only narrowed slightly, suggesting that the adoption of smoke control areas had a smaller impact on levels of SO$_2$. 

In addition to defining treated pollution measurement stations as those located \textit{inside} a SCA, we also construct an indicator for stations that are \textit{downwind} of a SCA. Using the historical wind direction data from ERA5, we construct vectors of SCAs' prevailing wind directions in the two years prior to their submission, and use them to simulate the boundaries of pollution dispersion from the SCAs. We then classify as downwind any stations that are located within the dispersion boundary but outside the boundary of the originating SCA.  We detail this procedure, and explore the robustness of our estimates to alternative approaches, in \autoref{sec:appendix_area_and_downwind}.

\subsection{Individuals}
To examine the long-term impact of the introduction of SCAs on individuals' human capital and health outcomes, we use the UK Biobank: a large prospective, population-based cohort living in the United Kingdom. Baseline information on approximately 500,000 individuals was collected between 2006 and 2010, when they were 40--69 years old. Participants are born between 1938 and 1971, with the majority in the 1940s to mid 1960s. The UK Biobank is not representative, with individuals being on average healthier and wealthier than the general UK population \citep{Fry2017}. The data include detailed information on demographics, physical and mental health, cognition, and economic outcomes, obtained via interviews, questionnaires, and measurements taken by nurses. The data have also been linked to hospital and mortality records. 

We use individuals' location of birth (i.e., eastings and northings) to assign them to SCAs and CBs by projecting their birth locations onto our digitised SCA boundaries and the 1971 CB boundaries, respectively. We classify individuals born in CBs that never introduced an SCA as `never-treated' (controls), and those born inside SCA boundaries as treated, depending on their dates of birth. Analogous to the sample construction for the pollution panel, when a CB is only partially covered by one (or more) SCAs, we define individuals born \textit{inside} the CB but \textit{outside} the SCA boundaries as never treated and explore the robustness of our results to dropping the latter group altogether.

To study the consequences of smoke control on human capital and health production, we focus on four broad outcomes motivated by the existing literature on the early-life pollution environment. First, building on the literature on pollution exposure and individuals’ health \citep[see e.g.,][]{currie2011traffic}, we rely on birth weight and adult height as two general indicators of early- and later-life health and development, allowing us to examine both the short and long-term impacts of pollution along these dimensions. Second, following the literature on the effects of pollution exposure and human capital production \citep[see e.g.,][]{isen2017every, Ball2018, persico2020can, von2023long}, we explore individuals’ years of education and fluid intelligence. 

Our measure of birth weight is self-reported and therefore likely to be measured with error, while adult height was measured by a nurse following a standardised protocol.\footnote{Despite birth weight being self-reported at a later age, it has been shown to correlate with a range of covariates in the expected direction (e.g., with non-singleton pregnancies, gender, maternal smoking) and it has good reliability \citep{tyrrell2013parental, zhang2021birth}.} We define years of education using individuals’ qualifications, and measure fluid intelligence using a battery of questions designed to measure logic and reasoning ability, independent of acquired knowledge. We standardise the latter with mean zero and standard deviation one.\footnote{\autoref{tab:years_educ_definition} shows how we map qualifications to years of education using a definition similar to that of, e.g., \citet{rietveld2013, okbay2022}.}

Starting with \pvarConstructIndividualsObsIndividualsAll individuals with non-missing data on birth location and year-month of birth, we restrict our estimation sample as follows. First, we drop individuals who are born before September 1957, as this coincides with the UK education reform that raised the minimum school leaving age, affecting individuals' education, income and potentially health outcomes \citep{harmon1995estimates, clark2013effect, davies2018causal}.\footnote{Note that although the pollution analysis is restricted to the years 1962 onwards due to the absence of pollution data in never-treated CBs pre-1962, we do not apply this restriction to the individual-level analysis. Indeed, we obtain intention-to-treat estimates for the latter, examining the impact of the introduction of SCAs, as opposed to estimating the direct effect of a reduction in pollution. Hence, we use those born between September 1957 and the end of 1969 for the individual-level analysis.} We also drop the last two birth cohorts born in 1970 and 1971 as they are more selected and very small compared to earlier cohorts \citep{berg2023early}. This leaves us with \pvarConstructIndividualsObsIndividualsPostRosla participants. Second, since we observe \textit{exact} Smoke Control Area boundaries in County Boroughs only, we restrict our sample to individuals born in adopting or non-adopting CBs, reducing our sample to \pvarConstructIndividualsObsIndividualsPostRoslaTC individuals. Third, we restrict our sample to individuals with a precisely measured birth location, leaving \pvarConstructIndividualsObsIndividualsPostRoslaTCNoBunch individuals.\footnote{UK Biobank participants who could not report their birth location with any precision (e.g., areas \textit{within} a larger town or city) were assigned a catch-all location in the approximate centre of the town or city. We drop these individuals since we rely on relatively precise birth location reports, e.g., distinguishing between two individuals born in the same CB but one inside and the other outside an SCA. 
However, our results are robust to including them in our estimation sample (see \autoref{sec:robustness_bunching}). Note that birth locations are reported with 1km resolution.} Our final sample size then ranges between \pvarConstructIndividualsObsIndividualsPostRoslaTCLeast and \pvarConstructIndividualsObsIndividualsPostRoslaTCMost depending on the number of missing values for our outcome of interest. In the robustness analysis, we also use an auxiliary sample where we take our data on the \textit{universe} of SCAs to identify individuals born in districts that \textit{never} introduced any SCA and include these participants in the never-treated group, increasing the sample size to between 9,868 and 25,410 individuals, depending on the frequency of missing data on the outcome.

\begin{table}
\centering
\caption{\label{tab:descriptives}Descriptives}
\centering
\fontsize{9}{11}\selectfont
\begin{threeparttable}
\begin{tabular}[t]{lrrrrrrrrr}
\toprule
\multicolumn{1}{c}{\em{}} & \multicolumn{3}{c}{\em{Full sample}} & \multicolumn{3}{c}{\em{Treated}} & \multicolumn{3}{c}{\em{Control}} \\
\cmidrule(l{3pt}r{3pt}){2-4} \cmidrule(l{3pt}r{3pt}){5-7} \cmidrule(l{3pt}r{3pt}){8-10}
\multicolumn{1}{c}{} & \multicolumn{1}{c}{Mean} & \multicolumn{1}{c}{SD} & \multicolumn{1}{c}{Obs.} & \multicolumn{1}{c}{Mean} & \multicolumn{1}{c}{SD} & \multicolumn{1}{c}{Obs.} & \multicolumn{1}{c}{Mean} & \multicolumn{1}{c}{SD} & \multicolumn{1}{c}{Obs.}\\
\midrule
\addlinespace[0.3em]
\multicolumn{10}{l}{\textit{Panel A -- Individuals}}\\
\midrule\hspace{1em}Male & 0.453 & 0.498 & 11,974 & 0.455 & 0.498 & 3,968 & 0.452 & 0.498 & 8,006\\
\hspace{1em}Birth weight, kg & 3.317 & 0.612 & 8,528 & 3.320 & 0.623 & 2,883 & 3.316 & 0.607 & 5,645\\
\hspace{1em}Adult height, cm & 169.919 & 9.260 & 11,952 & 169.918 & 9.270 & 3,962 & 169.919 & 9.256 & 7,990\\
\hspace{1em}Educational attainment, years & 13.235 & 2.123 & 11,719 & 13.124 & 2.099 & 3,827 & 13.289 & 2.132 & 7,892\\
\hspace{1em}Fluid intelligence, sd & 6.233 & 2.158 & 4,115 & 6.140 & 2.114 & 1,533 & 6.288 & 2.183 & 2,582\\
\addlinespace[0.3em]
\multicolumn{10}{l}{\textit{Panel B -- Pollution}}\\
\midrule\hspace{1em}Black smoke, mcg/m$^3$ & 103.767 & 99.870 & 26,302 & 110.920 & 100.874 & 8,855 & 100.137 & 99.162 & 17,447\\
\hspace{1em}Sulphur dioxide, mcg/m$^3$ & 132.721 & 90.312 & 26,195 & 148.888 & 94.864 & 8,804 & 124.536 & 86.779 & 17,391\\
\bottomrule
\end{tabular}
\begin{tablenotes}
\item Panels: (a) descriptives calculated on individual-level data, (b) descriptives calculated on pollution data.
\end{tablenotes}
\end{threeparttable}
\end{table}

Panel~A of \autoref{tab:descriptives} presents the individual-level descriptive statistics from the UK Biobank, showing that approximately 45\% of the sample is male, and individuals have just over 13 years of schooling, on average. The average birth weight is 3.3 kg, with an average height of 170 cm. 
We do not find strong differences between the treated and control groups, defined as individuals born within the SCA boundaries and those outside, respectively.

Panel~B of \autoref{tab:descriptives} presents the descriptive statistics of the pollution measurements, indicating average black smoke and sulphur dioxide concentrations of 104 mcg/m$^3$ and 133 mcg/m$^3$ respectively. Consistent with \autoref{fig:pollution_timeseries_treated_control}, pollution measurements are higher in treated compared to control areas, defined as measurement stations that are located in and outside the SCA boundaries, respectively.

\FloatBarrier
\section{Empirical strategy}
\label{sec:empirical_strategy}
The first part of our analysis explores whether smoke control areas served their intended purpose of reducing air pollution, especially black smoke emissions, by estimating the impact of the introduction of SCAs on local pollution levels. The second part then explores the long-term consequences of smoke control on individuals' human capital and health outcomes. The unit of interest in these two sections is the pollution station and the individual, respectively. For brevity, this Section refers to both as `unit'. Our identification exploits spatial and time variation in the roll-out of SCAs, comparing the outcomes of these `units' (i.e., pollution levels or individual-level outcomes) located in a SCA before and after its creation, relative to those of a control group of `never-treated' units. In our main analysis, we define the latter as units in \textit{non-adopting} CBs as well as units located outside SCA boundaries but in \textit{adopting} CBs. Since the latter may be affected by spillover (downwind) effects from neighbouring SCAs, we also run our analysis dropping these units altogether and find very similar results.

Our `event' of interest is the date on which SCAs are \emph{submitted} to the Ministry. We focus on this date for two reasons. First, the proposed operation date was published at the time of submission, indicating when the area is most likely to become smokeless. Second, local councils reimbursed costs associated with stove conversions only if they were incurred \textit{before} the operation date (see \autoref{sec:background}). Hence, households started requesting conversions immediately after the submission date. 

Our identification assumes that, conditional on controls and fixed effects, the timing of the introduction of smoke control areas was random, and that the areas selected to be smokeless were not systematically different from those not selected. We explore both of these assumptions in detail in \autoref{sec:appendix_timing}. First, we explore associations between pre-programme CB-level characteristics (i.e., population density, pollution levels, and socioeconomic composition) and the timing of SCA implementation, showing that apart from population density, these variables do not explain differences in the timing of SCA introduction, suggesting that the timing was largely random.
Second, we investigate whether stations and individuals  inside SCAs were systematically different from those outside SCAs, but within the same CB. Although \autoref{fig:pollution_timeseries_treated_control} shows that treated stations on average report higher levels of pollution (i.e., \textit{between}-CBs), we find that areas that are selected to become smokeless were slightly less polluted compared to those not selected within the same CB. Furthermore, our results suggest that residents in areas that were put under smoke control were of higher socio-economic status compared to those in control areas. To alleviate concerns about these differences driving our findings, we include station/area fixed effects as well as station/area-specific trends in our analyses, accounting for systematic differences between areas that did and did not become smokeless \textit{within} CBs. 

\subsection{Impact on pollution}
To estimate the impact of the introduction of SCAs on local pollution levels, we estimate a two-way fixed effects model of the form:
\begin{equation}
    y_{st} = \delta_s + \gamma_t + \tau_s t + D_{st} + \epsilon_{st}
    \label{eq:pollution_event}
\end{equation}
where $y_{st}$ denotes the average pollution measurement (i.e., black smoke or sulphur dioxide) at station $s$ in year-month $t$. The parameters $\delta_s$ and $\gamma_{t}$ denote station and year-month fixed effects to account for systematic variation in pollution (or population density; see \autoref{sec:appendix_timing}) between stations, across time and for seasonality, and $\tau_{s}t$ denotes station-specific linear trends. We consider three specifications of the term $D_{st}$. First, a dynamic event study specification, where $D_{st} = \sum_{\tau \in \mathcal{T}} \beta_{\tau} \,\SCA_{st}^{\tau}$, and $\SCA_{st}^{\tau}$ is an indicator denoting whether station $s$ in year-month $t$ is $\tau$ months away from being treated (i.e., submitted to the Ministry) and where the month before submission ($\tau=-1$) is the reference month.\footnote{We trim our sample to five years (60 months) before and after treatment for each station, though our results are robust to not trimming the sample or trimming more/fewer years (see \autoref{sec:robustness_trimming}).}

Second, a static difference-in-difference approach, where $D_{st} = \beta_{\text{Adj}} \, \allowbreak \text{Inside}_{s} \allowbreak \times \allowbreak \text{Adj}_{st} + \allowbreak \beta_{\text{Post}} \, \allowbreak \text{Inside}_{s}\allowbreak \times \text{Post}_{st}$. We define Inside$_{s}$ to be an indicator that is equal to one if the location of the measurement station is \textit{inside} the exact SCA boundary, and zero otherwise.\footnote{Note that Inside$_{s}$ only enters our specification interacted with other dummies, as the main term would be absorbed into the station fixed effects that we include in all specifications.} $\text{Adj}_{st}$ is a dummy that is equal to one when the SCA in which pollution station $s$ is located has been submitted to the Ministry, but not yet entered operation (we refer to this as the `adjustment period'), and zero otherwise. Its coefficient $\beta_{\text{Adj}}$ captures any immediate drops in pollution levels following submission of the SCA as well as gradual reductions in pollution in the period between submission and operation when appliances were being upgraded to allow for smokeless fuel. $\text{Post}_{st}$ is a dummy equal to one when the SCA in which pollution station $s$ is located started operating, and zero otherwise. Its coefficient $\beta_{\text{Post}}$ then captures the overall impact of a fully operating SCA on local pollution levels, relative to other stations that are not located within an SCA. This specification allows for differential impacts of the adjustment period and that post-operation.

Many CBs introduced multiple SCAs at different points in time. Although a station can only be within \textit{one} SCA, they can be downwind of \textit{multiple} SCAs given that pollution disperses along the wind direction vector. To investigate the importance of such spillovers, our third specification therefore allows for downwind stations to be differentially affected such that $D_{st} = \beta_{\text{Adj}} \, \text{Inside}_{s} \times \text{Adj}_{st} + \beta_{\text{Post}} \, \text{Inside}_{s} \times \text{Post}_{st} + \beta_{\text{Adj,DW}} \, \text{Downwind}_{s} \times \text{Adj}^{\text{DW}}_{st} + \beta_{\text{Post,DW}} \, \text{Downwind}_{s} \times \text{Post}^{\text{DW}}_{st}$. The terms Adj$_{st}$ and Post$_{st}$ are defined in the same way as in the previous specification, and Downwind$_{s}$ is a dummy for the station being downwind of any SCA. The binary indicator Adj$^{\text{DW}}_{st}$ (Post$^{\text{DW}}_{st}$) is equal to one when the upwind SCA enters the adjustment period (becomes operational) and zero otherwise.

We estimate \autoref{eq:pollution_event} using OLS, and -- considering the recent literature on staggered treatments \citep{sunabraham2021, callawaySantanna2021, borusyakJaravelSpiess2024} -- we also report the group-time average effects in \autoref{sec:appendix_staggered_did}, finding very similar results. We cluster our standard errors by station throughout.

\subsection{Individual-level analysis}
To estimate the effect of the introduction of SCAs on individual human capital and health outcomes, we follow an approach analogous to the previous section but at the individual-level:
\begin{equation}
    y_{ijt} = \theta_{j} + \gamma_t + \tau_{j} t + D_{i\! jt} + \boldsymbol\zeta \mathbf{x}_{i\! jt} + \epsilon_{i\! jt}
    \label{eq:individuals_event}
\end{equation}
where $y_{ijt}$ is the outcome for individual $i$, born in area $j$ in year-month $t$. The area is defined as [CB $\times$ Inside], allowing for differences in areas that did and did not become SCAs \textit{within} a CB. Hence, $\theta_{j}$ are [CB $\times$ Inside] fixed effects, controlling for spatial variation in the outcomes of interest (and accounting for systematic differences in socio-economic composition of treated and control areas within a CB; see \autoref{sec:appendix_timing}), $\gamma_t$ are year-month fixed effects controlling for differences across cohorts and seasonality, and $\tau_{j} t$ are (CB $\times$ Inside)-specific linear (yearly) time trends. The latter account for differential dynamics in outcomes across CBs, as well as \textit{within} CBs across areas that did and did not become smokeless.\footnote{We show in \autoref{sec:robustness_trends} that our results are robust to using CB-specific trends or not including a trend. We trim our sample to birth cohorts within five years (60 months) of treatment to be consistent with the pollution analysis, though our results are similar to not trimming the sample or trimming more/fewer years (see \autoref{sec:robustness_trimming}).} The vector 
$\mathbf{x}_{ijt}$ denotes additional covariates (sex, ethnicity and weather conditions in utero and in childhood) capturing further variation within CBs. 

We explore two definitions of the term $D_{\ijt}$. First, a dynamic event specification where we set $D_{\ijt} = \sum_{\tau \in \mathcal{T}} \beta_{\tau} \,\SCA_{\ijt}^{\tau}$, and define $\SCA_{\ijt}^{\tau}$ to be an indicator that is equal to one if the relative number of six months intervals between conception and SCA submission is equal to $\tau$, and the individual is born within the exact boundaries of an SCA.\footnote{For example, if an individual is conceived in an SCA six months before its submission date, $\SCA_{\ijt}^{\tau=-1} = 1$ with $\SCA_{\ijt}^{\tau} = 0$ for all other $\tau$. Similarly, if an individual is conceived in an SCA 12 months after its submission, $\SCA_{\ijt}^{\tau=2} = 1$ and $\SCA_{\ijt}^{\tau} = 0$ for all other values of $\tau$.} The parameters $\beta_\tau$ then compare the outcomes of interest for individuals born in SCAs and conceived at time $\tau$ relative to SCA submission, to those conceived at the same calendar time but outside smoke control areas. For $\tau < 0$ the parameters $\beta_\tau$ capture differences in the outcome of interest for individuals conceived \textit{prior to} SCA submission. As $\tau$ grows in negative direction, individuals are exposed to increasing durations of pollution exposure \textit{in early childhood} as they were born before SCA submission. This means that, depending on the dose-response relationship, as well as the timing of exposure relative to age, there may be heterogeneity in the estimates across different (negative) values of $\tau$. For positive values of $\tau$, the parameters $\beta_\tau$ capture post-SCA submission differences in outcomes, relative to those conceived outside SCAs.

Second, we consider a static difference-in-difference specification where $D_{i\! jt} = \beta_{\text{Adj}} \, \text{Inside}_{i\! j} \times \text{Adj}_{i\! jt} + \beta_{\text{Post}} \, \text{Inside}_{i\! j} \times \text{Post}_{i\! jt}$. We define $\text{Adj}_{i\! jt}$ to be a dummy that is equal to one when the SCA surrounding the individual has been submitted but not yet entered operation (i.e., the `adjustment period'), and zero otherwise. Similarly we define $\text{Post}_{i\! jt}$ as a dummy equal to one after the SCA has started operating, zero otherwise. This allow us to estimate the overall impact of smoke control on individuals' outcomes, differentiating between the impact of being conceived during the adjustment period (when appliances are gradually replaced and pollution declining) captured by $\beta_{Adj}$, and the impact during the period after SCAs are in operation, captured by $\beta_{Post}$.\footnote{We do not separately report the estimates for individuals being \textit{in} and \textit{downwind} of an SCA. Indeed, as we show below, we find no strong impact of the introduction of SCAs on pollution measurements in stations that are downwind from the SCA, suggesting that the introduction of smoke control areas mainly affected local pollution.  } We cluster our standard errors by CB; our results are robust to clustering by CB $\times$ Inside.

\FloatBarrier
\section{Results}
\label{sec:results}
We first use our pollution panel to examine the impact of the introduction of SCAs on local pollution levels. Next, we use individual-level data to investigate the immediate and long-term effects on individuals' health and human capital outcomes.

\subsection{Impact on pollution}
\begin{figure}[!h]\caption{\label{fig:event_estimates_ols_main}Event estimates -- Impact on local pollution levels.}\centering\includegraphics[width=0.95\textwidth]{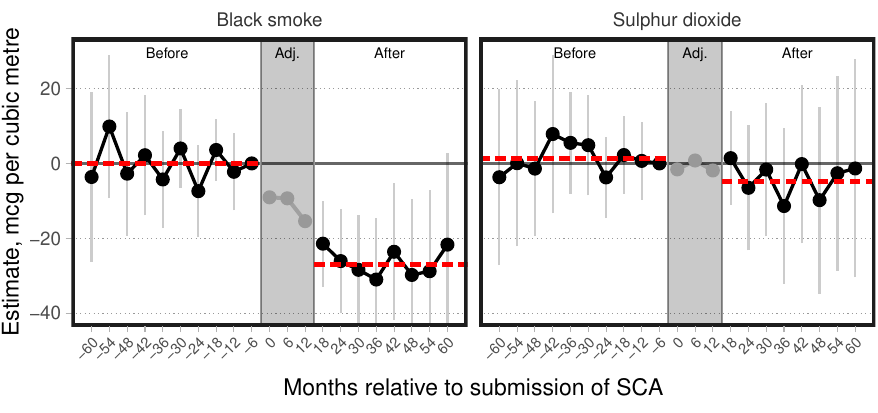}\caption*{\emph{OLS event estimates showing the impact of smoke control on individuals. OLS specification includes year-by-month and station fixed effects, and includes a station-specific yearly linear time-trend to capture differences in linear dynamics between stations. Control group consists of never-treated stations from both adopting and non-adopting county boroughs. Trims the sample to 5 years before and after the SCA submission date, and restricts the sample to pollution data for years 1962 to 1973. The dashed red lines indicate the average of the pre- and post-adjustment period estimates, where the former are averaged to zero. The grey part refers to the adjustment period. Clusters standard errors by station.}}\end{figure}

\autoref{fig:event_estimates_ols_main} plots the dynamic event study estimates of $\beta_\tau$ from \autoref{eq:pollution_event}, showing the time-varying impacts of the introduction of SCAs on local black smoke (left) and sulphur dioxide (right) levels. The first vertical line denotes our event of interest: the SCA submission date. The average duration between submission and operation (what we refer to as the `adjustment period') in our sample is \pvarAnalysisStationsEventObsAverageWait months; this is reflected by the second vertical line (the end of the `adjustment period').\footnote{This period differs across SCAs; we plot the distribution of months between the submission and operation date in our data in \autoref{fig:stations_waiting-times}. } 

The figures suggest no evidence of differential pre-trends in pollution levels across treated and control stations for both black smoke and sulphur dioxide. With the submission of an SCA, we find an immediate drop in levels of black smoke of about 10 mcg/m$^3$, reducing further over time and settling at just under 30 mcg/m$^3$. The drop remains visible until at least five years after submission.
The submission (and subsequent operation) of a SCA shows no visible impact on sulphur dioxide levels. Although this may seem surprising, it is consistent with the descriptive analysis in \autoref{fig:pollution_timeseries_treated_control} as well as with expectations at the time. First, smoke control areas targeted visible (black) smoke emitted from chimneys rather than specific pollutants such as sulfur dioxide that could only be observed using suitable instrumentation (and which was not the local government focus). Second, traditional and solid smokeless fuels had similar sulfur contents and thereby released similar amounts of sulphur dioxide when burned.

\begin{table}[!h]
\centering\centering\centering
\caption{\label{tab:stations_did_bs-so2}Difference-in-difference estimates -- Impact on pollution.}
\centering
\begin{threeparttable}
\fontsize{10}{12}\selectfont
\setlength{\tabcolsep}{1.5pt}
\begin{tabular}[t]{ldddd}
\toprule
\multicolumn{1}{c}{\em{}} & \multicolumn{4}{c}{\em{Depedent variable:}} \\
\cmidrule(l{3pt}r{3pt}){2-5}
\multicolumn{1}{c}{} & \multicolumn{1}{c}{(1)} & \multicolumn{1}{c}{(2)} & \multicolumn{1}{c}{(3)} & \multicolumn{1}{c}{(4)} \\
\multicolumn{1}{c}{ } & \multicolumn{1}{c}{{\specialcell[b]{Black \\ smoke}}} & \multicolumn{1}{c}{{\specialcell[b]{Black \\ smoke} }} & \multicolumn{1}{c}{{\specialcell[b]{Sulphur \\ dioxide}}} & \multicolumn{1}{c}{{\specialcell[b]{Sulphur \\ dioxide} }}\\
\midrule
Inside $\times$ Adj. & -8.053^{ *** } & -7.866^{ *** } & -1.180^{  } & -1.148^{  }\\
 & (3.010) & (3.031) & (3.740) & (3.747)\\
Inside $\times$ Post & -19.737^{ *** } & -19.811^{ *** } & -3.480^{  } & -3.677^{  }\\
 & (4.380) & (4.425) & (5.337) & (5.347)\\
Downwind $\times$ Adj. & {} & -1.032^{  } & {} & -6.245^{  }\\
 & {} & (3.727) & {} & (4.743)\\
Downwind $\times$ Post & {} & -8.225^{  } & {} & -7.317^{  }\\
 & {} & (5.631) & {} & (6.441)\\
\midrule
Observations & \multicolumn{1}{D{,}{,}{-3}}{26,302} & \multicolumn{1}{D{,}{,}{-3}}{26,302} & \multicolumn{1}{D{,}{,}{-3}}{26,195} & \multicolumn{1}{D{,}{,}{-3}}{26,195}\\
Mean dep. var. & \multicolumn{1}{d}{103.767} & \multicolumn{1}{d}{103.767} & \multicolumn{1}{d}{132.721} & \multicolumn{1}{d}{132.721}\\
$R^2$ & \multicolumn{1}{d}{0.81} & \multicolumn{1}{d}{0.81} & \multicolumn{1}{d}{0.79} & \multicolumn{1}{d}{0.79}\\
\bottomrule
\end{tabular}
\begin{tablenotes}
\item Columns: (1-2) level of black smoke, (3-4) level of sulphur dioxide. Control group conists of never-treated from both adopting and non-adopting county boroughs. Clusters standard errors by station. Includes year-by-month and station fixed effects, and includes a station-specific yearly linear time-trend to capture differences in linear dynamics between stations. Trims the sample to 5 years before and after the SCA submission date, and restricts the sample to pollution data for years 1962 to 1973. (*): $p < 0.1$, (**): $p<0.05$, (***): $p<0.01$.
\end{tablenotes}
\end{threeparttable}
\end{table}

To quantify the impact on pollution, we also present the difference-in-difference estimates from \autoref{eq:pollution_event} with a binary indicator for the pollution station being inside an SCA. \autoref{tab:stations_did_bs-so2} reports the OLS estimates of the introduction of SCAs on the local levels of black smoke (Columns~1--2) and sulphur dioxide (Columns~3--4) aggregated across all post-periods. Note here that the variable $\text{Post}_{st}$ is specific to the measurement station (rather than taking the average duration of the adjustment period of 16 months that is shown in grey in \autoref{fig:event_estimates_ols_main}). Columns~1 and 3 show that the submission of SCAs caused black smoke concentrations to drop by 8 mcg/m$^3$ on average, with no significant changes in levels of sulfur dioxide. Given mean black smoke levels of 104 mcg/m$^3$ prior to the introduction of SCAs, this is an $\sim$8\% reduction. Black smoke concentrations drop by approximately 19 mcg/m$^3$ (18\%) once the SCA is in operation. Distinguishing between pollution measurements that are taken inside SCAs and those downwind, Column~2 shows that the treatment effect for black smoke is primarily driven by the former, with no significant reductions in pollution for downwind areas, though the sign of both estimates is negative.\footnote{\autoref{tab:stations_area_bs-so2} in \autoref{sec:additional} explores the robustness of these findings by specifying alternative treatment variables: (1) the number of square kilometres (km$^2$) of SCA \emph{surrounding} a pollution station and the number of km$^2$ of SCA \emph{upwind} of the station, and (2) the sum of the two. These specifications are likely to be less precise, since the SCA km$^2$ is not necessarily a good proxy for the number of dwellings affected (and therefore of actual pollution exposure). While the estimates have large standard errors and are not statistically significant, they suggest that an additional km$^2$ of SCA is negatively related to black smoke concentrations, with more mixed results for sulphur dioxide.}

\FloatBarrier
\subsection{Impact on individuals}\label{sec:results_indiv}
\begin{figure}[!h]\caption{\label{fig:individuals_event_bw-height}Event estimates -- Impact on individuals.}\centering\includegraphics[width=0.95\textwidth]{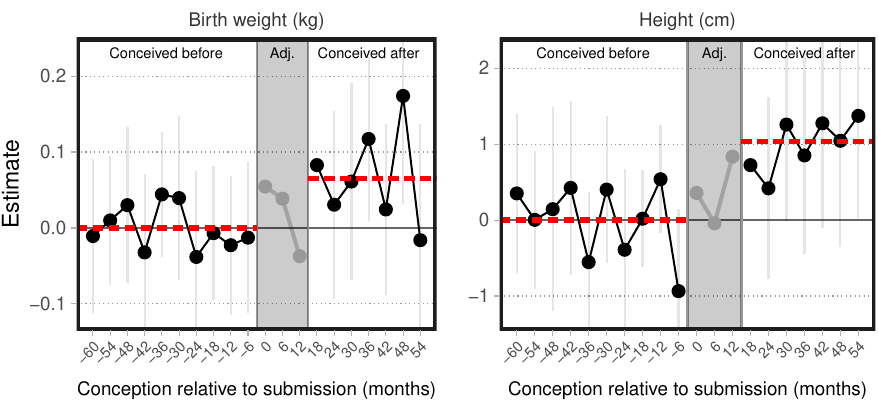}\caption*{\emph{OLS event estimates showing the impact of smoke control on individuals. OLS specification includes year-by-month and (CB $\times$ Inside) fixed effects, and includes (CB $\times$ Inside)-specific linear time trends Control group consists of never-treated individuals from both adopting and non-adopting county boroughs. Trims the sample to 5 years before and after the SCA submission date, and restricts the sample to birth cohorts in years 1958 to 1969. The dashed red lines indicate the average of the pre- and post-adjustment period estimates, where the former are averaged to zero. The grey part refers to the adjustment period. Clusters standard errors by CB.}}\end{figure}

\begin{figure}[!h]\caption{\label{fig:individuals_event_ea-fi}Event estimates -- Impact on individuals.}\centering\includegraphics[width=0.95\textwidth]{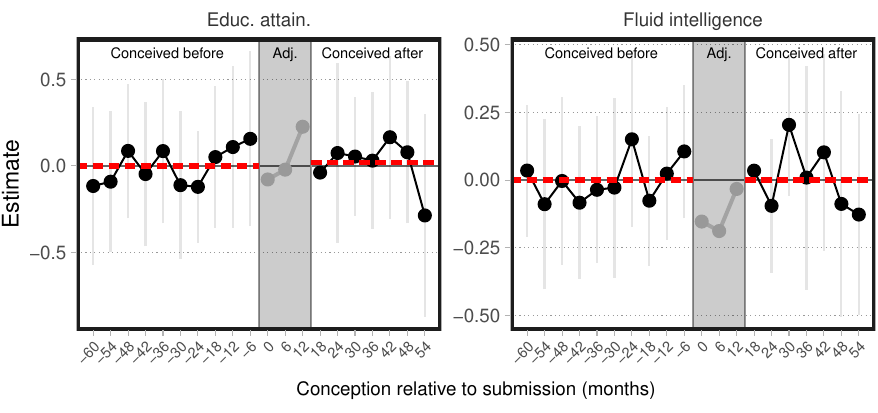}\caption*{\emph{OLS event estimates showing the impact of smoke control on individuals. OLS specification includes year-by-month and (CB $\times$ Inside) fixed effects, and includes (CB $\times$ Inside)-specific linear time trends Control group consists of never-treated individuals from both adopting and non-adopting county boroughs. Trims the sample to 5 years before and after the SCA submission date, and restricts the sample to birth cohorts in years 1958 to 1969. The dashed red lines indicate the average of the pre- and post-adjustment period estimates, where the former are averaged to zero. The grey part refers to the adjustment period. Clusters standard errors by CB.}}\end{figure}

We next examine how the introduction of smoke control areas and the subsequent reduction in local pollution translates into individual outcomes by plotting the dynamic event study estimates of $\beta_\tau$ from \autoref{eq:individuals_event}. \autoref{fig:individuals_event_bw-height} shows the impact of smoke control areas on individuals' birth weight and adult height, while \autoref{fig:individuals_event_ea-fi} reports the estimates for years of schooling and (standardised) fluid intelligence. The red dashed line before and after the adjustment period denotes the average of the pre- and post-adjustment period estimates.

\autoref{fig:individuals_event_bw-height} shows no evidence of trends in birth weight or adult height for individuals conceived \textit{prior to} the SCA submission date. This is expected for birth weight, since these children were only exposed to the SCA in childhood and birth weight cannot be affected by changes in pollution after birth. 
Instead, we find that the birth weights of individuals conceived after the SCA operation date (therefore experiencing reduced prenatal as well as childhood black smoke exposure) are on average 70-80g higher than those conceived before. Similarly, they are approximately 1 cm taller in adulthood.

\autoref{fig:individuals_event_ea-fi} presents the same estimates but for years of education and fluid intelligence. This shows no evidence that the introduction of SCAs -- whether during the in utero or in childhood period -- impacts on these human capital outcomes, with average estimates around zero. 

To further quantify the average impact on individuals' human capital and health, we also present the difference-in-difference estimates from \autoref{eq:individuals_event}, using a binary indicator that equals one if the individual is conceived within the boundaries of an SCA during the adjustment or operation period, relative to those conceived outside the SCA boundary at these times. \autoref{tab:individuals_did_bw-height-ea-fi} confirms the event study estimates, showing that -- compared to those who are conceived \textit{prior to} SCA submission -- the birth weights of those conceived when the SCA is in operation are 60g higher (column (1)), and these individuals are just under 1cm taller in adulthood (column (2)).\footnote{In \autoref{tab:individuals_did_low-bw} we show the impact on the probability of being born with low birth weight ($<2,500$g). While these point estimates are negative, suggesting slight reductions in this risk of about 2 percentage points, they are not significant. This suggest that the impact we find on birth weight is not driven specifically by improvements among low birth weight individuals, but rather a general shift of the birth weight distribution.} Relative to the average birth weight and adult height in the sample, this corresponds to a 2\% and 0.6\% increase, respectively. We again find no impact on years of education, nor on fluid intelligence.\footnote{\autoref{tab:individuals_did_qualifications} decomposes the total years of education into binary indicators for specific qualifications. This shows suggestive evidence that the introduction of SCAs decreased the probability of exiting the education system with lower secondary qualifications, while it increased the probability of obtaining an upper secondary degree, though the latter is not significant.} 

\begin{table}[!h]
\centering\centering\centering
\caption{\label{tab:individuals_did_bw-height-ea-fi}Difference-in-difference estimates -- Impact on individuals.}
\centering
\begin{threeparttable}
\fontsize{11}{13}\selectfont
\setlength{\tabcolsep}{1.5pt}
\begin{tabular}[t]{ldddd}
\toprule
\multicolumn{1}{c}{\em{}} & \multicolumn{4}{c}{\em{Dependent variable:}} \\
\cmidrule(l{3pt}r{3pt}){2-5}
\multicolumn{1}{c}{} & \multicolumn{1}{c}{(1)} & \multicolumn{1}{c}{(2)} & \multicolumn{1}{c}{(3)} & \multicolumn{1}{c}{(4)} \\
\multicolumn{1}{c}{ } & \multicolumn{1}{c}{{\specialcell[b]{Birth \\ weight}}} & \multicolumn{1}{c}{{\specialcell[b]{Adult \\ height}}} & \multicolumn{1}{c}{{\specialcell[b]{Educ. \\ attain.}}} & \multicolumn{1}{c}{{\specialcell[b]{Fluid \\ intelligence}}}\\
\midrule
Inside $\times$ Adj. & 0.039^{  } & 0.365^{  } & 0.040^{  } & -0.152^{  }\\
 & (0.034) & (0.343) & (0.178) & (0.131)\\
Inside $\times$ Post & 0.058^{ ** } & 0.942^{ *** } & -0.025^{  } & 0.035^{  }\\
 & (0.027) & (0.180) & (0.135) & (0.065)\\
\midrule
Observations & \multicolumn{1}{D{,}{,}{-3}}{8,510} & \multicolumn{1}{D{,}{,}{-3}}{11,922} & \multicolumn{1}{D{,}{,}{-3}}{11,689} & \multicolumn{1}{D{,}{,}{-3}}{4,106}\\
Mean dep. var. & \multicolumn{1}{d}{3.317} & \multicolumn{1}{d}{169.917} & \multicolumn{1}{d}{13.236} & \multicolumn{1}{d}{0.000}\\
$R^2$ & \multicolumn{1}{d}{0.066} & \multicolumn{1}{d}{0.543} & \multicolumn{1}{d}{0.107} & \multicolumn{1}{d}{0.149}\\
\bottomrule
\end{tabular}
\begin{tablenotes}
\item Columns: (1) birth weight in kilograms, (2) adult height in centimeters, (3) years of education, (4) standardised fluid intelligence score. OLS specification includes year-by-month and (CB $\times$ Inside) fixed effects, and includes (CB $\times$ Inside)-specific linear time trends. Controls for sex, ethnicity, and weather in utero and during childhood. Control group consists of never-treated individuals from both adopting and non-adopting county boroughs. Trims the sample to five years before and after the SCA submission date, and restricts the sample to birth cohorts in years 1958 to 1969. Clusters standard errors by CB. (*): $p < 0.1$, (**): $p<0.05$, (***): $p<0.01$.
\end{tablenotes}
\end{threeparttable}
\end{table}

\FloatBarrier
\subsubsection{Heterogeneity by gender}
The existing literature suggests that males are more susceptible to environmental insults in early life. Hence, we next explore whether there are gender differences in the impacts of smoke control. A priori it is not clear whether we would expect to find different impacts of smoke control for men and women. On the one hand, if a reduction in pollution has a larger impact on male infant mortality due to them being weaker on average, it may increase the probability of survival of weaker boys, leading to a drop in average male birth weights. On the other hand, holding the survival rate constant, a reduction in pollution may help vulnerable populations more, increasing male birth weights.

\autoref{tab:individuals_did_hetero-sex_bw-height-ea-fi} shows the estimates when we split the sample by gender. We find that the introduction of SCAs benefits males' birth weights more than females', with the former seeing an increase of 126 grams on average, while females see a modest rise of 21 grams.\footnote{This is consistent with the above example that holds survival rates constant. To explore this empirically, we examine the impact of smoke control on the probability of being female, shown in \autoref{tab:individuals_did_sex}. These results indicate that there are no gender-specific mortality impacts, and therefore provide suggestive evidence that the gender-specific estimates for our main outcomes are not driven by mortality selection. One plausible explanation for the bigger increase in male birth weights is therefore that the reduction in pollution disproportionally helps male (as opposed to female) growth.} We find no strong gender differences for adult height with both genders experiencing an average increase of about 1 cm. Similarly, and consistent with the main analyses, we do not find strong evidence of impacts on educational attainment and fluid intelligence.

\begin{table}[!h]
\centering\centering\centering
\caption{\label{tab:individuals_did_hetero-sex_bw-height-ea-fi}Difference-in-difference estimates -- Impact on individuals.}
\centering
\begin{threeparttable}
\fontsize{11}{13}\selectfont
\setlength{\tabcolsep}{1.5pt}
\begin{tabular}[t]{ldddd}
\toprule
\multicolumn{1}{c}{\em{}} & \multicolumn{4}{c}{\em{Dependent variable:}} \\
\cmidrule(l{3pt}r{3pt}){2-5}
\multicolumn{1}{c}{} & \multicolumn{1}{c}{(1)} & \multicolumn{1}{c}{(2)} & \multicolumn{1}{c}{(3)} & \multicolumn{1}{c}{(4)} \\
\multicolumn{1}{c}{ } & \multicolumn{1}{c}{{\specialcell[b]{Birth \\ weight}}} & \multicolumn{1}{c}{{\specialcell[b]{Adult \\ height}}} & \multicolumn{1}{c}{{\specialcell[b]{Educ. \\ attain.}}} & \multicolumn{1}{c}{{\specialcell[b]{Fluid \\ intelligence}}}\\
\midrule
\addlinespace[0.5em]
\multicolumn{5}{l}{\textit{Panel A -- Female}}\\
\midrule \hspace{1em}Inside $\times$ Adj. & 0.011^{  } & 0.618^{  } & -0.006^{  } & -0.187^{  }\\
\hspace{1em} & (0.037) & (0.470) & (0.268) & (0.158)\\
\hspace{1em}Inside $\times$ Post & 0.021^{  } & 1.097^{ *** } & 0.087^{  } & 0.023^{  }\\
\hspace{1em} & (0.038) & (0.309) & (0.180) & (0.095)\\
\hspace{1em}Observations & \multicolumn{1}{D{,}{,}{-3}}{5,031} & \multicolumn{1}{D{,}{,}{-3}}{6,522} & \multicolumn{1}{D{,}{,}{-3}}{6,390} & \multicolumn{1}{D{,}{,}{-3}}{2,280}\\
\hspace{1em}Mean dep. var. & \multicolumn{1}{d}{3.244} & \multicolumn{1}{d}{163.845} & \multicolumn{1}{d}{13.248} & \multicolumn{1}{d}{0.000}\\
\hspace{1em}$R^2$ & \multicolumn{1}{d}{0.079} & \multicolumn{1}{d}{0.071} & \multicolumn{1}{d}{0.118} & \multicolumn{1}{d}{0.213}\\
\addlinespace[0.5em]
\multicolumn{5}{l}{\textit{Panel B -- Male}}\\
\midrule \hspace{1em}Inside $\times$ Adj. & 0.090^{  } & 0.091^{  } & 0.120^{  } & -0.075^{  }\\
\hspace{1em} & (0.058) & (0.591) & (0.093) & (0.079)\\
\hspace{1em}Inside $\times$ Post & 0.126^{ ** } & 0.948^{ ** } & -0.138^{  } & 0.041^{  }\\
\hspace{1em} & (0.050) & (0.431) & (0.175) & (0.069)\\
\hspace{1em}Observations & \multicolumn{1}{D{,}{,}{-3}}{3,479} & \multicolumn{1}{D{,}{,}{-3}}{5,400} & \multicolumn{1}{D{,}{,}{-3}}{5,299} & \multicolumn{1}{D{,}{,}{-3}}{1,826}\\
\hspace{1em}Mean dep. var. & \multicolumn{1}{d}{3.424} & \multicolumn{1}{d}{177.250} & \multicolumn{1}{d}{13.221} & \multicolumn{1}{d}{0.000}\\
\hspace{1em}$R^2$ & \multicolumn{1}{d}{0.102} & \multicolumn{1}{d}{0.092} & \multicolumn{1}{d}{0.154} & \multicolumn{1}{d}{0.233}\\
\bottomrule
\end{tabular}
\begin{tablenotes}
\item Columns: (1) birth weight in kilograms, (2) adult height in centimeters, (3) years of education, (4) standardised fluid intelligence score. Panels show estimates using subsamples of (A) females and (B) males. OLS specification includes year-by-month and (CB $\times$ Inside) fixed effects, and includes (CB $\times$ Inside)-specific linear time trends. Controls for sex, ethnicity, and weather in utero and during childhood. Control group consists of never-treated individuals from both adopting and non-adopting county boroughs. Trims the sample to five years before and after the SCA submission date, and restricts the sample to birth cohorts in years 1958 to 1969. Clusters standard errors by CB. (*): $p < 0.1$, (**): $p<0.05$, (***): $p<0.01$.
\end{tablenotes}
\end{threeparttable}
\end{table}

\FloatBarrier
\subsubsection{Heterogeneity by individuals' genetic ``endowments''}
Next, we examine heterogeneity by individuals' genetic ``endowments'', as measured by one's polygenic score (or polygenic index) that is specific to the outcome of interest. As highlighted in the introduction, this analysis not only sheds light on the impact of smoke control on population inequalities, but is also informative about the existence of complementarities between endowments and public health investments and contributes to the debate about the importance of effort and circumstance in shaping individuals' outcomes.

We construct proxies for individuals’ genetic ``endowment'' by running a Genome-Wide Association Study (GWAS) for each of our main outcomes in the UK Biobank, using only individuals who are not in (or related to individuals in) our main analysis sample. We then use these summary statistics to construct polygenic scores (PGS) for each outcome for the individuals in our analysis sample. We standardise all polygenic scores to have zero mean and unit variance in the analysis sample. We discuss the polygenic score construction in more detail in \autoref{sec:appendix_genetics}, and report their predictive power in \autoref{tab:individuals_assoc_pgi-prediction}. 

\begin{table}[!h]
\centering\centering\centering
\caption{\label{tab:individuals_did_hetero-pgi-continuous_bw-height-ea-fi}Difference-in-difference estimates -- Impact on individuals.}
\centering
\begin{threeparttable}
\fontsize{9}{11}\selectfont
\setlength{\tabcolsep}{1.5pt}
\begin{tabular}[t]{ldddd}
\toprule
\multicolumn{1}{c}{\em{}} & \multicolumn{4}{c}{\em{Dependent variable:}} \\
\cmidrule(l{3pt}r{3pt}){2-5}
\multicolumn{1}{c}{} & \multicolumn{1}{c}{(1)} & \multicolumn{1}{c}{(2)} & \multicolumn{1}{c}{(3)} & \multicolumn{1}{c}{(4)} \\
\multicolumn{1}{c}{ } & \multicolumn{1}{c}{{\specialcell[b]{Birth \\ weight}}} & \multicolumn{1}{c}{{\specialcell[b]{Adult \\ height}}} & \multicolumn{1}{c}{{\specialcell[b]{Educ. \\ attain.}}} & \multicolumn{1}{c}{{\specialcell[b]{Fluid \\ intelligence}}}\\
\midrule
Inside $\times$ Adj. & 0.039^{  } & 0.561^{ * } & 0.050^{  } & -0.114^{  }\\
 & (0.032) & (0.294) & (0.152) & (0.108)\\
Inside $\times$ Post & 0.044^{ * } & 0.778^{ *** } & -0.011^{  } & 0.071^{  }\\
 & (0.025) & (0.221) & (0.103) & (0.077)\\
Inside $\times$ Adj. $\times$ PGS & -0.017^{  } & 0.574^{ ** } & -0.158^{ * } & 0.106^{ * }\\
 & (0.040) & (0.250) & (0.088) & (0.056)\\
Inside $\times$ Post $\times$ PGS & 0.043^{ * } & -0.029^{  } & 0.033^{  } & 0.034^{  }\\
 & (0.025) & (0.168) & (0.055) & (0.060)\\
PGS & 0.102^{ *** } & 3.706^{ *** } & 0.562^{ *** } & 0.236^{ *** }\\
 & (0.008) & (0.056) & (0.020) & (0.020)\\
\midrule
Observations & \multicolumn{1}{D{,}{,}{-3}}{7,982} & \multicolumn{1}{D{,}{,}{-3}}{11,108} & \multicolumn{1}{D{,}{,}{-3}}{11,056} & \multicolumn{1}{D{,}{,}{-3}}{3,825}\\
Mean dep. var. & \multicolumn{1}{d}{3.323} & \multicolumn{1}{d}{170.052} & \multicolumn{1}{d}{13.234} & \multicolumn{1}{d}{0.000}\\
$R^2$ & \multicolumn{1}{d}{0.1} & \multicolumn{1}{d}{0.702} & \multicolumn{1}{d}{0.177} & \multicolumn{1}{d}{0.2}\\
\bottomrule
\end{tabular}
\begin{tablenotes}
\item Columns: (1) birth weight in kilograms, (2) adult height in centimeters, (3) years of education, (4) standardised fluid intelligence score. OLS specification includes year-by-month and (CB $\times$ Inside) fixed effects, and includes (CB $\times$ Inside)-specific linear time trends. Controls for sex, genetic principal components, and weather in utero and during childhood. All covariates have been normalised to mean zero. Control group consists of never-treated individuals from both adopting and non-adopting county boroughs. Trims the sample to five years before and after the SCA submission date, and restricts the sample to birth cohorts in years 1958 to 1969. Clusters standard errors by CB. (*): $p < 0.1$, (**): $p<0.05$, (***): $p<0.01$.
\end{tablenotes}
\end{threeparttable}
\end{table}

For each outcome, we add interactions between the two variables of interest and the PGS, capturing the extent to which the introduction of SCAs differentially affected individuals with different genetic ``endowments''.\footnote{Following \citet{Keller2014}, we also add interactions between the PGS and all covariates and principal components.}
\autoref{tab:individuals_did_hetero-pgi-continuous_bw-height-ea-fi} reports the results. We find that the impact of smoke control on birth weight is larger for individuals with a high genetic endowment for birth weight, suggesting that this local government pollution reduction programme exacerbated genetic inequalities in birth weight. It is also consistent with the existence of complementarities between endowments and (public health) investments in producing child health: the returns to the investment are larger for those with higher endowments. For height, we see some evidence of complementarities during the adjustment period, but little evidence of heterogeneity in the effect size post SCA operation. Similar to the main analysis, we find no main effect on years of education and intelligence, though some evidence of genetic heterogeneity during the SCA adjustment period. Although these are of opposite sign for education and intelligence, they are only marginally significant, with no impacts post-SCA operation. We are therefore cautious not to overinterpret these.

\subsubsection{Heterogeneity by socioeconomic status}
Finally, we examine whether the policy had differential impacts on individuals depending on the socio-economic composition of their local area. Since the UK Biobank does not have information on the socio-economic status of participants (or their parents), we merge district-level information on occupation from the UK Census to the UK Biobank to classify CBs as either high or low SES. To do this, we calculate the share of CB residents that are in professional, managerial, or technical occupations, and define CBs as high SES if they are above the median share of this distribution across all CBs, or low if they are below the median. 

We estimate our main specification with additional interactions between the variables of interest and the dummy for being born in a low SES area. The results are presented  in \autoref{tab:individuals_did_hetero-ses-continuous_bw-height-ea-fi}. We do not find strong evidence of differential effects by SES for birth weight and height. 
In contrast, the results suggest that high SES areas experienced an increase in education after SCA submission, with no impacts for lower SES areas, or even a reduction in intelligence relative to those born in higher SES neighbourhoods. 

\begin{table}[!h]
\centering\centering\centering
\caption{\label{tab:individuals_did_hetero-ses-continuous_bw-height-ea-fi}Difference-in-difference estimates -- Impact on individuals.}
\centering
\begin{threeparttable}
\fontsize{9}{11}\selectfont
\setlength{\tabcolsep}{1.5pt}
\begin{tabular}[t]{ldddd}
\toprule
\multicolumn{1}{c}{\em{}} & \multicolumn{4}{c}{\em{Dependent variable:}} \\
\cmidrule(l{3pt}r{3pt}){2-5}
\multicolumn{1}{c}{} & \multicolumn{1}{c}{(1)} & \multicolumn{1}{c}{(2)} & \multicolumn{1}{c}{(3)} & \multicolumn{1}{c}{(4)} \\
\multicolumn{1}{c}{ } & \multicolumn{1}{c}{{\specialcell[b]{Birth \\ weight}}} & \multicolumn{1}{c}{{\specialcell[b]{Adult \\ height}}} & \multicolumn{1}{c}{{\specialcell[b]{Educ. \\ attain.}}} & \multicolumn{1}{c}{{\specialcell[b]{Fluid \\ intelligence}}}\\
\midrule
Inside $\times$ Adj. & 0.059^{  } & -0.054^{  } & 0.304^{ *** } & 0.091^{  }\\
 & (0.036) & (0.314) & (0.077) & (0.066)\\
Inside $\times$ Post & 0.019^{  } & 1.025^{ *** } & 0.157^{  } & 0.045^{  }\\
 & (0.027) & (0.201) & (0.101) & (0.074)\\
Inside $\times$ Adj. $\times$ Is low SES & -0.028^{  } & 0.691^{  } & -0.477^{ * } & -0.422^{ *** }\\
 & (0.065) & (0.609) & (0.255) & (0.157)\\
Inside $\times$ Post $\times$ Is low SES & 0.070^{  } & -0.104^{  } & -0.359^{ * } & -0.035^{  }\\
 & (0.048) & (0.333) & (0.196) & (0.139)\\
\midrule
Observations & \multicolumn{1}{D{,}{,}{-3}}{8,510} & \multicolumn{1}{D{,}{,}{-3}}{11,922} & \multicolumn{1}{D{,}{,}{-3}}{11,689} & \multicolumn{1}{D{,}{,}{-3}}{4,106}\\
Mean dep. var. & \multicolumn{1}{d}{3.317} & \multicolumn{1}{d}{169.917} & \multicolumn{1}{d}{13.236} & \multicolumn{1}{d}{0.000}\\
$R^2$ & \multicolumn{1}{d}{0.068} & \multicolumn{1}{d}{0.544} & \multicolumn{1}{d}{0.109} & \multicolumn{1}{d}{0.152}\\
\bottomrule
\end{tabular}
\begin{tablenotes}
\item Columns: (1) birth weight in kilograms, (2) adult height in centimeters, (3) years of education, (4) standardised fluid intelligence score. OLS specification includes year-by-month and (CB $\times$ Inside) fixed effects, and includes (CB $\times$ Inside)-specific linear time trends. Controls for sex, genetic principal components, and weather in utero and during childhood. All covariates have been normalised to mean zero. Control group consists of never-treated individuals from both adopting and non-adopting county boroughs. Trims the sample to five years before and after the SCA submission date, and restricts the sample to birth cohorts in years 1958 to 1969. Clusters standard errors by CB. (*): $p < 0.1$, (**): $p<0.05$, (***): $p<0.01$.
\end{tablenotes}
\end{threeparttable}
\end{table}

\FloatBarrier
\section{Robustness}\label{sec:overview_robustness}
Our main analysis shows that the the introduction of SCAs substantially and persistently reduced black smoke emissions, but did not impact local sulphur dioxide concentrations. This in turn improved child health outcomes (in particular birth weight), with longer-term impacts on adult height. However, we find no consistent evidence of impacts on economic outcomes, including years of education and intelligence. 

\begin{landscape}
    \begin{figure}[!h]\caption{\label{fig:robustness_all}Overview of robustness checks.}\centering\includegraphics[width=\linewidth]{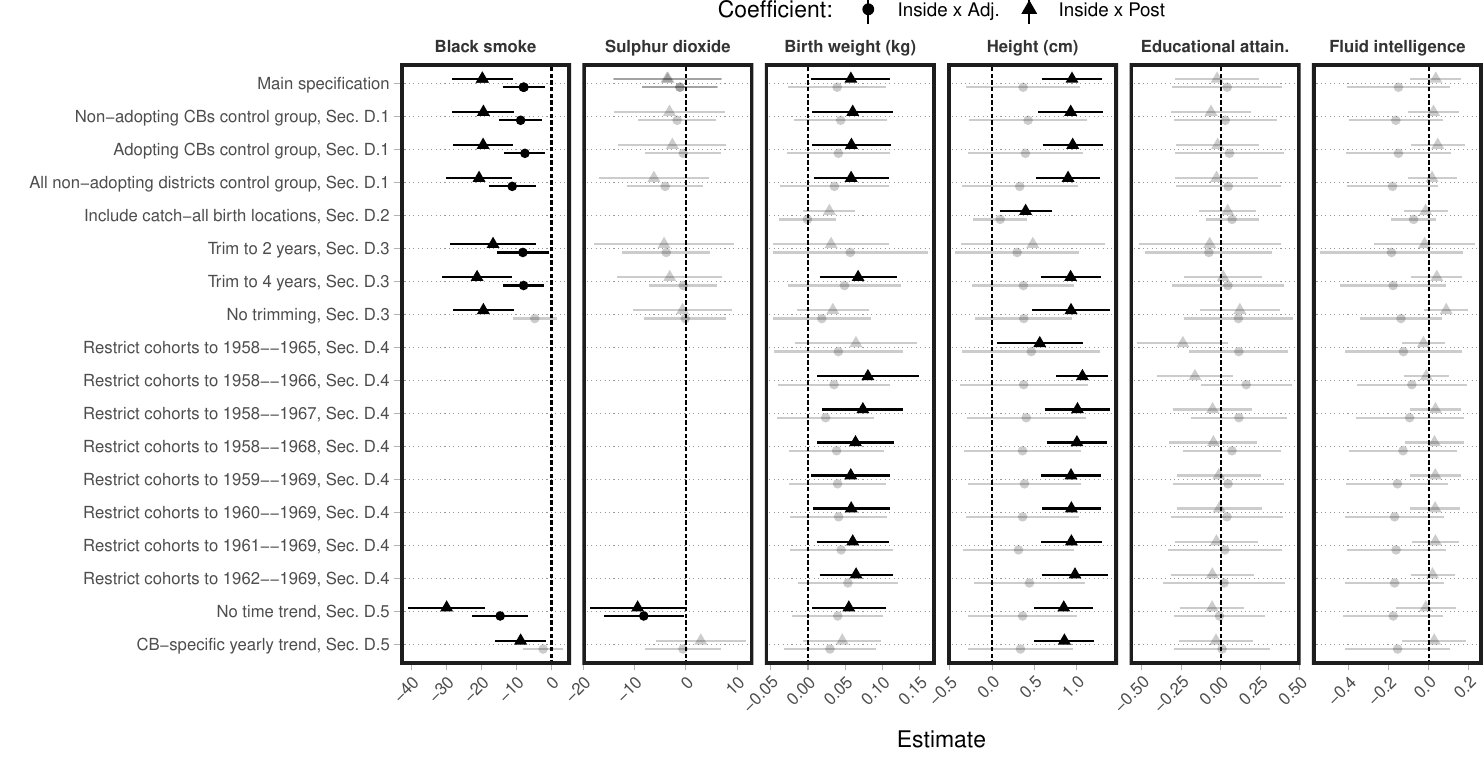}\caption*{\emph{Plots the point estimates and their 0.95 confidence bands for our main coefficients and outcomes across our various robustness checks. First two columns show estimates from the pollution sample, remaining four columns show estimates from individual sample. For checks that are specific to the individual analysis, we do not report the estimates for the first two columns, but only in the final four (individual-level) columns.}}\end{figure}
\end{landscape}

This section highlights that our findings are generally robust to a range of different specifications, samples and assumptions. We discuss these sensitivity checks in detail in \autoref{sec:robustness}, but summarise them graphically in \autoref{fig:robustness_all}. The figure shows six panels for our main outcomes of interest: the two pollution measures and four individual-level outcomes. The dots present the estimates of the impact of exposure during the `adjustment period'; the triangles present the estimates for the SCA being in operation. Both are shown with 95\% confidence intervals, with opaque colours indicating that they are significantly different from zero. The first row in each panel replicates the main specification, showing significant impacts on black smoke concentrations, as well as birth weight and height (the latter two only for those exposed after the operation date). 
%%%%%%%%%%%%%%%%%%%%%%%%%%%%%%%%%%%%%%%%%%
% Order in appendix
% Definition of control group D.1
% Precision of birth locations D.2
% Trimming of samples D.3
% Choice of birth cohorts D.4
% Time Trends D.5
% Time fixed effects D.6 - dropped
% Spatial fixed effects D.7 - dropped
%%%%%%%%%%%%%%%%%%%%%%%%%%%%%%%%%%%%%%%%%%
Each of the following rows correspond to different robustness checks, where the row refers to the specific section in \autoref{sec:robustness}. Rows 2-4 use alternative definitions of the control group (\autoref{sec:definition_of_control_group}). Row 5 is specific to the individual analysis, and uses alternative definitions of the sample depending on individuals' geolocation within County Boroughs (\autoref{sec:robustness_bunching}). Rows 6-8 explore the sensitivity to different bandwidths around the event time (\autoref{sec:robustness_trimming}). Rows 10-17 are specific to the individual analysis and show different restrictions of the relevant birth cohorts (\autoref{sec:robustness_birth-cohorts}). Rows 18-19 show different specifications of the time trend, specifying no time trend, or allowing for a CB-specific annual time trend (\autoref{sec:robustness_trends}). 

The main take-away from \autoref{fig:robustness_all} is that the estimates are very robust to the use of different specifications, samples, or model assumptions. In almost all specifications do we see a reduction in black smoke concentrations ranging from $\sim$10--30 mcg/m$^3$, followed by an approximately 60g increase in birth weight and 1 cm increase in adult height. We again find no consistent evidence of impacts on years of education, nor on fluid intelligence.

\FloatBarrier
\section{Conclusion}\label{sec:conclusion}
This paper examines the long-term effects of a national, large-scale pollution reduction programme that was rolled out in the UK in the late 1950s. We exploit temporal and spatial variation in the introduction of so-called ``Smoke Control Areas'' (SCAs); areas that banned all smoke emissions within its boundaries. 
Our identification compares pollution measurement stations (individuals) located (born) in or downwind of SCAs before and after their introduction, relative to those of a control group of never treated stations (individuals), controlling for weather variation, local area fixed effects, area-specific trends (and individual-level controls). Our main analysis specifies an event study approach, but we show that our results are robust to using group-time average effects that account for staggered treatment.

Using digitised historical pollution measurements, we provide one of the first empirical analyses of the dynamic and time-varying impact of the introduction of SCAs on local pollution levels \citep[see also][]{fukushima2021uk}. We show that they substantially and persistently reduced black smoke (but not sulphur dioxide) concentrations for at least five years post-introduction. This, in turn, affected individuals who were born in areas of smoke control, relative to those born elsewhere. We show that those exposed to SCAs had -- on average -- 60g higher birth weights and are 1cm taller in adulthood. Relative to the mean, these represent increases of 2\% and 0.6\% respectively. In contrast to much of the existing literature, we find no evidence of impacts on years of education and intelligence, with estimates that are relatively close to zero throughout, though with some suggestive evidence of heterogeneity by the socio-economic composition of individuals' district of birth. 

We highlight two potential reasons for our null result on education and intelligence. First, we examine the long-term impacts of small but persistent changes in pollution caused by the phased introduction of SCAs, whereas the existing literature that investigates the very long-term effects of pollution exposure have generally examined one large transitory pollution spike: the London smog. This may suggest that small, permanent reductions differentially impact long-term outcomes compared to extreme, but one-off events. Although this is possible, it is less likely given that the existing literature also shows that early life pollution exposure adversely affects human capital and labour market outcomes already in early adulthood \citep[e.g.,][]{isen2017every, persico2020can}. Unless these individuals catch up later, or those unexposed drop down the distribution, this is perhaps unlikely to explain our findings. 

Second, the fact that we do not find impacts on human capital outcomes may suggest that these are driven by exposure to pollutants other than those targeted by SCAs. Indeed, the quasi-experiments exploited in the existing economics literature (the London smog, but also e.g., toxic release inventory sites, and the US 1970 Clean Air Act Amendments) are likely to have affected a range of pollutants. In contrast, the introduction of SCAs targeted black smoke only, and the evidence suggests that moving from bituminous coal to smokeless fuel mainly affects black smoke/particulate matter, rather than other pollutants such as nitrogen oxides, sulphur oxides and carbon monoxide \citep{mitchell2016impact}. This is also what we find, in that we show reductions in black smoke concentrations, but not in sulphur dioxide. Hence, this is consistent with the idea that pollutants other than black smoke are responsible for the adverse effects on human capital, whereas black smoke reduces fetal as well as child growth. Unfortunately, since black smoke and sulphur dioxide are the only two pollutants that are measured throughout our observation period, our data do not allow us to explore this possible explanation in more detail.

We also examine heterogeneity in our estimates by individuals' genetic ``endowments'', obtained from the molecular genetic data available in the UK Biobank. In addition to this shedding light on whether the introduction of SCAs affected inequalities in our outcomes of interest, it is also informative about the existence of complementarities between endowments and public health investments. Our findings suggest that the introduction of SCAs increased inequalities in population health but not economic outcomes, with larger increases in birth weight and height among those with higher ``endowments''. 

Our analyses estimate intention to treat (ITT) effects, identifying the impact of the introduction of SCAs on individuals' outcomes, rather than the impact of pollution. Indeed, the latter is endogenous, with lower social class individuals more likely to live in highly polluted areas. Given our newly digitised and rare historical pollution data, one option is to use an instrumental variable approach, instrumenting local pollution levels with the phased introduction of SCAs. We are reluctant to do so however, since the policy itself may have led councils to change their spending patterns more generally. For example, by increasing their spending on pollution reduction policies, they may have had to reduce local expenditures on health and education, with potential direct impacts on our outcomes of interest. 

The 1956 Clean Air Act that allowed for the introduction of SCAs marked a significant step in the government's aim to reduce air pollution and improve public health. With its requirement that coal be replaced with smokeless fuels, it was among the first that aimed to reduce pollution emitted from \textit{residential} dwellings, with most previous policies aimed at industrial pollution. The move to smokeless fuels required adaptations of heating appliances, the cost of which was partially reimbursed by local authorities, which in turn would receive a contribution from the exchequer. By 1973, over 1,000 SCAs were introduced in English CBs, highlighting the widespread awareness of the health impacts of pollution as well as willingness to pay to reduce these not only by the government and local authorities, but also among the population. 

This in turn may suggest the presence of selective migration in response to local levels of air pollution. Indeed, individuals who were living in areas characterised by high levels of pollution (and little prospect of local authority intervention) may have been more likely to move into less polluted areas. Such potential avoidance behaviour is a limitation of our research; with no data on house moves in early life, and a sibling sample that is too small for any within-family analysis (n$\sim$160), we cannot explore this. However, the existence of such avoidance behaviours would likely lead to an underestimate of our effect of interest. Indeed, we define individuals who are conceived in non-SCAs, but who moved during the prenatal period to a SCA, as treated, despite them being exposed to higher levels of pollution in the early gestational period. Similarly, individuals who were born outside a SCA but who moved into one in early childhood would be defined as control (since we only observe the location \textit{at birth}), despite being exposed to less pollution in childhood. Both cases would lead to underestimating the difference between treated and control individuals. 

There are other reasons to believe that our coefficients underestimate the impacts of the introduction of SCAs. First, since pollution is linked to infant mortality and foetal loss, the higher levels of pollution in non-SCAs may have led to increased mortality \citep{fukushima2021uk}. Assuming that these individuals were more vulnerable than those who survived, our estimates are likely to be a lower bound.  Similarly, the fact that we only observe UK Biobank participants when they enter the data collection in 2006--10, we implicitly condition on survival until then. These survivors are likely to be stronger than those who did not make it, potentially attenuating our estimates of interest. 

Nevertheless, our analysis highlights the importance of a healthy environment in early life, showing immediate as well as long term impacts on individuals exposed to pollution in the early childhood period. This has direct implications for policy, suggesting that interventions that aim to reduce pollution not only have contemporaneous health benefits, but also improve individuals' outcomes in older age. Ignoring such longer-term effects underestimates the total welfare effects caused by pollution reduction.

% References
\newpage 
\singlespacing
\printbibliography
\newpage 
\doublespacing

% Appendix
\clearpage
\section*{\bfseries Online Appendix}
\appendix
\counterwithin{figure}{section}
\counterwithin{table}{section}

\FloatBarrier
\section{Additional Tables and Figures}
\label{sec:additional}

\begin{figure}[!h]\caption{\label{uni_sca_spatial}Number of smoke control areas in operation in England by 1973.}\centering\includegraphics[width=0.9\textwidth]{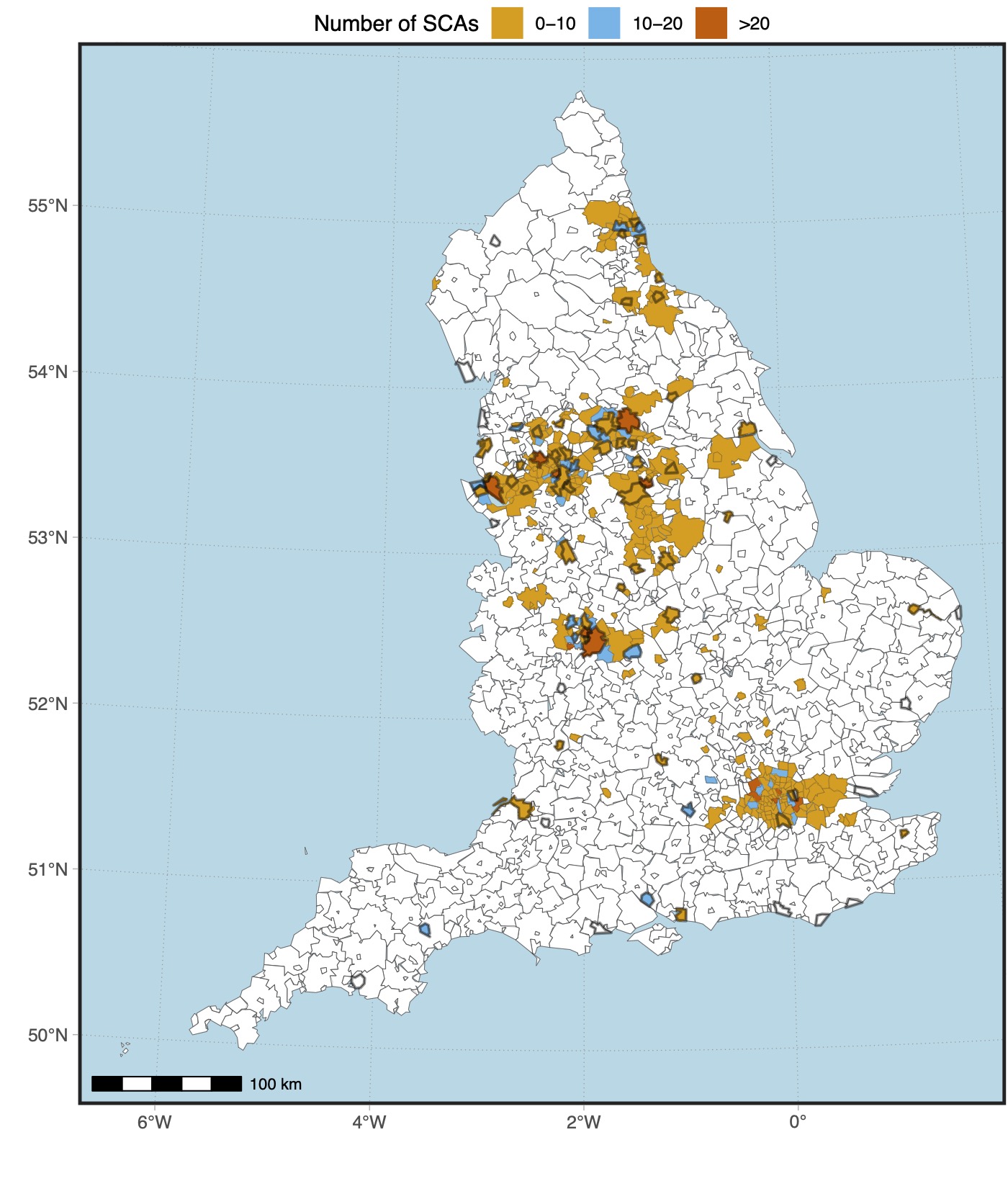}\caption*{\emph{Shows the spatial distribution of smoke control areas in operation in England by the end of 1973. The bold outlines indicate the boundaries of county boroughs.}}\end{figure}
\begin{figure}[!h]\caption{\label{fig:stations_waiting-times} Waiting times from submission to operation.}\centering\includegraphics[width=0.575\textwidth]{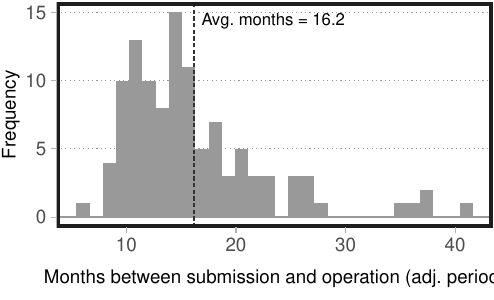}\caption*{\emph{Histogram of the waiting times between submission and operation dates. Only shows waiting times for smoke control areas in which we have at least one station in our pollution analysis. The average waiting time is 16.2 months.}}\end{figure}

% manually generated table showing mapping of qualifications to educational attainment
\begin{table}[h]
\caption{\label{tab:years_educ_definition}Mapping between qualifications and years of education.}
\centering
\begin{threeparttable}
\begin{tabular}{lc}
\toprule
Qualifications & Years of education \\ \midrule
College or university degree & 16 \\
A/AS levels $+$ NVQ/HND/HNC & 14 \\
A/AS levels $+$ Other professional qualifications & 15 \\
NVQ/HND/HNC & 13 \\
Other professional qualifications & 12 \\
A/AS levels & 13 \\
CSEs, GCSEs, or O levels & 11 \\
No qualifications & 10 \\ \bottomrule
\end{tabular}
\begin{tablenotes}
\item Columns: (1) the qualifications recorded in the UK Biobank, (2) the assigned years of education. A plus indicates that the individual must hold both of the specified qualifications simultaneously.
\end{tablenotes}
\end{threeparttable}
\end{table}

% manually generated table showing the adoption statuses of CBs in our sample
\begin{table}[!h]
\caption{\label{tab:cb_classification}County boroughs -- Smoke control area adoption status.}
\centering
\begin{threeparttable}
\fontsize{11}{13}\selectfont
\setlength{\tabcolsep}{2em}
\begin{tabular}[t]{lp{0.65\textwidth}}
\toprule
& \multicolumn{1}{c}{County boroughs}\\
\midrule
Adopting & \pvarConstructGroupsNamesOfCBsAdopting. \\
Non-adopting & \pvarConstructGroupsNamesOfCBsNonAdopting. \\
Dropped & \pvarConstructGroupsNamesOfCBsWithBorderIrregularity \\
\bottomrule
\end{tabular}
\begin{tablenotes}
\item County boroughs used in our analyses by SCA adoption status by end of 1973. Also lists the county boroughs dropped from the analyses due to border irregularities (see \autoref{sec:data_pollution}).%.
\end{tablenotes}
\end{threeparttable}
\end{table}

\begin{table}[!h]
\centering\centering\centering
\caption{\label{tab:stations_area_bs-so2}Difference-in-difference estimates with continuous treatments -- Impact on pollution.}
\centering
\begin{threeparttable}
\fontsize{11}{13}\selectfont
\setlength{\tabcolsep}{1.5pt}
\begin{tabular}[t]{ldddd}
\toprule
\multicolumn{1}{c}{\em{}} & \multicolumn{4}{c}{\em{Depedent variable:}} \\
\cmidrule(l{3pt}r{3pt}){2-5}
\multicolumn{1}{c}{} & \multicolumn{1}{c}{(1)} & \multicolumn{1}{c}{(2)} & \multicolumn{1}{c}{(3)} & \multicolumn{1}{c}{(4)} \\
\multicolumn{1}{c}{ } & \multicolumn{1}{c}{{\specialcell[b]{Black \\ smoke}}} & \multicolumn{1}{c}{{\specialcell[b]{Black \\ smoke} }} & \multicolumn{1}{c}{{\specialcell[b]{Sulphur \\ dioxide}}} & \multicolumn{1}{c}{{\specialcell[b]{Sulphur \\ dioxide} }}\\
\midrule
Area, surrounding + upwind, km$^2$ & -0.432^{  } & {} & -0.412^{  } & {}\\
 & (0.535) & {} & (0.566) & {}\\
Area, surrounding, km$^2$ & {} & -0.575^{  } & {} & 0.251^{  }\\
 & {} & (0.562) & {} & (0.564)\\
Area, upwind, km$^2$ & {} & -0.229^{  } & {} & -1.359^{  }\\
 & {} & (0.901) & {} & (1.088)\\
\midrule
Observations & \multicolumn{1}{D{,}{,}{-3}}{26,302} & \multicolumn{1}{D{,}{,}{-3}}{26,302} & \multicolumn{1}{D{,}{,}{-3}}{26,195} & \multicolumn{1}{D{,}{,}{-3}}{26,195}\\
Mean dep. var. & \multicolumn{1}{d}{103.767} & \multicolumn{1}{d}{103.767} & \multicolumn{1}{d}{132.721} & \multicolumn{1}{d}{132.721}\\
$R^2$ & \multicolumn{1}{d}{0.81} & \multicolumn{1}{d}{0.81} & \multicolumn{1}{d}{0.79} & \multicolumn{1}{d}{0.79}\\
\bottomrule
\end{tabular}
\begin{tablenotes}
\item Columns: (1-2) level of black smoke, (3-4) level of sulphur dioxide. Control group consists of never-treated stations from both adopting and non-adopting county boroughs. Includes year-by-month and station fixed effects, and includes a station-specific yearly linear time-trend to capture differences in linear dynamics between stations. Trims the sample to 5 years before and after the SCA submission date, drops always treated stations, and restricts the sample to pollution data for years 1962 to 1973. Clusters standard errors by station. (*): $p < 0.1$, (**): $p<0.05$, (***): $p<0.01$.
\end{tablenotes}
\end{threeparttable}
\end{table}

\begin{table}[!h]
\centering\centering\centering
\caption{\label{tab:individuals_did_low-bw}Difference-in-difference estimates -- Impact on individuals.}
\centering
\begin{threeparttable}
\fontsize{11}{13}\selectfont
\setlength{\tabcolsep}{1.5pt}
\begin{tabular}[t]{ld}
\toprule
\multicolumn{1}{c}{\em{}} & \multicolumn{1}{c}{\em{Dependent variable:}} \\
\cmidrule(l{3pt}r{3pt}){2-2}
\multicolumn{1}{c}{} & \multicolumn{1}{c}{(1)} \\
\multicolumn{1}{c}{ } & \multicolumn{1}{c}{{\specialcell[b]{Low \\ birth weight}}}\\
\midrule
Inside $\times$ Adj. & -0.018^{  }\\
 & (0.019)\\
Inside $\times$ Post & -0.024^{ * }\\
 & (0.013)\\
\midrule
Observations & \multicolumn{1}{D{,}{,}{-3}}{8,510}\\
Mean dep. var. & \multicolumn{1}{d}{3.317}\\
$R^2$ & \multicolumn{1}{d}{0.051}\\
\bottomrule
\end{tabular}
\begin{tablenotes}
\item Columns: (1) binary indicator of whether birth weight is low, that is, under 2,500 grams. OLS specification includes year-by-month and (CB $\times$ Inside) fixed effects, and includes (CB $\times$ Inside)-specific linear time trends. Controls for sex, ethnicity, and weather in utero and during childhood. Control group consists of never-treated individuals from both adopting and non-adopting county boroughs. Trims the sample to five years before and after the SCA submission date, and restricts the sample to birth cohorts in years 1958 to 1969. Clusters standard errors by CB. (*): $p < 0.1$, (**): $p<0.05$, (***): $p<0.01$.
\end{tablenotes}
\end{threeparttable}
\end{table}

\begin{table}[!h]
\centering\centering\centering
\caption{\label{tab:individuals_did_qualifications}Difference-in-difference estimates -- Impact on individuals.}
\centering
\begin{threeparttable}
\fontsize{11}{13}\selectfont
\setlength{\tabcolsep}{1.5pt}
\begin{tabular}[t]{ldddd}
\toprule
\multicolumn{1}{c}{\em{}} & \multicolumn{4}{c}{\em{Exits with highest qualification:}} \\
\cmidrule(l{3pt}r{3pt}){2-5}
\multicolumn{1}{c}{} & \multicolumn{1}{c}{(1)} & \multicolumn{1}{c}{(2)} & \multicolumn{1}{c}{(3)} & \multicolumn{1}{c}{(4)} \\
\multicolumn{1}{c}{ } & \multicolumn{1}{c}{{\specialcell[b]{Degree}}} & \multicolumn{1}{c}{{\specialcell[b]{Upper secondary}}} & \multicolumn{1}{c}{{\specialcell[b]{Lower secondary}}} & \multicolumn{1}{c}{{\specialcell[b]{None}}}\\
\midrule
Inside $\times$ Adj. & 0.004^{  } & 0.018^{  } & -0.020^{  } & -0.003^{  }\\
 & (0.036) & (0.024) & (0.023) & (0.013)\\
Inside $\times$ Post & -0.006^{  } & 0.033^{  } & -0.039^{ * } & 0.012^{  }\\
 & (0.033) & (0.028) & (0.021) & (0.011)\\
\midrule
Observations & \multicolumn{1}{D{,}{,}{-3}}{11,689} & \multicolumn{1}{D{,}{,}{-3}}{11,689} & \multicolumn{1}{D{,}{,}{-3}}{11,689} & \multicolumn{1}{D{,}{,}{-3}}{11,689}\\
Mean dep. var. & \multicolumn{1}{d}{0.306} & \multicolumn{1}{d}{0.399} & \multicolumn{1}{d}{0.243} & \multicolumn{1}{d}{0.052}\\
$R^2$ & \multicolumn{1}{d}{0.105} & \multicolumn{1}{d}{0.042} & \multicolumn{1}{d}{0.051} & \multicolumn{1}{d}{0.047}\\
\bottomrule
\end{tabular}
\begin{tablenotes}
\item Columns: (1) exits with university/college, (2) exits at upper secondary level (A/AS-levels, professional/vocational training), (3) exits at lower secondary level (CSEs, GCSEs, O-levels), (4) exits with no qualifications. OLS specification includes year-by-month and (CB $\times$ Inside) fixed effects, and includes (CB $\times$ Inside)-specific linear time trends. Controls for sex, ethnicity, and weather in utero and during childhood. Control group consists of never-treated individuals from both adopting and non-adopting county boroughs. Trims the sample to five years before and after the SCA submission date, and restricts the sample to birth cohorts in years 1958 to 1969. Clusters standard errors by CB. (*): $p < 0.1$, (**): $p<0.05$, (***): $p<0.01$.
\end{tablenotes}
\end{threeparttable}
\end{table}

\begin{table}[!h]
\centering\centering\centering
\caption{\label{tab:individuals_did_sex}Difference-in-difference estimates -- Impact on the probability of being born female.}
\centering
\begin{threeparttable}
\fontsize{11}{13}\selectfont
\setlength{\tabcolsep}{1.5pt}
\begin{tabular}[t]{ld}
\toprule
\multicolumn{1}{c}{\em{}} & \multicolumn{1}{c}{\em{Dependent variable:}} \\
\cmidrule(l{3pt}r{3pt}){2-2}
\multicolumn{1}{c}{} & \multicolumn{1}{c}{(1)} \\
\multicolumn{1}{c}{ } & \multicolumn{1}{c}{{\specialcell[b]{Pr(Female)}}}\\
\midrule
Inside $\times$ Adj. & -0.009^{  }\\
 & (0.019)\\
Inside $\times$ Post & 0.019^{  }\\
 & (0.016)\\
\midrule
Observations & \multicolumn{1}{D{,}{,}{-3}}{11,944}\\
Mean dep. var. & \multicolumn{1}{d}{0.453}\\
$R^2$ & \multicolumn{1}{d}{0.031}\\
\bottomrule
\end{tabular}
\begin{tablenotes}
\item Columns: (1) impact on probability of being born female. OLS specification includes year-by-month and (CB $\times$ Inside) fixed effects, and includes (CB $\times$ Inside)-specific linear time trends. Controls for sex, ethnicity, and weather in utero and during childhood. Control group consists of never-treated individuals from both adopting and non-adopting county boroughs. Trims the sample to five years before and after the SCA submission date, and restricts the sample to birth cohorts in years 1958 to 1969. Clusters standard errors by CB. (*): $p < 0.1$, (**): $p<0.05$, (***): $p<0.01$.
\end{tablenotes}
\end{threeparttable}
\end{table}

\clearpage
\FloatBarrier
\section{Randomness in timing and selection of smokeless areas}
\label{sec:appendix_timing} 
Our identification makes two assumptions. First, it assumes that the timing of the introduction of smoke control areas is random. Second, that conditional on controls and fixed effects, the areas selected to be smokeless were not systematically different from those that were not selected. We explore both of these assumptions below.

\paragraph{Timing.} To investigate determinants of the timing of SCA implementation, we construct a time index that counts the number of months elapsed since 1 January 1950 and define our dependent variable as the indexed time when 10\% of the CB's area was under smoke control. We then plot and regress this variable against CB-level pre-SCA characteristics to examine whether the timing is systematically related to such pre-programme covariates. Hence, this analysis relies on between-CB variation in timing and covariates.

\begin{figure}[!h]\caption{\label{fig:sca_randomness_10pctcoverage_scatterplot_density}Population density and time till 10 pct. of district area covered}\centering\includegraphics[width=0.65\textwidth]{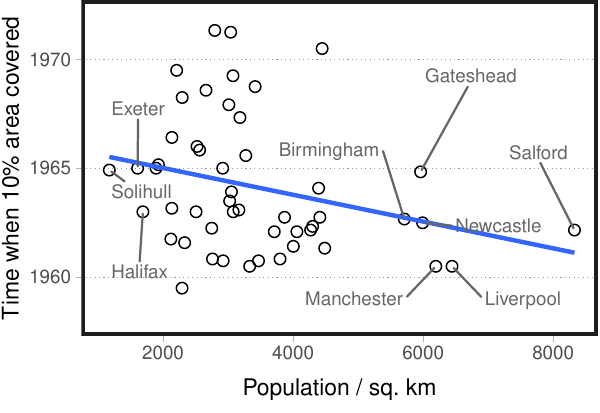}\caption*{\emph{Plots districts' population per sq. km in 1951 against the time when they reached 10 pct. of area covered by smoke control areas.}}\end{figure}

As the 1952 smog in the capital was the driver of the Clean Air Act, we first explore the relationship between the timing of SCA introduction and population density, where we expect more densely populated cities to implement SCAs earlier than less densely populated cities. To do this, we merge in CB-level population counts from the 1951 UK Census. \autoref{fig:sca_randomness_10pctcoverage_scatterplot_density} plots the time to reach 10\% coverage against the population density, showing that more densely populated areas such as Manchester, Birmingham and Liverpool implemented SCAs earlier than less densely populated areas.\footnote{We focus on the time associated with a 10\% coverage rate rather than the time of (e.g.) the first SCA because almost 75\% of all adopting CBs implement their first SCA within the same 2-3 years (i.e., between 1957-1960). However, we show robustness of these analysis to different thresholds in \autoref{tab:sca_randomness_regressions_extended} below.} Column 1 in \autoref{tab:sca_randomness_regressions} quantifies these estimates, showing that $1,000$ more residents per $km^2$ is associated with reaching the 10\% coverage approximately half a year earlier.\footnote{We focus only on treated CBs in this analysis, and we drop \pvarAnalysisRandomnessNumberCbsNotReachTenPctCoverage CBs (\pvarAnalysisRandomnessNamesCbsNotReachTenPctCoverage) that do not reach the 10\% milestone by the end of our sampling window.  } Although this is an interesting observation, it is not a problem for our analyses, since we control for station and $\text{CB} \times \text{Inside}$ fixed effects as well as station and $\text{CB} \times \text{Inside}$-specific annual trends in the pollution and individual-level analysis respectively.

We next turn to pre-programme CB-level characteristics obtained from the 1951 UK Census as well as our historical pollution data to explore the extent to which these variables explain the timing of SCA introduction. To account for the fact that larger cities implement SCAs earlier than smaller ones, the following analyses all control for population density. 
First, using the CB-level shares of residents by socio-economic class in the 1951 census, we see no strong relationship between CBs' socio-economic composition and the timing of SCA implementation (see columns (2) and (3) of \autoref{tab:sca_randomness_regressions}). 
Second, using average CB-levels of black smoke and sulphur dioxide over the period 1954--1956, we find that pre-programme pollution levels have negligible associations with SCA timings. This suggests that CBs that were more polluted prior to 1957 did not adopt SCAs systematically earlier compared to less polluted CBs (see columns (4) and (5) of \autoref{tab:sca_randomness_regressions}).

\begin{table}[!h]
\centering\centering\centering
\caption{\label{tab:sca_randomness_regressions}OLS estimates -- Pre-treatment characteristics and timing of implementation.}
\centering
\begin{threeparttable}
\fontsize{12}{14}\selectfont
\setlength{\tabcolsep}{3pt}
\begin{tabular}[t]{lddddd}
\toprule
\multicolumn{1}{c}{\em{}} & \multicolumn{5}{c}{\em{Specification}} \\
\cmidrule(l{3pt}r{3pt}){2-6}
\multicolumn{1}{c}{} & \multicolumn{1}{c}{(1)} & \multicolumn{1}{c}{(2)} & \multicolumn{1}{c}{(3)} & \multicolumn{1}{c}{(4)} & \multicolumn{1}{c}{(5)} \\
\multicolumn{1}{c}{ } & \multicolumn{1}{c}{{\specialcell[b]{10\% \\ covered}}} & \multicolumn{1}{c}{{\specialcell[b]{10\% \\ covered} }} & \multicolumn{1}{c}{{\specialcell[b]{10\% \\ covered}  }} & \multicolumn{1}{c}{{\specialcell[b]{10\% \\ covered}   }} & \multicolumn{1}{c}{{\specialcell[b]{10\% \\ covered}    }}\\
\midrule
Population per sq. km & -0.007^{ ** } & -0.008^{ ** } & -0.009^{ ** } & -0.008^{  } & -0.006^{  }\\
 & (0.004) & (0.004) & (0.004) & (0.005) & (0.005)\\
Pct. low SES in 1951 & {} & 0.488^{  } & {} & {} & {}\\
 & {} & (0.972) & {} & {} & {}\\
Pct. high SES in 1951 & {} & {} & -1.396^{  } & {} & {}\\
 & {} & {} & (1.420) & {} & {}\\
Pre-1957 black smoke & {} & {} & {} & -0.016^{  } & {}\\
 & {} & {} & {} & (0.063) & {}\\
Pre-1957 sulphur dioxide & {} & {} & {} & {} & -0.018^{  }\\
 & {} & {} & {} & {} & (0.087)\\
\midrule
Observations & \multicolumn{1}{D{,}{,}{-3}}{50} & \multicolumn{1}{D{,}{,}{-3}}{50} & \multicolumn{1}{D{,}{,}{-3}}{50} & \multicolumn{1}{D{,}{,}{-3}}{20} & \multicolumn{1}{D{,}{,}{-3}}{18}\\
$R^2$ & \multicolumn{1}{d}{0.079} & \multicolumn{1}{d}{0.084} & \multicolumn{1}{d}{0.097} & \multicolumn{1}{d}{0.166} & \multicolumn{1}{d}{0.122}\\
\bottomrule
\end{tabular}
\begin{tablenotes}
\item Dependent variable is the time index when 10 \% of the district was covered by SCAs. The time index starts at one on 1 January 1950 and increases by one unit per month. SES and population data are from the 1951 census (i.e. prior to the SCA programme) and pollution levels are the average from 1954-1956 (i.e. prior to the SCA programme) using our pollution panel. (*): $p < 0.1$, (**): $p<0.05$, (***): $p<0.01$.
\end{tablenotes}
\end{threeparttable}
\end{table}

\begin{table}[!h]
\centering\centering\centering
\caption{\label{tab:sca_randomness_regressions_extended}OLS estimates -- Pre-treatment characteristics and timing of implementation.}
\centering
\begin{threeparttable}
\fontsize{7}{9}\selectfont
\setlength{\tabcolsep}{1.5pt}
\begin{tabular}[t]{lddddd}
\toprule
\multicolumn{1}{c}{\em{}} & \multicolumn{5}{c}{\em{Specification}} \\
\cmidrule(l{3pt}r{3pt}){2-6}
\multicolumn{1}{c}{} & \multicolumn{1}{c}{(1)} & \multicolumn{1}{c}{(2)} & \multicolumn{1}{c}{(3)} & \multicolumn{1}{c}{(4)} & \multicolumn{1}{c}{(5)} \\
\multicolumn{1}{c}{ } & \multicolumn{1}{c}{{\specialcell[b]{Time}}} & \multicolumn{1}{c}{{\specialcell[b]{Time\hspace{0pt}}}} & \multicolumn{1}{c}{{\specialcell[b]{Time\hspace{0pt}\hspace{0pt}}}} & \multicolumn{1}{c}{{\specialcell[b]{Time\hspace{0pt}\hspace{0pt}\hspace{0pt}}}} & \multicolumn{1}{c}{{\specialcell[b]{Time\hspace{0pt}\hspace{0pt}\hspace{0pt}\hspace{0pt}}}}\\
\midrule
\addlinespace[0.5em]
\multicolumn{6}{l}{\textit{Panel A -- Time till first SCA}}\\
\midrule \hspace{1em}Population per sq. km & -0.002^{  } & 0.000^{  } & -0.003^{  } & 0.000^{  } & 0.007^{  }\\
\hspace{1em} & (0.004) & (0.005) & (0.005) & (0.006) & (0.006)\\
\hspace{1em}Pct. low SES in 1951 & {} & -0.767^{  } & {} & {} & {}\\
\hspace{1em} & {} & (1.085) & {} & {} & {}\\
\hspace{1em}Pct. high SES in 1951 & {} & {} & -1.344^{  } & {} & {}\\
\hspace{1em} & {} & {} & (1.504) & {} & {}\\
\hspace{1em}Pre-1957 black smoke & {} & {} & {} & 0.015^{  } & {}\\
\hspace{1em} & {} & {} & {} & (0.085) & {}\\
\hspace{1em}Pre-1957 sulphur dioxide & {} & {} & {} & {} & -0.178^{ * }\\
\hspace{1em} & {} & {} & {} & {} & (0.093)\\
\hspace{1em}Observations & \multicolumn{1}{D{,}{,}{-3}}{53} & \multicolumn{1}{D{,}{,}{-3}}{53} & \multicolumn{1}{D{,}{,}{-3}}{53} & \multicolumn{1}{D{,}{,}{-3}}{21} & \multicolumn{1}{D{,}{,}{-3}}{19}\\
\hspace{1em}$R^2$ & \multicolumn{1}{d}{0.003} & \multicolumn{1}{d}{0.013} & \multicolumn{1}{d}{0.018} & \multicolumn{1}{d}{0.002} & \multicolumn{1}{d}{0.197}\\
\addlinespace[0.5em]
\multicolumn{6}{l}{\textit{Panel B -- Time till 10\% coverage}}\\
\midrule \hspace{1em}Population per sq. km & -0.007^{ ** } & -0.008^{ ** } & -0.009^{ ** } & -0.008^{  } & -0.006^{  }\\
\hspace{1em} & (0.004) & (0.004) & (0.004) & (0.005) & (0.005)\\
\hspace{1em}Pct. low SES in 1951 & {} & 0.488^{  } & {} & {} & {}\\
\hspace{1em} & {} & (0.972) & {} & {} & {}\\
\hspace{1em}Pct. high SES in 1951 & {} & {} & -1.396^{  } & {} & {}\\
\hspace{1em} & {} & {} & (1.420) & {} & {}\\
\hspace{1em}Pre-1957 black smoke & {} & {} & {} & -0.016^{  } & {}\\
\hspace{1em} & {} & {} & {} & (0.063) & {}\\
\hspace{1em}Pre-1957 sulphur dioxide & {} & {} & {} & {} & -0.018^{  }\\
\hspace{1em} & {} & {} & {} & {} & (0.087)\\
\hspace{1em}Observations & \multicolumn{1}{D{,}{,}{-3}}{50} & \multicolumn{1}{D{,}{,}{-3}}{50} & \multicolumn{1}{D{,}{,}{-3}}{50} & \multicolumn{1}{D{,}{,}{-3}}{20} & \multicolumn{1}{D{,}{,}{-3}}{18}\\
\hspace{1em}$R^2$ & \multicolumn{1}{d}{0.079} & \multicolumn{1}{d}{0.084} & \multicolumn{1}{d}{0.097} & \multicolumn{1}{d}{0.166} & \multicolumn{1}{d}{0.122}\\
\addlinespace[0.5em]
\multicolumn{6}{l}{\textit{Panel C -- Time till 25\% coverage}}\\
\midrule \hspace{1em}Population per sq. km & -0.013^{ *** } & -0.013^{ *** } & -0.015^{ *** } & -0.013^{ ** } & -0.008^{  }\\
\hspace{1em} & (0.004) & (0.005) & (0.005) & (0.005) & (0.006)\\
\hspace{1em}Pct. low SES in 1951 & {} & -0.077^{  } & {} & {} & {}\\
\hspace{1em} & {} & (1.200) & {} & {} & {}\\
\hspace{1em}Pct. high SES in 1951 & {} & {} & -1.481^{  } & {} & {}\\
\hspace{1em} & {} & {} & (2.156) & {} & {}\\
\hspace{1em}Pre-1957 black smoke & {} & {} & {} & -0.008^{  } & {}\\
\hspace{1em} & {} & {} & {} & (0.073) & {}\\
\hspace{1em}Pre-1957 sulphur dioxide & {} & {} & {} & {} & -0.070^{  }\\
\hspace{1em} & {} & {} & {} & {} & (0.102)\\
\hspace{1em}Observations & \multicolumn{1}{D{,}{,}{-3}}{45} & \multicolumn{1}{D{,}{,}{-3}}{45} & \multicolumn{1}{D{,}{,}{-3}}{45} & \multicolumn{1}{D{,}{,}{-3}}{18} & \multicolumn{1}{D{,}{,}{-3}}{16}\\
\hspace{1em}$R^2$ & \multicolumn{1}{d}{0.183} & \multicolumn{1}{d}{0.183} & \multicolumn{1}{d}{0.192} & \multicolumn{1}{d}{0.298} & \multicolumn{1}{d}{0.251}\\
\addlinespace[0.5em]
\multicolumn{6}{l}{\textit{Panel D -- Time till 50\% coverage}}\\
\midrule \hspace{1em}Population per sq. km & -0.002^{  } & -0.002^{  } & -0.002^{  } & 0.000^{  } & 0.003^{  }\\
\hspace{1em} & (0.004) & (0.005) & (0.005) & (0.005) & (0.005)\\
\hspace{1em}Pct. low SES in 1951 & {} & -0.043^{  } & {} & {} & {}\\
\hspace{1em} & {} & (1.579) & {} & {} & {}\\
\hspace{1em}Pct. high SES in 1951 & {} & {} & -0.048^{  } & {} & {}\\
\hspace{1em} & {} & {} & (2.519) & {} & {}\\
\hspace{1em}Pre-1957 black smoke & {} & {} & {} & -0.041^{  } & {}\\
\hspace{1em} & {} & {} & {} & (0.090) & {}\\
\hspace{1em}Pre-1957 sulphur dioxide & {} & {} & {} & {} & -0.109^{  }\\
\hspace{1em} & {} & {} & {} & {} & (0.131)\\
\hspace{1em}Observations & \multicolumn{1}{D{,}{,}{-3}}{20} & \multicolumn{1}{D{,}{,}{-3}}{20} & \multicolumn{1}{D{,}{,}{-3}}{20} & \multicolumn{1}{D{,}{,}{-3}}{10} & \multicolumn{1}{D{,}{,}{-3}}{9}\\
\hspace{1em}$R^2$ & \multicolumn{1}{d}{0.011} & \multicolumn{1}{d}{0.011} & \multicolumn{1}{d}{0.011} & \multicolumn{1}{d}{0.028} & \multicolumn{1}{d}{0.112}\\
\bottomrule
\end{tabular}
\begin{tablenotes}
\item The time index starts at one on 1 January 1950 and increases by one unit per month. SES and population data are from the 1951 census (i.e. prior to the SCA programme) and pollution levels are the average from 1954-1956 (i.e. prior to the SCA programme) using our pollution panel. (*): $p < 0.1$, (**): $p<0.05$, (***): $p<0.01$.
\end{tablenotes}
\end{threeparttable}
\end{table}

These analyses compare \textit{across} CBs and support the assumption that the timing of SCA introduction is largely unrelated to CB-level pre-programme characteristics. 
We next explore the timing of SCA introduction \textit{within} CBs to investigate whether the timing of one SCA is correlated to its pre-programme characteristics relative to another SCA within the same CB. To do this, we construct a sample of CBs with at least two pollution stations located in different SCAs. In \autoref{tab:stations_assoc_pretreatment-timing}, we then regress the pre-treatment black smoke and sulphur dioxide concentrations on the time index, conditional on CB fixed effects (all columns) and year fixed effects (Columns 2 and 4). This suggests that \textit{within} a CB, stations in SCAs that were introduced earlier (i.e., with a `lower' time of submission) had significantly higher levels of black smoke pre-programme. However, this relationship disappears when we account for time fixed effects. In summary, these analyses therefore support the assumption that the timing of the introduction of SCAs is as good as random.

\begin{table}[!h]
\centering\centering\centering
\caption{\label{tab:stations_assoc_pretreatment-timing}OLS estimates -- SCA-level pre-treatment pollution levels and timing of implementation.}
\centering
\begin{threeparttable}
\fontsize{10}{12}\selectfont
\setlength{\tabcolsep}{1.5pt}
\begin{tabular}[t]{ldddd}
\toprule
\multicolumn{1}{c}{\em{}} & \multicolumn{4}{c}{\em{Dependent variable:}} \\
\cmidrule(l{3pt}r{3pt}){2-5}
\multicolumn{1}{c}{} & \multicolumn{1}{c}{(1)} & \multicolumn{1}{c}{(2)} & \multicolumn{1}{c}{(3)} & \multicolumn{1}{c}{(4)} \\
\multicolumn{1}{c}{ } & \multicolumn{1}{c}{{\specialcell[b]{Black \\ smoke}}} & \multicolumn{1}{c}{{\specialcell[b]{Black \\ smoke} }} & \multicolumn{1}{c}{{\specialcell[b]{Sulphur \\ dioxide}}} & \multicolumn{1}{c}{{\specialcell[b]{Sulphur \\ dioxide} }}\\
\midrule
Time of submission, years & -12.556^{ *** } & -1.935^{  } & -4.032^{  } & 1.321^{  }\\
 & (2.771) & (2.553) & (3.963) & (3.692)\\
\midrule
Observations & \multicolumn{1}{D{,}{,}{-3}}{3,269} & \multicolumn{1}{D{,}{,}{-3}}{3,269} & \multicolumn{1}{D{,}{,}{-3}}{3,264} & \multicolumn{1}{D{,}{,}{-3}}{3,264}\\
CB FE & \multicolumn{1}{c}{Yes} & \multicolumn{1}{c}{Yes} & \multicolumn{1}{c}{Yes} & \multicolumn{1}{c}{Yes}\\
Year FE & \multicolumn{1}{c}{No} & \multicolumn{1}{c}{Yes} & \multicolumn{1}{c}{No} & \multicolumn{1}{c}{Yes}\\
$R^2$ & \multicolumn{1}{d}{0.283} & \multicolumn{1}{d}{0.309} & \multicolumn{1}{d}{0.311} & \multicolumn{1}{d}{0.329}\\
\bottomrule
\end{tabular}
\begin{tablenotes}
\item Columns: (1-2) level of black smoke in mcg/m3, (3-4) level of sulphur dioxide in mcg/m3. Regresses outcomes onto the time of submission of the stations' SCAs and fixed effects. Time of submission is measured as years elapsed from 1 January 1950 until the submission date. Restricts sample to measurements from stations in SCAs taken prior to the submission date, and at most 5 years before. Restricts the sample to pollution data for years 1962 to 1973. Clusters standard errors by station. (*): $p < 0.1$, (**): $p<0.05$, (***): $p<0.01$.
\end{tablenotes}
\end{threeparttable}
\end{table}

\FloatBarrier

\paragraph{Selection of areas.} The second identifying assumption is that, conditional on our controls and fixed effects, the areas within CBs that were selected to be smokeless were not systematically different from those that were not selected. Hence, this assumption is also based on \textit{within}-CB variation, distinguishing between areas that became smokeless and areas that did not within the same CB. To examine this, we regress a set of pre-determined characteristics onto an indicator that is equal to one if the station (individual) was located (born) inside a smoke control area, and zero otherwise. 

We start by using pre-\textit{treatment} pollution levels.\footnote{Note that this is slightly different from above, where we use pre-\textit{1957} pollution levels. This is because for a given CB, either all stations are inside or all stations are outside SCA boundaries prior to 1957, meaning there is no within-CB variation. Hence, we cannot compare pre-programme pollution levels for stations located inside vs outside SCAs \textit{within} a CB; we can only make this comparison using pre-\emph{treatment} (i.e., introduction of SCA) pollution levels.} 
\autoref{tab:stations_assoc_ever-in-sca} reports the results, showing that prior to the introduction of SCAs, areas selected to become SCAs were slightly less polluted compared to those not selected within the same CB. In particular, selected areas had on average about 12 mcg/m$^3$ lower pre-treatment levels of black smoke, and 9 mcg/m$^3$ lower levels of sulphur dioxide, compared to areas that were not selected, though only the former difference is statistically significant.

\begin{table}[!h]
\centering\centering\centering
\caption{\label{tab:stations_assoc_ever-in-sca}OLS estimates -- Associations between station being located in a SCA and pollution levels.}
\centering
\begin{threeparttable}
\fontsize{10}{12}\selectfont
\setlength{\tabcolsep}{1.5pt}
\begin{tabular}[t]{ldd}
\toprule
\multicolumn{1}{c}{\em{}} & \multicolumn{2}{c}{\em{Dependent variable:}} \\
\cmidrule(l{3pt}r{3pt}){2-3}
\multicolumn{1}{c}{} & \multicolumn{1}{c}{(1)} & \multicolumn{1}{c}{(2)} \\
\multicolumn{1}{c}{ } & \multicolumn{1}{c}{{\specialcell[b]{Black \\ smoke}}} & \multicolumn{1}{c}{{\specialcell[b]{Sulphur \\ dioxide}}}\\
\midrule
Inside & -12.419^{ ** } & -9.277^{  }\\
 & (5.908) & (8.748)\\
\midrule
Observations & \multicolumn{1}{D{,}{,}{-3}}{17,097} & \multicolumn{1}{D{,}{,}{-3}}{17,037}\\
Mean dep. var. & \multicolumn{1}{d}{113.565} & \multicolumn{1}{d}{140.585}\\
$R^2$ & \multicolumn{1}{d}{0.745} & \multicolumn{1}{d}{0.698}\\
\bottomrule
\end{tabular}
\begin{tablenotes}
\item Columns: (1) level of black smoke in mcg/m3, (2) level of sulphur dioxide in mcg/m3. Includes year-by-month and CB fixed effects, and includes CB-specific yearly trends. Restricts sample to stations in adopting CBs only, and only includes pre-treatment observations up to 5 years before SCA submission only. Restricts the sample to pollution data for years 1962 to 1973. Clusters standard errors by station. (*): $p < 0.1$, (**): $p<0.05$, (***): $p<0.01$.
\end{tablenotes}
\end{threeparttable}
\end{table}

Second, using the individual-level UK Biobank data, we focus on three pre-determined variables aiming to proxy socio-economic status: maternal smoking status around birth, whether one was breastfed, and the polygenic score for education.\footnote{\citet{von2023prevalence} highlights the social gradient in maternal smoking that appears in the UK after WWII using the UK Biobank. The proportion of UK Biobank participants that report to have been breastfed is higher in high SES areas, and the polygenic score for education is highly correlated with educational outcomes (see e.g. \autoref{sec:results_indiv}).} \autoref{tab:individuals_assoc_ever-in-sca_predetermined} reports the results. We find negative associations with maternal smoking, and positive associations with breast feeding and the polygenic score for educational attainment. This implies that individuals born in areas that would become smokeless were less likely to have a mother that smoked around birth, and more likely to have been breastfed and have a higher genetic ``predisposition'' for educational attainment. Taken together, this suggests that individuals conceived in areas that would become smokeless were generally of higher socio-economic status, compared to those conceived outside but in the same CB. To alleviate concerns about these differences driving the results in our main analysis, we include station fixed effects and station-specific annual trends in our pollution analysis, and (CB $\times$ Inside) fixed effects and (CB $\times$ Inside)-specific trends in our individual-level analysis. The latter accounts for systematic differences in means as well as trends between areas that did and did not become smokeless \textit{within} a CB.

\begin{table}[!h]
\centering\centering\centering
\caption{\label{tab:individuals_assoc_ever-in-sca_predetermined}OLS estimates -- Associations between being born inside a SCA boundary and individual pre-determined characteristics.}
\centering
\begin{threeparttable}
\fontsize{11}{13}\selectfont
\setlength{\tabcolsep}{1.5pt}
\begin{tabular}[t]{lddd}
\toprule
\multicolumn{1}{c}{\em{}} & \multicolumn{3}{c}{\em{Dependent variable:}} \\
\cmidrule(l{3pt}r{3pt}){2-4}
\multicolumn{1}{c}{} & \multicolumn{1}{c}{(1)} & \multicolumn{1}{c}{(2)} & \multicolumn{1}{c}{(3)} \\
\multicolumn{1}{c}{ } & \multicolumn{1}{c}{{\specialcell[b]{Maternal \\ smoking}}} & \multicolumn{1}{c}{{\specialcell[b]{Breastfed}}} & \multicolumn{1}{c}{{\specialcell[b]{PGS \\ Educ. attain.}}}\\
\midrule
Inside & -0.049^{ ** } & 0.038^{ ** } & 0.056^{ * }\\
 & (0.021) & (0.015) & (0.029)\\
\midrule
Observations & \multicolumn{1}{D{,}{,}{-3}}{9,056} & \multicolumn{1}{D{,}{,}{-3}}{8,586} & \multicolumn{1}{D{,}{,}{-3}}{9,258}\\
Mean dep. var. & \multicolumn{1}{d}{0.349} & \multicolumn{1}{d}{0.562} & \multicolumn{1}{d}{0.000}\\
$R^2$ & \multicolumn{1}{d}{0.051} & \multicolumn{1}{d}{0.075} & \multicolumn{1}{d}{0.054}\\
\bottomrule
\end{tabular}
\begin{tablenotes}
\item Columns: (1) whether mother was smoking around time of birth, (2) whether individual was breastfed, (3) standardised PGI for eductional attainment. OLS specification includes year-by-month and CB fixed effects, and includes CB-specific linear time trends. Controls for sex, ethnicity, and weather in utero and during childhood. Trims the sample. Restricts samples to individuals born in adopting CBs. Clusters standard errors by CB. (*): $p < 0.1$, (**): $p<0.05$, (***): $p<0.01$.
\end{tablenotes}
\end{threeparttable}
\end{table}

\clearpage
\FloatBarrier
\section{Construction of downwind variables}
\label{sec:appendix_area_and_downwind}
To explore potential downwind effects of new smoke control areas, we identify stations that are downwind of each SCA. We do this in three alternative ways, each of which exploiting the wind direction data from \citet{era5} to estimate the geographic boundaries of downwind areas and then identifying all stations within these boundaries. We describe these below, and examine the robustness of our results to the method used. Our preferred method, used in our main analysis, is the first of the following three.

\begin{figure}[h]
\caption{Illustrations of the downwind definitions.}
\includegraphics[width=\textwidth]{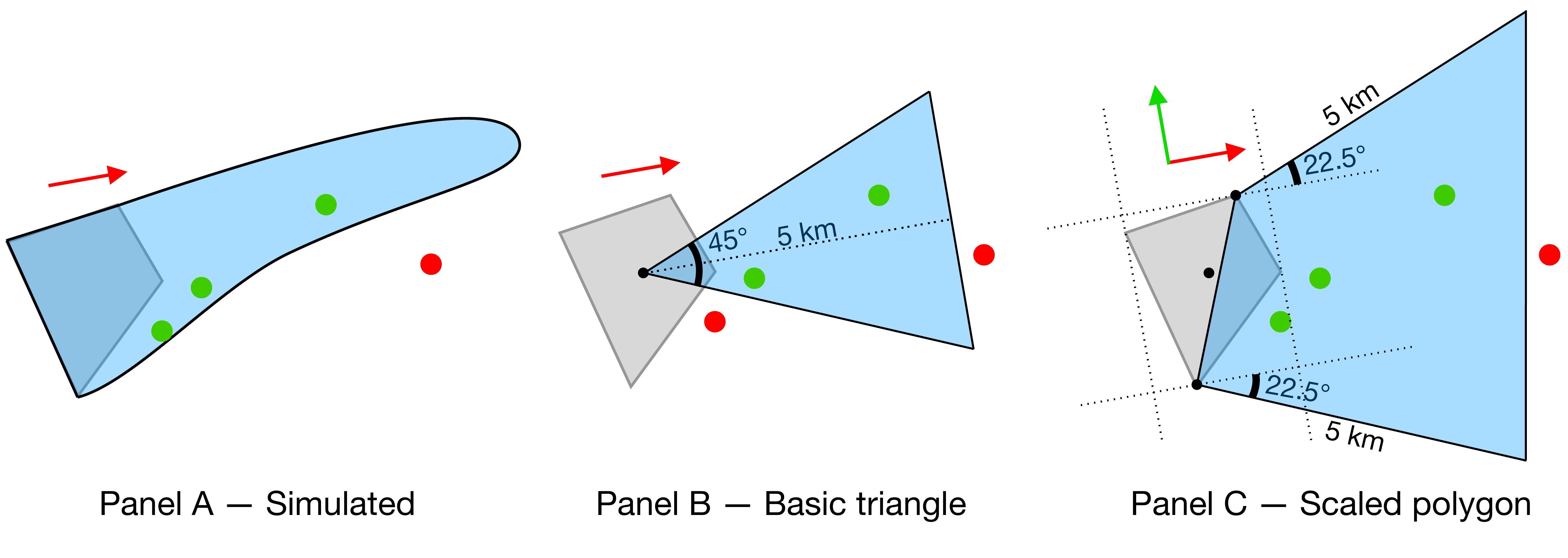}
\caption*{\emph{The three definitions of being downwind of a smoke control area. The smoke control area boundary is shown as the polygon filled with gray and the centroid of the boundary is marked by a black dot. The red arrow shows the wind vector. The green points show the units classified as `downwind', while the red points represent units classified as not being downwind.}}
\label{fig:downwind_definition_illustration}
\end{figure}

\paragraph{Simulated.} Panel~A of \autoref{fig:downwind_definition_illustration} illustrates the first, and preferred, method where we use a basic Gaussian dispersion model to estimate the boundary of the pollution plume for each SCA. Since we do not have data on actual emissions, the model does not give us concentrations that we can interpret in an absolute sense. We can however use the model to assess the approximate shapes of SCAs' pollution plumes and with that identify `downwind' pollution stations. We assume that the emission $S_{\text{SCA}}$ of a SCA is proportional to the SCA's area relative to the surrounding CB's area times the total population of the CB in 1951. This is a strong assumption as population density is not uniformly distributed within CBs. 
We place a number $n_c$ of `chimneys' in a uniform grid across the SCA and set the emission of each chimney to $S = S_{\text{SCA}} / n_c$. Based on the Gaussian dispersion model in \citet[Chp. 6]{seigneur2019air}, the pollution concentration $C$ at ground level at location $(x, y)$ downwind from a 4.5 meter tall chimney located at origin $(0, 0)$ is then calculated as:
$$C(x, y) = \frac{S}{2 \pi u \sigma_y(x) \sigma_z(x)} \exp\left( - \frac{y^2}{2 \sigma_y(x)^2} - \frac{4.5^2}{2 \sigma_z(x)^2} \right),$$
where $u$ is the wind speed in meters per second in the direction of $x$ (illustrated by the red arrow in \autoref{fig:downwind_definition_illustration}, we get this from our weather data), and $(\sigma_z, \sigma_y)$ are the Pasquill-Gifford dispersion coefficients:
$$
\sigma_z(x) = \exp\left(a_z + b_z \ln(x) + c_z \ln(x)^2\right), \quad \sigma_y(x) = \exp\left(a_y + b_y \ln(x) + c_y \ln(x)^2\right).
$$
We set $a_z = -2.341$, $b_z = 0.9477$, $c_z = -0.0020$, $a_y = -2.054$, $b_y = 1.0231$ and $c_y = -0.0076$, corresponding to wind regime C in Table~6.1 in \citet{seigneur2019air}.
To calculate the pollution concentration at a point $(x, y)$ downwind from the SCA, we take the sum of the concentrations at $(x, y)$ across all chimneys in the SCA. We then set a concentration threshold and use this to trace out the contour of the SCA's dispersed pollution.

\paragraph{Basic triangle.} Panel~B of \autoref{fig:downwind_definition_illustration} illustrates the second method which constructs isosceles triangles ('basic triangle') aligned to SCAs' predominant wind directions. We draw each triangle such that the symmetry axis (dotted line) is parallel to the SCA's average wind vector (red arrow, the wind vector is averaged component-wise over the two years prior to the SCA's submission date), and the apex (black dot) is at the SCA's centroid. We set the height of the triangle to five kilometers and the vertex angle to 45 degrees.

\paragraph{Scaled polygon.} Panel~C of \autoref{fig:downwind_definition_illustration} shows the third method, constructing polygons scaled to the size of the SCA (`scaled polygon'). For each SCA, we calculate its average wind vector (red arrow), find an orthogonal vector (green arrow), and project the SCA's boundary polygon onto the axis identified by the orthogonal vector to determine two anchor points (black dots). To construct the downwind polygon (solid black lines) we (1) connect the anchor points with a line segment, (2) at each anchor point draw line segments 5 km in length and with an angle of 22.5 degrees relative to the wind vector, and (3) close the polygon by connecting the endpoints of these two line segments. This procedure scales the downwind polygons to account for the dimensions of SCAs. Though, it fails in a number of edge cases, e.g., if the SCA is long and narrow, which is why we prefer the simulated downwind boundary in Panel~A.

\paragraph{Robustness.} To assess the sensitivity of our results, we use our pollution panel to estimate our main specification separately for each `downwind definition' and report the results in \autoref{tab:stations_did_robustness_downwind}. Panel~A replicates our preferred specification from \autoref{tab:stations_did_bs-so2}, showing an immediate reduction in levels of black smoke during the adjustment period, which increases 2-3 fold when the SCA is in operation. We find no impact on stations that are downwind of the SCA. Our estimates are very similar when we use the two alternative definitions, again showing large impacts on stations inside SCAs during and after the adjustment period, but no impacts for those downwind. We find no evidence of any effects on level sof sulphur dioxide. Taken together, these results suggest that our main estimates are robust to the downwind definition used. 

\begin{table}[!h]
\centering\centering\centering
\caption{\label{tab:stations_did_robustness_downwind}Difference-in-difference estimates -- Impact on pollution. Robustness to choice of downwind definition.}
\centering
\begin{threeparttable}
\fontsize{9}{11}\selectfont
\setlength{\tabcolsep}{5pt}
\begin{tabular}[t]{ldd}
\toprule
\multicolumn{1}{c}{\em{}} & \multicolumn{2}{c}{\em{Specification:}} \\
\cmidrule(l{3pt}r{3pt}){2-3}
\multicolumn{1}{c}{} & \multicolumn{1}{c}{(1)} & \multicolumn{1}{c}{(2)} \\
\multicolumn{1}{c}{ } & \multicolumn{1}{c}{{\specialcell[b]{Black \\ smoke}}} & \multicolumn{1}{c}{{\specialcell[b]{Sulphur \\ dioxide}}}\\
\midrule
\addlinespace[0.5em]
\multicolumn{3}{l}{\textit{Panel A -- Simulated:}}\\
\midrule \hspace{1em}Inside $\times$ Adj. & -7.866^{ *** } & -1.148^{  }\\
\hspace{1em} & (3.031) & (3.747)\\
\hspace{1em}Inside $\times$ Post & -19.811^{ *** } & -3.677^{  }\\
\hspace{1em} & (4.425) & (5.347)\\
\hspace{1em}Downwind $\times$ Adj. & -1.032^{  } & -6.245^{  }\\
\hspace{1em} & (3.727) & (4.743)\\
\hspace{1em}Downwind $\times$ Post & -8.225^{  } & -7.317^{  }\\
\hspace{1em} & (5.631) & (6.441)\\
\addlinespace[0.5em]
\multicolumn{3}{l}{\textit{Panel B -- Basic triangle:}}\\
\midrule \hspace{1em}Inside $\times$ Adj. & -8.172^{ *** } & -1.031^{  }\\
\hspace{1em} & (3.017) & (3.750)\\
\hspace{1em}Inside $\times$ Post & -19.950^{ *** } & -3.231^{  }\\
\hspace{1em} & (4.383) & (5.315)\\
\hspace{1em}Downwind $\times$ Adj. & -0.982^{  } & 0.118^{  }\\
\hspace{1em} & (3.473) & (5.716)\\
\hspace{1em}Downwind $\times$ Post & -4.096^{  } & 3.594^{  }\\
\hspace{1em} & (5.400) & (7.543)\\
\addlinespace[0.5em]
\multicolumn{3}{l}{\textit{Panel C -- Scaled polygon:}}\\
\midrule \hspace{1em}Inside $\times$ Adj. & -8.030^{ *** } & -1.182^{  }\\
\hspace{1em} & (3.000) & (3.751)\\
\hspace{1em}Inside $\times$ Post & -19.766^{ *** } & -3.506^{  }\\
\hspace{1em} & (4.387) & (5.346)\\
\hspace{1em}Downwind $\times$ Adj. & 3.143^{  } & 2.630^{  }\\
\hspace{1em} & (3.339) & (4.567)\\
\hspace{1em}Downwind $\times$ Post & -1.189^{  } & 0.536^{  }\\
\hspace{1em} & (5.392) & (5.594)\\
\midrule
\hspace{1em}Observations & \multicolumn{1}{D{,}{,}{-3}}{26,302} & \multicolumn{1}{D{,}{,}{-3}}{26,195}\\
\hspace{1em}Mean dep. var. & \multicolumn{1}{d}{103.767} & \multicolumn{1}{d}{132.721}\\
\hspace{1em}$R^2$ & \multicolumn{1}{d}{0.81} & \multicolumn{1}{d}{0.79}\\
\bottomrule
\end{tabular}
\begin{tablenotes}
\item Columns: (1) black smoke, (2) sulphur dioxide. Panels show the different definitions of `being downwind'. Panel~A is our main specification. Control group includes never-treated from both adopting and non-adopting county boroughs. Clusters standard errors by station. Includes year-by-month and station fixed effects, and includes a station-specific yearly linear time-trend to capture differences in linear dynamics between stations. Trims the sample to 5 years before and after the SCA order date, and restricts the sample to pollution data for years 1962 to 1973. (*): $p < 0.1$, (**): $p<0.05$, (***): $p<0.01$.
\end{tablenotes}
\end{threeparttable}
\end{table}

\FloatBarrier
\section{Robustness to sample selection and model specification}\label{sec:robustness}
This Appendix shows the robustness of our main findings to different sample selection criteria, assumptions, and model specifications. We explore the sensitivity of our results to alternative definitions of the control group (\autoref{sec:definition_of_control_group}), alternative definitions of the sample depending on how individuals are geolocated within county boroughs (\autoref{sec:robustness_bunching}), to the use of different bandwidths (\autoref{sec:robustness_trimming}), birth cohorts (\autoref{sec:robustness_birth-cohorts}), and the specification of time trends  (\autoref{sec:robustness_trends}).

\FloatBarrier
\subsection{Definition of control group}
\label{sec:definition_of_control_group}
Our main analysis specifies the control group as units located in CBs that never applied for an SCA (i.e., the non-adopting CBs), as well as units located \textit{outside} SCAs, but \textit{inside} CBs that introduced other SCAs (i.e., within adopting CBs). First, we investigate the sensitivity of our findings to omitting either group. \autoref{tab:stations_did_robustness_control-group} reports the results for pollution. Panel~A replicates our main estimates, Panel~B drops stations outside SCA boundaries but inside adopting CB, and Panel~C drops stations in non-adopting CBs. This shows that the effects on black smoke and sulphur dioxide concentrations are similar across the alternative definitions of the control group, and close to our main estimates. Similarly, for the impacts on individuals, Panels B and C of \autoref{tab:individuals_did_robustness_control-group} show negligible differences in the estimates that specify alternative control groups when comparing to our main estimates in Panel~A.

Second, we explore the robustness of our results to increasing the never-treated group by incorporating data on the \textit{universe} of SCAs in England and Wales (see \autoref{sec:Data_SCA}), considering all units located in areas that never applied for a smoke control area.\footnote{Since we do not observe the boundaries and shapes of SCAs \textit{outside} CBs, we cannot identify whether units (i.e., stations or individuals) are \textit{inside} or \textit{outside} an SCA in other (non-CB) adoption areas. We can, however, identify districts that \textit{never} introduced any SCAs and add the units located in these districts to the control group.} This increases the number of observations by a factor of $\sim$1.7 for the pollution analysis, and over 2 for individuals. We report the estimates for pollution and individuals using this alternative control group in Panel~D of \autoref{tab:stations_did_robustness_control-group} and \autoref{tab:individuals_did_robustness_control-group}, respectively. This again shows very similar estimates to those reported in our main analysis.

\begin{table}[!h]
\centering\centering\centering
\caption{\label{tab:stations_did_robustness_control-group}Difference-in-difference estimates -- Impact on pollution. Robustness to choice of control group.}
\centering
\begin{threeparttable}
\fontsize{9}{11}\selectfont
\setlength{\tabcolsep}{10pt}
\begin{tabular}[t]{ldd}
\toprule
\multicolumn{1}{c}{\em{}} & \multicolumn{2}{c}{\em{Specification:}} \\
\cmidrule(l{3pt}r{3pt}){2-3}
\multicolumn{1}{c}{} & \multicolumn{1}{c}{(1)} & \multicolumn{1}{c}{(2)} \\
\multicolumn{1}{c}{ } & \multicolumn{1}{c}{{\specialcell[b]{Black \\ smoke}}} & \multicolumn{1}{c}{{\specialcell[b]{Sulphur \\ dioxide}}}\\
\midrule
\addlinespace[0.5em]
\multicolumn{3}{l}{\textit{Panel A -- Main sample:}}\\
\midrule \hspace{1em}Inside $\times$ Adj. & -8.053^{ *** } & -1.180^{  }\\
\hspace{1em} & (3.010) & (3.740)\\
\hspace{1em}Inside $\times$ Post & -19.737^{ *** } & -3.480^{  }\\
\hspace{1em} & (4.380) & (5.337)\\
\hspace{1em}Observations & \multicolumn{1}{D{,}{,}{-3}}{26,302} & \multicolumn{1}{D{,}{,}{-3}}{26,195}\\
\hspace{1em}Mean dep. var. & \multicolumn{1}{d}{103.767} & \multicolumn{1}{d}{132.721}\\
\hspace{1em}$R^2$ & \multicolumn{1}{d}{0.81} & \multicolumn{1}{d}{0.79}\\
\addlinespace[0.5em]
\multicolumn{3}{l}{\textit{Panel B -- Control group from non-adopting CBs only:}}\\
\midrule \hspace{1em}Inside $\times$ Adj. & -8.821^{ *** } & -1.694^{  }\\
\hspace{1em} & (3.088) & (3.852)\\
\hspace{1em}Inside $\times$ Post & -19.467^{ *** } & -3.198^{  }\\
\hspace{1em} & (4.512) & (5.498)\\
\hspace{1em}Observations & \multicolumn{1}{D{,}{,}{-3}}{12,474} & \multicolumn{1}{D{,}{,}{-3}}{12,422}\\
\hspace{1em}Mean dep. var. & \multicolumn{1}{d}{97.096} & \multicolumn{1}{d}{128.629}\\
\hspace{1em}$R^2$ & \multicolumn{1}{d}{0.813} & \multicolumn{1}{d}{0.796}\\
\addlinespace[0.5em]
\multicolumn{3}{l}{\textit{Panel C -- Control group from adopting CBs only:}}\\
\midrule \hspace{1em}Inside $\times$ Adj. & -7.594^{ ** } & -0.507^{  }\\
\hspace{1em} & (2.992) & (3.775)\\
\hspace{1em}Inside $\times$ Post & -19.557^{ *** } & -2.633^{  }\\
\hspace{1em} & (4.418) & (5.377)\\
\hspace{1em}Observations & \multicolumn{1}{D{,}{,}{-3}}{22,683} & \multicolumn{1}{D{,}{,}{-3}}{22,577}\\
\hspace{1em}Mean dep. var. & \multicolumn{1}{d}{110.228} & \multicolumn{1}{d}{141.276}\\
\hspace{1em}$R^2$ & \multicolumn{1}{d}{0.811} & \multicolumn{1}{d}{0.783}\\
\addlinespace[0.5em]
\multicolumn{3}{l}{\textit{Panel D -- Control group from all non-adopting districts:}}\\
\midrule \hspace{1em}Inside $\times$ Adj. & -11.228^{ *** } & -4.034^{  }\\
\hspace{1em} & (3.400) & (3.806)\\
\hspace{1em}Inside $\times$ Post & -20.710^{ *** } & -6.233^{  }\\
\hspace{1em} & (4.752) & (5.485)\\
\hspace{1em}Observations & \multicolumn{1}{D{,}{,}{-3}}{44,565} & \multicolumn{1}{D{,}{,}{-3}}{42,679}\\
\hspace{1em}Mean dep. var. & \multicolumn{1}{d}{55.611} & \multicolumn{1}{d}{81.000}\\
\hspace{1em}$R^2$ & \multicolumn{1}{d}{0.795} & \multicolumn{1}{d}{0.778}\\
\bottomrule
\end{tabular}
\begin{tablenotes}
\item Columns: (1) black smoke, (2) sulphur dioxide. Panel A replicates our main estimates, while the remaining panels show three alternative control groups. Includes year-by-month and station fixed effects, and includes a station-specific yearly linear time-trend to capture differences in linear dynamics between stations. Trims the sample to 5 years before and after the SCA submission date, and restricts the sample to pollution data for years 1962 to 1973. Clusters standard errors by station. (*): $p < 0.1$, (**): $p<0.05$, (***): $p<0.01$.
\end{tablenotes}
\end{threeparttable}
\end{table}

\begin{table}[!h]
\centering\centering\centering
\caption{\label{tab:individuals_did_robustness_control-group}Difference-in-difference estimates -- Impact on individuals. Robustness to choice of control group.}
\centering
\begin{threeparttable}
\fontsize{10}{12}\selectfont
\setlength{\tabcolsep}{1.5pt}
\begin{tabular}[t]{ldddd}
\toprule
\multicolumn{1}{c}{\em{}} & \multicolumn{4}{c}{\em{Specification:}} \\
\cmidrule(l{3pt}r{3pt}){2-5}
\multicolumn{1}{c}{} & \multicolumn{1}{c}{(1)} & \multicolumn{1}{c}{(2)} & \multicolumn{1}{c}{(3)} & \multicolumn{1}{c}{(4)} \\
\multicolumn{1}{c}{ } & \multicolumn{1}{c}{{\specialcell[b]{Birth \\ weight}}} & \multicolumn{1}{c}{{\specialcell[b]{Adult \\ height}}} & \multicolumn{1}{c}{{\specialcell[b]{Educ. \\ attain.}}} & \multicolumn{1}{c}{{\specialcell[b]{Fluid \\ intelligence}}}\\
\midrule
\addlinespace[0.5em]
\multicolumn{5}{l}{\textit{Panel A -- Main sample:}}\\
\midrule \hspace{1em}Inside $\times$ Adj. & 0.039^{  } & 0.365^{  } & 0.040^{  } & -0.152^{  }\\
\hspace{1em} & (0.034) & (0.343) & (0.178) & (0.131)\\
\hspace{1em}Inside $\times$ Post & 0.058^{ ** } & 0.942^{ *** } & -0.025^{  } & 0.035^{  }\\
\hspace{1em} & (0.027) & (0.180) & (0.135) & (0.065)\\
\hspace{1em}Observations & \multicolumn{1}{D{,}{,}{-3}}{8,510} & \multicolumn{1}{D{,}{,}{-3}}{11,922} & \multicolumn{1}{D{,}{,}{-3}}{11,689} & \multicolumn{1}{D{,}{,}{-3}}{4,106}\\
\hspace{1em}Mean dep. var. & \multicolumn{1}{d}{3.317} & \multicolumn{1}{d}{169.917} & \multicolumn{1}{d}{13.236} & \multicolumn{1}{d}{0.000}\\
\hspace{1em}$R^2$ & \multicolumn{1}{d}{0.066} & \multicolumn{1}{d}{0.543} & \multicolumn{1}{d}{0.107} & \multicolumn{1}{d}{0.149}\\
\addlinespace[0.5em]
\multicolumn{5}{l}{\textit{Panel B -- Control group from non-adopting CBs only:}}\\
\midrule \hspace{1em}Inside $\times$ Adj. & 0.044^{  } & 0.424^{  } & 0.027^{  } & -0.165^{  }\\
\hspace{1em} & (0.032) & (0.355) & (0.166) & (0.120)\\
\hspace{1em}Inside $\times$ Post & 0.060^{ ** } & 0.927^{ *** } & -0.063^{  } & 0.023^{  }\\
\hspace{1em} & (0.028) & (0.196) & (0.129) & (0.066)\\
\hspace{1em}Observations & \multicolumn{1}{D{,}{,}{-3}}{4,076} & \multicolumn{1}{D{,}{,}{-3}}{5,653} & \multicolumn{1}{D{,}{,}{-3}}{5,515} & \multicolumn{1}{D{,}{,}{-3}}{2,225}\\
\hspace{1em}Mean dep. var. & \multicolumn{1}{d}{3.329} & \multicolumn{1}{d}{170.256} & \multicolumn{1}{d}{13.444} & \multicolumn{1}{d}{0.000}\\
\hspace{1em}$R^2$ & \multicolumn{1}{d}{0.091} & \multicolumn{1}{d}{0.56} & \multicolumn{1}{d}{0.121} & \multicolumn{1}{d}{0.172}\\
\addlinespace[0.5em]
\multicolumn{5}{l}{\textit{Panel C -- Control group from adopting CBs only:}}\\
\midrule \hspace{1em}Inside $\times$ Adj. & 0.041^{  } & 0.393^{  } & 0.054^{  } & -0.152^{  }\\
\hspace{1em} & (0.035) & (0.346) & (0.176) & (0.133)\\
\hspace{1em}Inside $\times$ Post & 0.058^{ ** } & 0.951^{ *** } & -0.021^{  } & 0.045^{  }\\
\hspace{1em} & (0.027) & (0.180) & (0.134) & (0.068)\\
\hspace{1em}Observations & \multicolumn{1}{D{,}{,}{-3}}{7,312} & \multicolumn{1}{D{,}{,}{-3}}{10,223} & \multicolumn{1}{D{,}{,}{-3}}{9,993} & \multicolumn{1}{D{,}{,}{-3}}{3,410}\\
\hspace{1em}Mean dep. var. & \multicolumn{1}{d}{3.313} & \multicolumn{1}{d}{169.730} & \multicolumn{1}{d}{13.079} & \multicolumn{1}{d}{0.000}\\
\hspace{1em}$R^2$ & \multicolumn{1}{d}{0.066} & \multicolumn{1}{d}{0.537} & \multicolumn{1}{d}{0.083} & \multicolumn{1}{d}{0.142}\\
\addlinespace[0.5em]
\multicolumn{5}{l}{\textit{Panel D -- Control group from all non-adopting districts:}}\\
\midrule \hspace{1em}Inside $\times$ Adj. & 0.035^{  } & 0.325^{  } & 0.046^{  } & -0.181^{  }\\
\hspace{1em} & (0.037) & (0.351) & (0.170) & (0.116)\\
\hspace{1em}Inside $\times$ Post & 0.058^{ ** } & 0.896^{ *** } & -0.028^{  } & 0.016^{  }\\
\hspace{1em} & (0.026) & (0.194) & (0.134) & (0.063)\\
\hspace{1em}Observations & \multicolumn{1}{D{,}{,}{-3}}{18,511} & \multicolumn{1}{D{,}{,}{-3}}{25,410} & \multicolumn{1}{D{,}{,}{-3}}{25,199} & \multicolumn{1}{D{,}{,}{-3}}{9,868}\\
\hspace{1em}Mean dep. var. & \multicolumn{1}{d}{3.351} & \multicolumn{1}{d}{170.776} & \multicolumn{1}{d}{13.805} & \multicolumn{1}{d}{0.000}\\
\hspace{1em}$R^2$ & \multicolumn{1}{d}{0.118} & \multicolumn{1}{d}{0.565} & \multicolumn{1}{d}{0.15} & \multicolumn{1}{d}{0.214}\\
\bottomrule
\end{tabular}
\begin{tablenotes}
\item Columns: (1) birth weight in kilograms, (2) height in centimeters, (3) years of education, (4) standardised fluid intelligence score. Panel A replicates our main estimates, while the remaining panels show three alternative control groups. OLS specification includes year-by-month and (CB $\times$ Inside) fixed effects, and includes (CB $\times$ Inside)-specific linear time trends. Controls for sex, ethnicity, and weather in utero and during childhood. Trims the sample to five years before and after the SCA submission date, and restricts the sample to birth cohorts in years 1958 to 1969. Clusters standard errors by CB. (*): $p < 0.1$, (**): $p<0.05$, (***): $p<0.01$.
\end{tablenotes}
\end{threeparttable}
\end{table}

\FloatBarrier
\subsection{Precision of birth locations}
\label{sec:robustness_bunching}
Our identification exploits the fact that variation in individuals' birth locations induces spatial variation in their exposure to smoke control. To assign treatment, we project individuals' birth locations over the SCA boundaries and define the treated group as those born within the SCA boundaries. As smoke control areas are geographical partitions of CBs, we must be able to reliably geolocate individuals \emph{within} CBs. 

The UK Biobank collects birth locations from all participants who indicated they were born in England, Wales or Scotland by asking ``\emph{What is the town or district you first lived in when you were born?}'' The interviewer selected the corresponding place from a detailed list of place names, which were converted to north and east coordinates with a 1km resolution. However, in cases where the participant was not sufficiently specific, their birth location was assigned to a catch-all location roughly in the center of the town/district. For individuals assigned to such locations, we can therefore not reliably distinguish whether they were born in- or outside a SCA \emph{within} a CB.\footnote{This issue plays much less of a role when defining individuals' CB of birth, as only 6\% of CBs in the UKB are estimated to be incorrectly reported \citep{von2024analysis}; it is more important \textit{within} CBs. Furthermore, this is relevant only for adopting CBs, since within-CB variation in birth location is irrelevant in non-adopting CBs. This issue also does not apply to the pollution analysis, as there is no measurement error in the location of the pollution station. We therefore only explore the robustness of the individual-level estimates.} We drop these individuals from our main analysis since including them is likely to introduce substantial measurement error, but explore the robustness of this selection criteria here.

\begin{table}[!h]
\centering\centering\centering
\caption{\label{tab:individuals_did_robustness_sample-selection}Difference-in-difference estimates -- Impact on individuals. Robustness to catch-all birth locations.}
\centering
\begin{threeparttable}
\fontsize{11}{13}\selectfont
\setlength{\tabcolsep}{1.5pt}
\begin{tabular}[t]{ldddd}
\toprule
\multicolumn{1}{c}{\em{}} & \multicolumn{4}{c}{\em{Dependent variable:}} \\
\cmidrule(l{3pt}r{3pt}){2-5}
\multicolumn{1}{c}{} & \multicolumn{1}{c}{(1)} & \multicolumn{1}{c}{(2)} & \multicolumn{1}{c}{(3)} & \multicolumn{1}{c}{(4)} \\
\multicolumn{1}{c}{ } & \multicolumn{1}{c}{{\specialcell[b]{Birth \\ weight}}} & \multicolumn{1}{c}{{\specialcell[b]{Adult \\ height}}} & \multicolumn{1}{c}{{\specialcell[b]{Educ. \\ attain.}}} & \multicolumn{1}{c}{{\specialcell[b]{Fluid \\ intelligence}}}\\
\midrule
\addlinespace[0.5em]
\multicolumn{5}{l}{\textit{Panel A -- Main sample}}\\
\midrule \hspace{1em}Inside $\times$ Adj. & 0.039^{  } & 0.365^{  } & 0.040^{  } & -0.152^{  }\\
\hspace{1em} & (0.034) & (0.343) & (0.178) & (0.131)\\
\hspace{1em}Inside $\times$ Post & 0.058^{ ** } & 0.942^{ *** } & -0.025^{  } & 0.035^{  }\\
\hspace{1em} & (0.027) & (0.180) & (0.135) & (0.065)\\
\hspace{1em}Observations & \multicolumn{1}{D{,}{,}{-3}}{8,510} & \multicolumn{1}{D{,}{,}{-3}}{11,922} & \multicolumn{1}{D{,}{,}{-3}}{11,689} & \multicolumn{1}{D{,}{,}{-3}}{4,106}\\
\hspace{1em}Mean dep. var. & \multicolumn{1}{d}{3.317} & \multicolumn{1}{d}{169.917} & \multicolumn{1}{d}{13.236} & \multicolumn{1}{d}{0.000}\\
\hspace{1em}$R^2$ & \multicolumn{1}{d}{0.066} & \multicolumn{1}{d}{0.543} & \multicolumn{1}{d}{0.107} & \multicolumn{1}{d}{0.149}\\
\addlinespace[0.5em]
\multicolumn{5}{l}{\textit{Panel B -- Main sample as well as births at catch-all locations}}\\
\midrule \hspace{1em}Inside $\times$ Adj. & -0.001^{  } & 0.094^{  } & 0.071^{  } & -0.076^{  }\\
\hspace{1em} & (0.020) & (0.165) & (0.085) & (0.057)\\
\hspace{1em}Inside $\times$ Post & 0.029^{  } & 0.395^{ ** } & 0.043^{  } & -0.016^{  }\\
\hspace{1em} & (0.018) & (0.156) & (0.091) & (0.057)\\
\hspace{1em}Observations & \multicolumn{1}{D{,}{,}{-3}}{22,381} & \multicolumn{1}{D{,}{,}{-3}}{32,195} & \multicolumn{1}{D{,}{,}{-3}}{31,892} & \multicolumn{1}{D{,}{,}{-3}}{12,567}\\
\hspace{1em}Mean dep. var. & \multicolumn{1}{d}{3.315} & \multicolumn{1}{d}{169.892} & \multicolumn{1}{d}{13.278} & \multicolumn{1}{d}{0.000}\\
\hspace{1em}$R^2$ & \multicolumn{1}{d}{0.041} & \multicolumn{1}{d}{0.524} & \multicolumn{1}{d}{0.058} & \multicolumn{1}{d}{0.098}\\
\bottomrule
\end{tabular}
\begin{tablenotes}
\item Columns: (1) birth weight in kilograms, (2) height in centimeters, (3) years of education, (4) standardised fluid intelligence score. Panels: (A) main sample, (B) main sample as well as individuals born at catch-all birth locations. OLS specification includes year-by-month and (CB $\times$ Inside) fixed effects, and includes (CB $\times$ Inside)-specific linear time trends. Controls for sex, ethnicity, and weather in utero and during childhood. Control group consists of never-treated individuals from both adopting and non-adopting county boroughs. Trims the sample to five years before and after the SCA submission date, and restricts the sample to birth cohorts in years 1958 to 1969. Clusters standard errors by CB. (*): $p < 0.1$, (**): $p<0.05$, (***): $p<0.01$.
\end{tablenotes}
\end{threeparttable}
\end{table}

Dropping the catch-all locations reduces our estimation sample by a factor of 0.32-0.38, depending on the outcome, indicating that the catch-all birth locations account for a large number of births. Panel~A of \autoref{tab:individuals_did_robustness_sample-selection} replicates our main estimates for comparison, while Panel B includes all individuals. As expected, including a large number of individuals for whom their birth location is measured with error reduces the point estimates for birth weight and height. The estimate for birth weight reduces from 58 grams to 29 grams, and the impact on height decreases to about 0.4 cm. We do not find impacts on educational attainment and fluid intelligence.

Across all estimates, the standard errors are 1.15-1.5 times larger in our estimation sample compared to the sample that does not drop the catch-all birth locations. Given the reduction in sample size by a factor of 0.32-0.38 when dropping the catch-all locations, the standard errors should generally be about 1.6 times larger in our estimation sample purely due to the $n^{-1/2}$ scale factor. As the standard errors increase by a factor less than 1.6, it suggests that the noise in our estimates have generally reduced by dropping the catch-all locations.

\subsection{Trimming of samples}
\label{sec:robustness_trimming}
We next explore the sensitivity of our estimates to different bandwidths around the SCA submission date, trimming the sample to those exposed up to two or four years before and after the introduction of the SCA, as well as to not trimming the sample at all. 

\autoref{tab:stations_did_robustness_trimming} presents the results for pollution, showing four different bandwidths with Panel~A replicating our main estimates where we trim to five years before and after. Panels B and C reduce the bandwidth to 4 and 2 years respectively, highlighting that our estimates are stable across narrower bandwidths, though with larger standard errors. Panel~D reports the results when we do not trim the sample, showing similar results to the other panels.

\autoref{tab:individuals_did_robustness_trimming} analogously reports the estimates for individuals. Trimming to 4 years, as shown in Panel~B, produces estimates similar to our main specification. Further trimming to 2 years, as in Panel~C, reduces our sample size and increases the standard errors, and cuts our post operation estimates to about half for birth weight and adult height. This attenuation may be due to the trimming being defined relative to the submission times of SCAs rather than operation times, whereby narrower bandwidths potentially trim off SCAs with longer waiting times between submission and operation. As shown in \autoref{fig:individuals_waiting-times}, a non-negligible share of SCAs have waiting times longer than two years. Finally, in Panel~D we report the estimates without trimming of the sample, showing that the estimate for height is similar to the main specification, while the estimate for birth weight is reduced to about half compared to the main specification.

\begin{table}[!h]
\centering\centering\centering
\caption{\label{tab:stations_did_robustness_trimming}Difference-in-difference estimates -- Impact on pollution. Sentivity to trimming around the event time.}
\centering
\begin{threeparttable}
\fontsize{10}{12}\selectfont
\setlength{\tabcolsep}{10pt}
\begin{tabular}[t]{ldd}
\toprule
\multicolumn{1}{c}{\em{}} & \multicolumn{2}{c}{\em{Dependent variable:}} \\
\cmidrule(l{3pt}r{3pt}){2-3}
\multicolumn{1}{c}{} & \multicolumn{1}{c}{(1)} & \multicolumn{1}{c}{(2)} \\
\multicolumn{1}{c}{ } & \multicolumn{1}{c}{{\specialcell[b]{Black \\ smoke}}} & \multicolumn{1}{c}{{\specialcell[b]{Sulphur \\ dioxide}}}\\
\midrule
\addlinespace[0.5em]
\multicolumn{3}{l}{\textit{Panel A -- Main sample:}}\\
\midrule \hspace{1em}Inside $\times$ Adj. & -7.928^{ *** } & -1.139^{  }\\
\hspace{1em} & (2.994) & (3.749)\\
\hspace{1em}Inside $\times$ Post & -19.604^{ *** } & -3.323^{  }\\
\hspace{1em} & (4.407) & (5.370)\\
\hspace{1em}Observations & \multicolumn{1}{D{,}{,}{-3}}{26,128} & \multicolumn{1}{D{,}{,}{-3}}{26,045}\\
\hspace{1em}Mean dep. var. & \multicolumn{1}{d}{103.887} & \multicolumn{1}{d}{132.660}\\
\hspace{1em}$R^2$ & \multicolumn{1}{d}{0.81} & \multicolumn{1}{d}{0.789}\\
\addlinespace[0.5em]
\multicolumn{3}{l}{\textit{Panel B -- Trim to 4 years before and after:}}\\
\midrule \hspace{1em}Inside $\times$ Adj. & -7.975^{ *** } & -0.498^{  }\\
\hspace{1em} & (3.024) & (3.385)\\
\hspace{1em}Inside $\times$ Post & -21.277^{ *** } & -3.148^{  }\\
\hspace{1em} & (5.051) & (5.218)\\
\hspace{1em}Observations & \multicolumn{1}{D{,}{,}{-3}}{24,288} & \multicolumn{1}{D{,}{,}{-3}}{24,206}\\
\hspace{1em}Mean dep. var. & \multicolumn{1}{d}{102.700} & \multicolumn{1}{d}{130.642}\\
\hspace{1em}$R^2$ & \multicolumn{1}{d}{0.809} & \multicolumn{1}{d}{0.788}\\
\addlinespace[0.5em]
\multicolumn{3}{l}{\textit{Panel C -- Trim to 2 years before and after:}}\\
\midrule \hspace{1em}Inside $\times$ Adj. & -8.187^{ ** } & -3.826^{  }\\
\hspace{1em} & (3.786) & (4.356)\\
\hspace{1em}Inside $\times$ Post & -16.707^{ *** } & -4.254^{  }\\
\hspace{1em} & (6.316) & (6.951)\\
\hspace{1em}Observations & \multicolumn{1}{D{,}{,}{-3}}{20,801} & \multicolumn{1}{D{,}{,}{-3}}{20,739}\\
\hspace{1em}Mean dep. var. & \multicolumn{1}{d}{101.587} & \multicolumn{1}{d}{127.765}\\
\hspace{1em}$R^2$ & \multicolumn{1}{d}{0.809} & \multicolumn{1}{d}{0.792}\\
\addlinespace[0.5em]
\multicolumn{3}{l}{\textit{Panel D -- No trimming:}}\\
\midrule \hspace{1em}Inside $\times$ Adj. & -4.783^{  } & -0.125^{  }\\
\hspace{1em} & (3.180) & (4.096)\\
\hspace{1em}Inside $\times$ Post & -19.473^{ *** } & -0.649^{  }\\
\hspace{1em} & (4.398) & (4.898)\\
\hspace{1em}Observations & \multicolumn{1}{D{,}{,}{-3}}{34,574} & \multicolumn{1}{D{,}{,}{-3}}{34,420}\\
\hspace{1em}Mean dep. var. & \multicolumn{1}{d}{102.647} & \multicolumn{1}{d}{137.406}\\
\hspace{1em}$R^2$ & \multicolumn{1}{d}{0.811} & \multicolumn{1}{d}{0.791}\\
\bottomrule
\end{tabular}
\begin{tablenotes}
\item Columns: (1) black smoke, (2) sulphur dioxide. Panels show different intervals of trimming of observations around the SCA submission date. Panel~A is our main sample, trimming to 5 years on both sides of the submission date. Control group includes never-treated from both adopting and non-adopting county boroughs.  Clusters standard errors by station. Includes station fixed effects, and includes a station-specific yearly linear time-trend to capture differences in linear dynamics between stations. Restricts the sample to pollution data for years 1962 to 1973. (*): $p < 0.1$, (**): $p<0.05$, (***): $p<0.01$."
\end{tablenotes}
\end{threeparttable}
\end{table}

\begin{table}[!h]
\centering\centering\centering
\caption{\label{tab:individuals_did_robustness_trimming}Difference-in-difference estimates -- Impact on individuals. Sentivity to trimming around the event time.}
\centering
\begin{threeparttable}
\fontsize{10}{12}\selectfont
\setlength{\tabcolsep}{1.5pt}
\begin{tabular}[t]{ldddd}
\toprule
\multicolumn{1}{c}{\em{}} & \multicolumn{4}{c}{\em{Dependent variable:}} \\
\cmidrule(l{3pt}r{3pt}){2-5}
\multicolumn{1}{c}{} & \multicolumn{1}{c}{(1)} & \multicolumn{1}{c}{(2)} & \multicolumn{1}{c}{(3)} & \multicolumn{1}{c}{(4)} \\
\multicolumn{1}{c}{ } & \multicolumn{1}{c}{{\specialcell[b]{Birth \\ weight}}} & \multicolumn{1}{c}{{\specialcell[b]{Adult \\ height}}} & \multicolumn{1}{c}{{\specialcell[b]{Educ. \\ attain.}}} & \multicolumn{1}{c}{{\specialcell[b]{Fluid \\ intelligence}}}\\
\midrule
\addlinespace[0.5em]
\multicolumn{5}{l}{\textit{Panel A -- Main sample:}}\\
\midrule \hspace{1em}Inside $\times$ Adj. & 0.039^{  } & 0.365^{  } & 0.040^{  } & -0.152^{  }\\
\hspace{1em} & (0.034) & (0.343) & (0.178) & (0.131)\\
\hspace{1em}Inside $\times$ Post & 0.058^{ ** } & 0.942^{ *** } & -0.025^{  } & 0.035^{  }\\
\hspace{1em} & (0.027) & (0.180) & (0.135) & (0.065)\\
\hspace{1em}Observations & \multicolumn{1}{D{,}{,}{-3}}{8,510} & \multicolumn{1}{D{,}{,}{-3}}{11,922} & \multicolumn{1}{D{,}{,}{-3}}{11,689} & \multicolumn{1}{D{,}{,}{-3}}{4,106}\\
\hspace{1em}Mean dep. var. & \multicolumn{1}{d}{3.317} & \multicolumn{1}{d}{169.917} & \multicolumn{1}{d}{13.236} & \multicolumn{1}{d}{0.000}\\
\hspace{1em}$R^2$ & \multicolumn{1}{d}{0.066} & \multicolumn{1}{d}{0.543} & \multicolumn{1}{d}{0.107} & \multicolumn{1}{d}{0.149}\\
\addlinespace[0.5em]
\multicolumn{5}{l}{\textit{Panel B -- Trim to 4 years before and after:}}\\
\midrule \hspace{1em}Inside $\times$ Adj. & 0.049^{  } & 0.368^{  } & 0.044^{  } & -0.179^{  }\\
\hspace{1em} & (0.038) & (0.308) & (0.181) & (0.135)\\
\hspace{1em}Inside $\times$ Post & 0.067^{ ** } & 0.926^{ *** } & 0.015^{  } & 0.040^{  }\\
\hspace{1em} & (0.026) & (0.181) & (0.126) & (0.066)\\
\hspace{1em}Observations & \multicolumn{1}{D{,}{,}{-3}}{8,017} & \multicolumn{1}{D{,}{,}{-3}}{11,235} & \multicolumn{1}{D{,}{,}{-3}}{11,022} & \multicolumn{1}{D{,}{,}{-3}}{3,858}\\
\hspace{1em}Mean dep. var. & \multicolumn{1}{d}{3.317} & \multicolumn{1}{d}{169.919} & \multicolumn{1}{d}{13.248} & \multicolumn{1}{d}{0.000}\\
\hspace{1em}$R^2$ & \multicolumn{1}{d}{0.068} & \multicolumn{1}{d}{0.542} & \multicolumn{1}{d}{0.111} & \multicolumn{1}{d}{0.153}\\
\addlinespace[0.5em]
\multicolumn{5}{l}{\textit{Panel C -- Trim to 2 years before and after}}\\
\midrule \hspace{1em}Inside $\times$ Adj. & 0.057^{  } & 0.295^{  } & -0.076^{  } & -0.187^{  }\\
\hspace{1em} & (0.053) & (0.374) & (0.205) & (0.183)\\
\hspace{1em}Inside $\times$ Post & 0.031^{  } & 0.480^{  } & -0.071^{  } & -0.021^{  }\\
\hspace{1em} & (0.040) & (0.432) & (0.229) & (0.129)\\
\hspace{1em}Observations & \multicolumn{1}{D{,}{,}{-3}}{6,855} & \multicolumn{1}{D{,}{,}{-3}}{9,643} & \multicolumn{1}{D{,}{,}{-3}}{9,489} & \multicolumn{1}{D{,}{,}{-3}}{3,238}\\
\hspace{1em}Mean dep. var. & \multicolumn{1}{d}{3.314} & \multicolumn{1}{d}{169.914} & \multicolumn{1}{d}{13.265} & \multicolumn{1}{d}{0.000}\\
\hspace{1em}$R^2$ & \multicolumn{1}{d}{0.072} & \multicolumn{1}{d}{0.548} & \multicolumn{1}{d}{0.124} & \multicolumn{1}{d}{0.174}\\
\addlinespace[0.5em]
\multicolumn{5}{l}{\textit{Panel D -- No trimming}}\\
\midrule \hspace{1em}Inside $\times$ Adj. & 0.019^{  } & 0.373^{  } & 0.111^{  } & -0.139^{  }\\
\hspace{1em} & (0.033) & (0.292) & (0.176) & (0.105)\\
\hspace{1em}Inside $\times$ Post & 0.033^{  } & 0.931^{ *** } & 0.120^{  } & 0.087^{  }\\
\hspace{1em} & (0.025) & (0.232) & (0.130) & (0.056)\\
\hspace{1em}Observations & \multicolumn{1}{D{,}{,}{-3}}{11,076} & \multicolumn{1}{D{,}{,}{-3}}{15,572} & \multicolumn{1}{D{,}{,}{-3}}{15,247} & \multicolumn{1}{D{,}{,}{-3}}{5,263}\\
\hspace{1em}Mean dep. var. & \multicolumn{1}{d}{3.318} & \multicolumn{1}{d}{169.806} & \multicolumn{1}{d}{13.169} & \multicolumn{1}{d}{0.000}\\
\hspace{1em}$R^2$ & \multicolumn{1}{d}{0.051} & \multicolumn{1}{d}{0.537} & \multicolumn{1}{d}{0.093} & \multicolumn{1}{d}{0.13}\\
\bottomrule
\end{tabular}
\begin{tablenotes}
\item Columns: (1) birth weight in kilograms, (2) height in centimeters, (3) years of education, (4) standardised fluid intelligence score. Panels shows different bandwidths of trimming of observations around the SCA submission date. Panel~A is our main specification. OLS specification includes year-by-month and (CB $\times$ Inside) fixed effects, and includes (CB $\times$ Inside)-specific linear time trends. Controls for sex, ethnicity, and weather in utero and during childhood. Control group consists of never-treated individuals from both adopting and non-adopting county boroughs. Restricts the sample to birth cohorts in years 1958 to 1969. Clusters standard errors by CB. (*): $p < 0.1$, (**): $p<0.05$, (***): $p<0.01$.
\end{tablenotes}
\end{threeparttable}
\end{table}

\FloatBarrier
\subsection{Choice of birth cohorts}
\label{sec:robustness_birth-cohorts}
\begin{figure}[!h]\caption{\label{fig:individuals_did_robustness_cohorts}Difference-in-difference estimates -- Impact on individuals. Sensitivity to choice of birth cohorts.}\centering\includegraphics[width=0.6\textwidth]{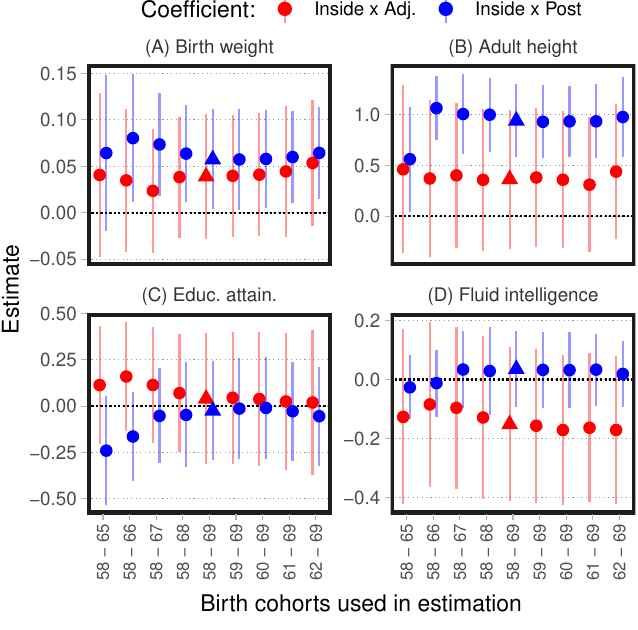}\caption*{\emph{Shows the Inside $\times$ Adj. (red) and Inside $\times$ Post (blue) estimates and 0.95 confidence intervals across outcomes and different choices of birth cohorts. Our main specification is marked with a triangle, which uses all the available cohorts between 1958 and 1969. OLS specification includes year-by-month and (CB $\times$ Inside) fixed effects, and includes (CB $\times$ Inside)-specific linear time trends. Controls for sex, ethnicity, and weather in utero and during childhood. Trims the sample to five years before and after the SCA submission date. Clusters standard errors by CB.}}\end{figure}

Our main sample for the individual analysis includes all birth cohorts born after the Raising of the School Leaving Age (RoSLA) and until 1969. Here we investigate alternative choice of birth cohorts. \autoref{fig:individuals_did_robustness_cohorts} shows the adjustment (red) and post operation (blue) estimates and their 95\% confidence intervals for the different choices of birth cohorts shown on the horisontal axis. Each panel corresponds to one of our four outcomes, and we mark our main estimates with a triangle.

Panel~A shows positive post operation impact of smoke control on birth weight across all choices of birth cohorts, though with a slightly larger effect for earlier cohorts. The adjustment period impact is insignificant throughout, though larger in positive direction for later cohorts. Panel~B highlights that the impacts on height are positive and stable across cohorts, except for a reduced post operation estimate for the earliest choice of cohorts (1958--1965).  Panel~C shows that our null findings for educational attainment are generally stable, and we see no statistical significant impact across any of the cohort choices. We do however find that for early cohorts the post operation estimates move downward in negative direction, while the adjustment period estimates move slightly in positive direction. Panel~D confirms that our null finding for fluid intelligence score is stable across cohorts, with all estimates being statistically insignificant. However, we do see larger negative estimates for the adjustment period for later cohorts compared to earlier.

\FloatBarrier
\subsection{Time trends}
\label{sec:robustness_trends}
Our main specification includes station-specific and CB $\times$ Inside-specific linear trends for the pollution and individual analysis respectively, allowing for differential linear dynamics in outcomes across space. We next explore the robustness of our findings with respect to alternative trend specifications. 

\autoref{tab:stations_did_robustness_trend} replicates the pollution estimates for our main specification in Panel~A, that is, with station-specific time trends. Panel~B reports the results with CB-specific trends and Panel~C drops the trends altogether (i.e., including only year-month fixed effects). This shows the largest estimates when omitting the trend, with a reduction of 30 mcg/m$^3$ post-adjustment period for black smoke and 9.4 mcg/m$^3$ for sulphur dioxide. Adding either CB-specific or station-specific trends reduces these estimates, suggesting that there are indeed spatial differences in pollution trends. 

\autoref{tab:individuals_did_robustness_trend} presents the analysis at the individual-level, with Panel~A presenting our preferred estimates with (CB $\times$ Inside)-specific time trends, controlling for CB-specific trends in Panel~B, and not controlling for any trend in Panel~C (i.e., only including year-month fixed effects). This indicates largely similar estimates across the alternative specifications.

\begin{table}[!h]
\centering\centering\centering
\caption{\label{tab:stations_did_robustness_trend}Difference-in-difference estimates -- Impact on pollution. Robustness to choice of time trend.}
\centering
\begin{threeparttable}
\fontsize{10}{12}\selectfont
\setlength{\tabcolsep}{0.5pt}
\begin{tabular}[t]{ldd}
\toprule
\multicolumn{1}{c}{\em{}} & \multicolumn{2}{c}{\em{Specification:}} \\
\cmidrule(l{3pt}r{3pt}){2-3}
\multicolumn{1}{c}{} & \multicolumn{1}{c}{(1)} & \multicolumn{1}{c}{(2)} \\
\multicolumn{1}{c}{ } & \multicolumn{1}{c}{{\specialcell[b]{Black \\ smoke}}} & \multicolumn{1}{c}{{\specialcell[b]{Sulphur \\ dioxide}}}\\
\midrule
\addlinespace[0.5em]
\multicolumn{3}{l}{\textit{Panel A -- Main specification:}}\\
\midrule \hspace{1em}Inside $\times$ Adj. & -8.053^{ *** } & -1.180^{  }\\
\hspace{1em} & (3.010) & (3.740)\\
\hspace{1em}Inside $\times$ Post & -19.737^{ *** } & -3.480^{  }\\
\hspace{1em} & (4.380) & (5.337)\\
\hspace{1em}$R^2$ & \multicolumn{1}{d}{0.81} & \multicolumn{1}{d}{0.79}\\
\addlinespace[0.5em]
\multicolumn{3}{l}{\textit{Panel B -- CB-specific trends:}}\\
\midrule \hspace{1em}Inside $\times$ Adj. & -2.399^{  } & -0.558^{  }\\
\hspace{1em} & (2.933) & (3.739)\\
\hspace{1em}Inside $\times$ Post & -8.832^{ ** } & 2.906^{  }\\
\hspace{1em} & (3.727) & (4.455)\\
\hspace{1em}$R^2$ & \multicolumn{1}{d}{0.799} & \multicolumn{1}{d}{0.777}\\
\addlinespace[0.5em]
\multicolumn{3}{l}{\textit{Panel C -- No trend:}}\\
\midrule \hspace{1em}Inside $\times$ Adj. & -14.679^{ *** } & -8.203^{ ** }\\
\hspace{1em} & (4.101) & (3.948)\\
\hspace{1em}Inside $\times$ Post & -29.976^{ *** } & -9.402^{ ** }\\
\hspace{1em} & (5.663) & (4.717)\\
\hspace{1em}$R^2$ & \multicolumn{1}{d}{0.778} & \multicolumn{1}{d}{0.759}\\
\midrule
\hspace{1em}Observations & \multicolumn{1}{D{,}{,}{-3}}{26,302} & \multicolumn{1}{D{,}{,}{-3}}{26,195}\\
\hspace{1em}Mean dep. var. & \multicolumn{1}{d}{103.767} & \multicolumn{1}{d}{132.721}\\
\bottomrule
\end{tabular}
\begin{tablenotes}
\item Columns: (1) black smoke, (2) sulphur dioxide. Panels show different choices of time trends. Panel~A is our main specification with station-specific time trends. Control group includes never-treated from both adopting and non-adopting county boroughs. Clusters standard errors by station. Includes year-by-month and station fixed effects. Trims the sample to 5 years before and after the SCA submission date, and restricts the sample to pollution data for years 1962 to 1973. (*): $p < 0.1$, (**): $p<0.05$, (***): $p<0.01$.
\end{tablenotes}
\end{threeparttable}
\end{table}

\begin{table}[!h]
\centering\centering\centering
\caption{\label{tab:individuals_did_robustness_trend}Difference-in-difference estimates -- Impact on individuals. Robustness to inclusion of time trend.}
\centering
\begin{threeparttable}
\fontsize{11}{13}\selectfont
\setlength{\tabcolsep}{1.5pt}
\begin{tabular}[t]{ldddd}
\toprule
\multicolumn{1}{c}{\em{}} & \multicolumn{4}{c}{\em{Dependent variable:}} \\
\cmidrule(l{3pt}r{3pt}){2-5}
\multicolumn{1}{c}{} & \multicolumn{1}{c}{(1)} & \multicolumn{1}{c}{(2)} & \multicolumn{1}{c}{(3)} & \multicolumn{1}{c}{(4)} \\
\multicolumn{1}{c}{ } & \multicolumn{1}{c}{{\specialcell[b]{Birth \\ weight}}} & \multicolumn{1}{c}{{\specialcell[b]{Adult \\ height}}} & \multicolumn{1}{c}{{\specialcell[b]{Educ. \\ attain.}}} & \multicolumn{1}{c}{{\specialcell[b]{Fluid \\ intelligence}}}\\
\midrule
\addlinespace[0.5em]
\multicolumn{5}{l}{\textit{Panel A -- Main specification:}}\\
\midrule \hspace{1em}Inside $\times$ Adj. & 0.039^{  } & 0.365^{  } & 0.040^{  } & -0.152^{  }\\
\hspace{1em} & (0.034) & (0.343) & (0.178) & (0.131)\\
\hspace{1em}Inside $\times$ Post & 0.058^{ ** } & 0.942^{ *** } & -0.025^{  } & 0.035^{  }\\
\hspace{1em} & (0.027) & (0.180) & (0.135) & (0.065)\\
\hspace{1em}$R^2$ & \multicolumn{1}{d}{0.066} & \multicolumn{1}{d}{0.543} & \multicolumn{1}{d}{0.107} & \multicolumn{1}{d}{0.149}\\
\addlinespace[0.5em]
\multicolumn{5}{l}{\textit{Panel B -- CB-specific yearly trend}}\\
\midrule \hspace{1em}Inside $\times$ Adj. & 0.029^{  } & 0.334^{  } & 0.007^{  } & -0.156^{  }\\
\hspace{1em} & (0.031) & (0.319) & (0.156) & (0.135)\\
\hspace{1em}Inside $\times$ Post & 0.046^{ * } & 0.853^{ *** } & -0.031^{  } & 0.027^{  }\\
\hspace{1em} & (0.027) & (0.181) & (0.119) & (0.081)\\
\hspace{1em}$R^2$ & \multicolumn{1}{d}{0.062} & \multicolumn{1}{d}{0.542} & \multicolumn{1}{d}{0.104} & \multicolumn{1}{d}{0.142}\\
\addlinespace[0.5em]
\multicolumn{5}{l}{\textit{Panel C -- No time trend}}\\
\midrule \hspace{1em}Inside $\times$ Adj. & 0.040^{  } & 0.359^{  } & -0.009^{  } & -0.178^{  }\\
\hspace{1em} & (0.031) & (0.329) & (0.147) & (0.128)\\
\hspace{1em}Inside $\times$ Post & 0.055^{ ** } & 0.845^{ *** } & -0.056^{  } & -0.015^{  }\\
\hspace{1em} & (0.025) & (0.177) & (0.103) & (0.076)\\
\hspace{1em}$R^2$ & \multicolumn{1}{d}{0.055} & \multicolumn{1}{d}{0.539} & \multicolumn{1}{d}{0.1} & \multicolumn{1}{d}{0.122}\\
\midrule
\hspace{1em}Observations & \multicolumn{1}{D{,}{,}{-3}}{8,510} & \multicolumn{1}{D{,}{,}{-3}}{11,922} & \multicolumn{1}{D{,}{,}{-3}}{11,689} & \multicolumn{1}{D{,}{,}{-3}}{4,106}\\
\hspace{1em}Mean dep. var. & \multicolumn{1}{d}{3.317} & \multicolumn{1}{d}{169.917} & \multicolumn{1}{d}{13.236} & \multicolumn{1}{d}{0.000}\\
\bottomrule
\end{tabular}
\begin{tablenotes}
\item Columns: (1) birth weight in kilograms, (2) height in centimeters, (3) years of education, (4) standardised fluid intelligence score. Panels show different choices of time trend. Panel~A is our main specification which includes (CB $\times$ Inside) fixed effects. Controls for sex, ethnicity, and weather in utero and during childhood. Control group consists of never-treated individuals from both adopting and non-adopting county boroughs. Trims the sample to five years before and after the SCA submission date, and restricts the sample to birth cohorts in years 1958 to 1969. Clusters standard errors by CB. (*): $p < 0.1$, (**): $p<0.05$, (***): $p<0.01$.
\end{tablenotes}
\end{threeparttable}
\end{table}

\clearpage
\FloatBarrier
\section{Accounting for staggered treatment}
\label{sec:appendix_staggered_did}
\begin{figure}[!h]\caption{\label{fig:event_estimates_ols_csa}OLS and group-time average effect estimates -- Impact on local pollution levels.}\centering\includegraphics[width=0.95\textwidth]{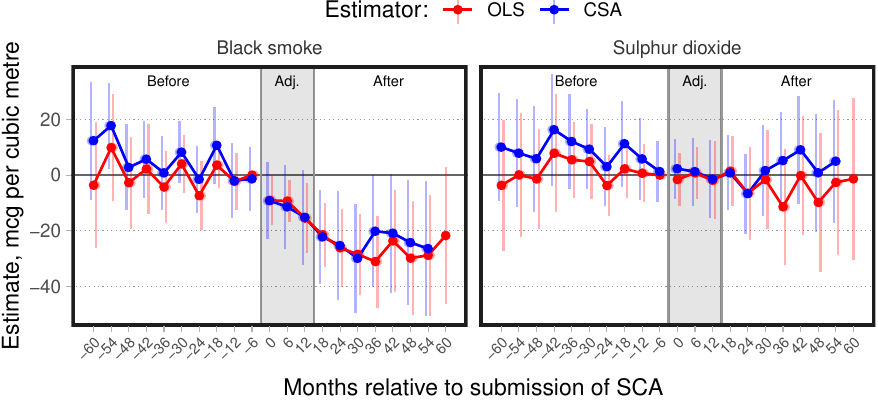}\caption*{\emph{OLS specification includes year-by-month and station fixed effects, and includes a station-specific yearly linear time-trend to capture differences in linear dynamics between stations. Control group consists of never-treated stations from both adopting and non-adopting county boroughs. Clusters standard errors by station. Trims the sample to 5 years before and after the SCA order date, and restricts the sample to pollution data for years 1962 to 1973.}}\end{figure}

The recent literature on the staggered introduction of treatment has shown that standard OLS estimates in event and difference-in-difference specifications can be arbitrarily biased \citep{sunabraham2021,callawaySantanna2021,borusyakJaravelSpiess2024}. To alleviate concerns about such biases impacting our main estimates, we use the estimation approach by \citet{callawaySantanna2021} to obtain robust estimates (group-time average effects, GTA) of, respectively, the dynamic (\autoref{fig:event_estimates_ols_csa}) and the overall (\autoref{tab:did_estimates_ols_csa}) impacts of introducing a SCA on pollution. The dynamic estimates in \autoref{fig:event_estimates_ols_csa} are largely similar for the GTA and OLS estimates, with the former showing slightly more pronounced differences in pollution between treated and control stations, in particular for the period long (4--5 years) before SCA submission. The GTA point estimate for black smoke (\autoref{tab:did_estimates_ols_csa}) is larger than the OLS estimate, suggesting a stronger reduction in black smoke following the introduction of a SCA, while the robust estimate for sulphur dioxide is close to the OLS estimate, showing no evidence of an effect on sulphur dioxide. In summary, our analyses suggest that the OLS estimates are comparable to those obtained using the \citet{callawaySantanna2021} approach.

\begin{table}[!h]

\caption{\label{tab:did_estimates_ols_csa}Group-time average effects -- Impact on pollution.}
\centering
\begin{threeparttable}
\fontsize{11}{13}\selectfont
\setlength{\tabcolsep}{2em}
\begin{tabular}[t]{ldd}
\toprule
\multicolumn{1}{c}{\em{}} & \multicolumn{2}{c}{\em{Specification:}} \\
\cmidrule(l{3pt}r{3pt}){2-3}
\multicolumn{1}{c}{} & \multicolumn{1}{c}{(1)} & \multicolumn{1}{c}{(2)} \\
\multicolumn{1}{c}{ } & \multicolumn{1}{c}{{\specialcell[b]{Black \\ smoke}}} & \multicolumn{1}{c}{{\specialcell[b]{Sulphur \\ dioxide}}}\\
\midrule
ATT & -19.173^{ *** } & -0.731^{  }\\
 & (6.331) & (6.216)\\
\midrule
Observations & \multicolumn{1}{D{,}{,}{-3}}{24,675} & \multicolumn{1}{D{,}{,}{-3}}{24,593}\\
Mean dep. var. & \multicolumn{1}{d}{102.022} & \multicolumn{1}{d}{130.446}\\
\bottomrule
\end{tabular}
\begin{tablenotes}
\item Columns: (1) black smoke, (2) sulphur dioxide. Shows the group-time average effects of the overall impact of being inside a smoke control area in the first 5 years after the introduction. Control group includes never-treated stations from both adopting and non-adopting county boroughs. Clusters standard errors by station. Trims the sample to 5 years before and after the SCA submission date, drops always treated stations, and restricts the sample to pollution data for years 1962 to 1973. (*): $p < 0.1$, (**): $p<0.05$, (***): $p<0.01$.
\end{tablenotes}
\end{threeparttable}
\end{table}

\clearpage
\FloatBarrier
\section{Genetics}
\label{sec:appendix_genetics}
Every human cell, except for sex cells, contains 46 chromosomes. These chromosomes are arranged in pairs, with each pair containing one maternal and one paternal copy, and consist of double-stranded deoxyribonucleic acid (DNA) with a large number of base pairs. The sequence of base pairs, built from the nucleotides adenine (A) binding with thymine (T), and guanine (G) binding with cytosine (C), form the human genome. Across a population there will be variation in the base pairs at specific locations. Such variation is called a single nucleotide polymorphism (SNPs) and it is one of the most commonly studied types of genetic variation. In cases where there are two possible base pairs at a given location (i.e., two alleles), the more frequent base pair is referred to as the major allele, while the less frequent pair is called the minor allele. Since humans possess two copies of each chromosome, an individual can have zero, one, or two copies of the minor allele at a given location.

Genome-Wide Association Studies (GWAS) aim to identify SNPs that robustly associate with particular outcomes of interest. As GWAS samples typically contain many more SNPs than individuals, multivariate regression models cannot jointly identify the SNP effects. Instead, a GWAS conducts univariate regressions of the outcome on each SNP. GWASes have revealed that many outcomes of interest in the social sciences are `polygenic', meaning they are influenced by numerous SNPs, each with a small effect size. Thus, to enhance the predictive power of SNPs with respect to an outcome, it is custom to aggregate the individual SNPs into polygenic scores (PGS). These scores, \( G_i \), are linear combinations of SNPs, where the weights (\( \beta_j \)) are obtained from a GWAS:
\[ G_i = \sum_{j} \beta_j X_{ij} \]
where \( X_{ij} \) denotes the minor allele count (0, 1, or 2) at SNP \( j \) for individual \( i \). This approach is motived by an additive genetic model where each SNP contributes additively to an individual's overall genetic predisposition \citep[see e.g.,][]{purcell2009prs}.

% GWASes and samples
We run a Genome-Wide Association Study (GWAS) for each of our outcomes to estimate the SNP effect sizes. To avoid overfitting when we later construct our PGSes, we run the GWASes on a hold-out sample consisting of individuals not in our analysis sample. That is, we partition the UK Biobank into two non-overlapping samples: a GWAS discovery sample, and the analysis sample. We use the former to run our GWASes and construct polygenic scores for the individuals in the latter that capture their genetic endowments with respect to each of our outcomes. The analysis sample is described in \autoref{sec:data}, while the GWAS discovery sample contains all individuals born prior to 1957, and thus not in the analysis sample. \autoref{tab:gwas-descriptives} reports descriptives statistics and sample sizes for the GWAS discovery samples. Our GWASes include controls for sex, genotyping array, first 20 genetic principal component, and cubic polynomials in year of birth by sex, allowing for sex-specific outcome dynamics over time. We quality control the genetic data following \citet{mitchell2019gwas}, removing genetic outliers and ensuring genotypes are well-measured. We use BOLT-LMM \citep{boltlmm} to run the GWASes, which is a linear mixed model approach that allows for some degree of relatedness among individuals and less restrictions on genetic ancestry (i.e., European ancestry instead of white British individuals only).

% Construction of PGSes
Instead of directly using the GWAS effect size to construct polygenic scores, we use them as inputs to LDPred2 \citep{prive2020ldpred2}, a bayesian method adjusting for linkage disequilibrium (SNP correlations). We assume an infinitesimal model, where all SNPs are causal, and consider only the subset of 1.6 million SNPs that are in Hapmap3 \citep{hapmap3}. We also filter the SNPs using a minor allele frequency threshold of $0.01$ and an info score threshold of $0.97$.

% PGS prediction checks
We standardise the polygenic scores to zero mean and unit standard deviation in the analysis sample. In \autoref{tab:individuals_assoc_pgi-prediction} we use our analysis sample to test the predictive power of the polygenic scores for their target outcomes. The regressions control for sex and the first 12 genetic principal, and we report the estimates together with the incremental $R^2$, defined as the increase in $R^2$ when the polygenic score is included as a covariate. Taken together, the results show that the polygenic scores are highly predictive of their respective outcomes, with incremental $R^2$ between 3\% and 16\%.

\begin{table}

\caption{\label{tab:gwas-descriptives}GWAS samples -- Descriptives}
\centering
\fontsize{10}{12}\selectfont
\begin{threeparttable}
\begin{tabular}[t]{lcdd}
\toprule
\multicolumn{1}{c}{} & \multicolumn{1}{c}{(1)} & \multicolumn{1}{c}{(2)} & \multicolumn{1}{c}{(3)} \\
\multicolumn{1}{c}{} & \multicolumn{1}{c}{Obs.} & \multicolumn{1}{c}{Mean} & \multicolumn{1}{c}{Std. dev.}\\
\midrule
Birth weight & 173,367 & 3.317 & 0.669\\
Adult height & 331,004 & 168.071 & 9.182\\
Educational attainment & 328,277 & 12.903 & 2.362\\
Fluid intelligence & 120,453 & 0.042 & 0.987\\
\bottomrule
\end{tabular}
\begin{tablenotes}
\item Columns: (1) Number of observations in the GWAS sample. (2) Sample mean in the GWAS sample. (3) Sample standard deviation in the GWAS sample. Fluid intelligence is standardised but during the GWAS routine a small number of individuals are discarded from the sample and this causes the mean and variance to differ slightly from zero and unity above. 
\end{tablenotes}
\end{threeparttable}
\end{table}

\begin{table}[!h]
\centering\centering\centering
\caption{\label{tab:individuals_assoc_pgi-prediction}OLS estimates -- Predictive power of polygenic scores for outcomes.}
\centering
\begin{threeparttable}
\fontsize{9}{11}\selectfont
\setlength{\tabcolsep}{1.5pt}
\begin{tabular}[t]{ldddd}
\toprule
\multicolumn{1}{c}{\em{}} & \multicolumn{4}{c}{\em{Dependent variable:}} \\
\cmidrule(l{3pt}r{3pt}){2-5}
\multicolumn{1}{c}{} & \multicolumn{1}{c}{(1)} & \multicolumn{1}{c}{(2)} & \multicolumn{1}{c}{(3)} & \multicolumn{1}{c}{(4)} \\
\multicolumn{1}{c}{ } & \multicolumn{1}{c}{{\specialcell[b]{Birth \\ weight}}} & \multicolumn{1}{c}{{\specialcell[b]{Adult \\ height}}} & \multicolumn{1}{c}{{\specialcell[b]{Educ. \\ attain.}}} & \multicolumn{1}{c}{{\specialcell[b]{Fluid \\ intelligence}}}\\
\midrule
PGS & 0.170^{ *** } & 0.402^{ *** } & 0.303^{ *** } & 0.600^{ *** }\\
 & (0.011) & (0.005) & (0.009) & (0.033)\\
\midrule
Observations & \multicolumn{1}{D{,}{,}{-3}}{8,204} & \multicolumn{1}{D{,}{,}{-3}}{11,494} & \multicolumn{1}{D{,}{,}{-3}}{11,439} & \multicolumn{1}{D{,}{,}{-3}}{3,993}\\
$R^2$ & \multicolumn{1}{d}{0.055} & \multicolumn{1}{d}{0.684} & \multicolumn{1}{d}{0.098} & \multicolumn{1}{d}{0.104}\\
Incremental $R^2$ & 0.029 & 0.158 & 0.090 & 0.076\\
\bottomrule
\end{tabular}
\begin{tablenotes}
\item Columns: (1) birth weight in kilograms, (2) adult height in centimeters, (3) years of education, (4) standardised fluid intelligence score. Controls for sex and the first 12 genetic principal components. Uses same sample as in main analysis, that is: Control group consists of never-treated individuals from both adopting and non-adopting county boroughs. Trims the sample to five years before and after the SCA submission date, and restricts the sample to birth cohorts in years 1958 to 1969. (*): $p < 0.1$, (**): $p<0.05$, (***): $p<0.01$.
\end{tablenotes}
\end{threeparttable}
\end{table}

\end{document}